\documentclass[%
 reprint,
 superscriptaddress,
 amsmath,amssymb,
 aps,
 pra,
 tightenlines,
 twocolumn,
]{revtex4-1}

\usepackage{graphicx}
\usepackage{dcolumn}
\usepackage{bm}

\begin{document}

\preprint{APS/123-QED}

\title{Magnetic structures, spin-flop transition and coupling of Eu and Mn magnetism in the Dirac semimetal EuMnBi$_2$}

\author{Fengfeng Zhu}
\email{f.zhu@fz-juelich.de}
\affiliation{J\"ulich Centre for Neutron Science (JCNS) at Heinz Maier-Leibnitz Zentrum (MLZ), Forschungszentrum J\"ulich, Lichtenbergstrasse 1, D-85747 Garching, Germany}
\affiliation{Department of Physics and Astronomy, Shanghai Jiao Tong University, Shanghai 200240, China}

\author{Xiao Wang}
\affiliation{J\"ulich Centre for Neutron Science (JCNS) at Heinz Maier-Leibnitz Zentrum (MLZ), Forschungszentrum J\"ulich, Lichtenbergstrasse 1, D-85747 Garching, Germany}

\author{Martin Meven}
\affiliation{J\"ulich Centre for Neutron Science (JCNS) at Heinz Maier-Leibnitz Zentrum (MLZ), Forschungszentrum J\"ulich, Lichtenbergstrasse 1, D-85747 Garching, Germany}
\affiliation{Institute of Crystallography, RWTH Aachen University, D-52056, Aachen, Germany}

\author{Junda Song}
\affiliation{J\"ulich Centre for Neutron Science (JCNS) at Heinz Maier-Leibnitz Zentrum (MLZ), Forschungszentrum J\"ulich, Lichtenbergstrasse 1, D-85747 Garching, Germany}
\affiliation{Hefei National Laboratory for Physical Sciences at Microscale, University of Science and Technology of China, Hefei, Anhui 230026, China}

\author{Thomas Mueller}
\affiliation{J\"ulich Centre for Neutron Science (JCNS) at Heinz Maier-Leibnitz Zentrum (MLZ), Forschungszentrum J\"ulich, Lichtenbergstrasse 1, D-85747 Garching, Germany}

\author{Changjiang Yi}
\affiliation{Institute of Physics and Beijing National Laboratory for Condensed Matter Physics, Chinese Academy of Sciences, Beijing 100190, China}

\author{Wenhai Ji}
\affiliation{J\"ulich Centre for Neutron Science JCNS and Peter Grünberg Institut PGI, JARA-FIT, Forschungszentrum J\"ulich, D-52425 J\"ulich, Germany}

\author{Youguo Shi}
\affiliation{Institute of Physics and Beijing National Laboratory for Condensed Matter Physics, Chinese Academy of Sciences, Beijing 100190, China}

\author{Jie Ma}
\affiliation{Department of Physics and Astronomy, Shanghai Jiao Tong University, Shanghai 200240, China}

\author{Karin Schmalzl}
\affiliation{J\"ulich Centre for Neutron Science (JCNS) at ILL, Forschungszentrum J\"ulich, F-38000 Grenoble, France}

\author{Wolfgang F. Schmidt}
\affiliation{J\"ulich Centre for Neutron Science (JCNS) at ILL, Forschungszentrum J\"ulich, F-38000 Grenoble, France}

\author{Yixi Su}
\email{y.su@fz-juelich.de}
\affiliation{J\"ulich Centre for Neutron Science (JCNS) at Heinz Maier-Leibnitz Zentrum (MLZ), Forschungszentrum J\"ulich, Lichtenbergstrasse 1, D-85747 Garching, Germany}

\author{Thomas Br\"uckel}
\affiliation{J\"ulich Centre for Neutron Science JCNS and Peter Grünberg Institut PGI, JARA-FIT, Forschungszentrum J\"ulich, D-52425 J\"ulich, Germany}
\date{\today}

\begin{abstract}
In recently emerging correlated topological materials, such as magnetic Dirac/Weyl semimetals, additional tunabilities of their novel transport and magnetic properties may be achieved by utilizing possible interaction between the exotic relativistic fermions and magnetic degree of freedom.
The two-dimensional antiferromagnetic (AFM) Dirac semimetal EuMnBi$_2$, in which an intricate interplay between multiple magnetic sublattices and Dirac fermions was suggested, provides an ideal platform to test this scenario.
We report here a comprehensive study of the AFM structures of the Eu and Mn magnetic sublattices as well as the interplay between Eu and Mn magnetism in this compound by using both polarized and non-polarized single-crystal neutron diffraction.
Magnetic susceptibility, specific heat capacity measurements and the temperature dependence of magnetic diffractions suggest that the AFM ordering temperature of the Eu and Mn moments is at 22 K and 337 K, respectively.
The magnetic moments of both Eu and Mn ions are oriented along the crystallographic $c$ axis, and the respective magnetic propagation vector is $\textbf{k}_{Eu} = (0,0,1)$ and $\textbf{k}_{Mn}=(0,0,0)$. With proper neutron absorption correction, the ordered moments are refined at 3 K as 7.7(1) $\mu_B$ and 4.1(1) $\mu_B$ for the Eu and Mn ions, respectively.
In addition, a spin-flop (SF) phase transition of the Eu moments in an applied magnetic field along the $c$ axis was confirmed to take place at a critical field of H$_c$ $\sim$ 5.3 T.
The antiferromagnetic exchange interaction and magnetic anisotropy parameters ($J=0.81$ meV, $K_u=0.18$ meV, $K_e=-0.11$ meV) are determined based on a subsequent quantitative analysis of the spin-flop transition.
The evolution of the Eu magnetic moment direction as a function of the applied magnetic field in the SF phase was also determined.
Clear kinks in both field and temperature dependence of the magnetic reflections ($\pm1$, 0, 1) of Mn were observed at the onset of the SF phase transition and the AFM order of the Eu moments, respectively.
This unambiguously indicates the existence of a strong coupling between Eu and Mn magnetism.
The interplay between two magnetic sublattices could bring new possibilities to tune Dirac fermions via changing magnetic structures by applied fields in this class of magnetic topological semimetals.

\end{abstract}

\maketitle


\section{\label{sec:introduction}Introduction}
Dirac/Weyl semimetals have attracted a great deal of recent research interests largely owing to their exotic quantum states and emergent phenomena as well as their high potentials for future technological applications \cite{Armitage2018}.
The linear dispersive electronic bands with gapless crossings near the Fermi level, that are protected by topology or symmetries, can be described as massless relativistic quasi-particles Dirac or Weyl fermions, which can give rise to novel transport behaviors such as high carrier mobility, immunity to disorder, ballistic electronic transport and quantum Hall effect \cite{Hasan2010,Pesin2010,Rau2016}.

Furthermore, a potential coupling of Dirac/Weyl fermions to other degrees of freedom such as magnetism \cite{Sapkota2020,Masuda2016} may open up a new avenue for the exploration and tuning of novel physical properties.
Recently, particular attention has been focused on magnetic Dirac/Weyl materials, in which it is possible to tune the electronic transport properties by utilizing the interaction between the relativistic quasi-particles and magnetism \cite{Smejkal2017,Smejkal2018,Guo2014,Zhang2016,Zhang2019}.
A few candidates of magnetic Dirac/Weyl materials have already been theoretically proposed or experimentally verified, like Co$_3$Sn$_2$S$_2$ \cite{Xu2018a,Liu2018d}, MnBi$_2$Te$_4$ \cite{Li2019c,Zhang2019d,Gong2019}, and the layered manganese pnictides AMnBi$_2$ (A = Rare/Alkaline earth) ``112'' system \cite{Park2011,Lee2013,Farhan2014,Wang2011,Wang2012,Borisenko2015a}.
Among them, the experimental evidence for the coexistence of Dirac fermions and AFM order was found in AMnBi$_2$ compounds by different methods \cite{Wang2012,He2012,Wang2012b,Feng2015,Zhang2016,Li2016,Wang2016c,Liu2017b,Masuda2016,Park2016,Liu2017,Zhang2019,Liu2016,Huang2017,Zhu2019,Kealhofer2018}, such as quantum oscillation, magneto-resistant behavior, angle-resolved photon emission spectroscopy, and optical conductivity etc.
In addition, due to coupling of the magnetic layer and Bi square-net layer with Dirac fermions, a strong influence of magnetic order on electronic transport properties was found \cite{Guo2014,Zhang2019,Masuda2016,Masuda2018,Rahn2017}, and Dirac fermions were also reported to enhance the exchange coupling between magnetic moments in AMnBi$_2$ \cite{Zhang2016}.
EuMnBi$_2$ and YbMnBi$_2$ were recently discovered as two of the possible candidates, especially YbMnBi$_2$ whose magnetic structures and excitations were studied by neutron scattering in previous works \cite{Soh2019b,Sapkota2020}, shows significant coupling of Dirac bands with spins \cite{Sapkota2020}.
By spontaneous or externally induced time-reversal symmetry breaking, EuMnBi$_2$ and YbMnBi$_2$ could also be driven to host Weyl physics.
Such magnetic Dirac materials where the magnetic and conducting layers are coupled but separated spatially provide an ideal platform to study the interplay between magnetic moments and Dirac carriers, which may find promising application potential in spintronics devices.

In the case of EuMnBi$_2$, novel physical properties like the half-integer Quantum Hall effect \cite{Masuda2016} and the magnetopiezoelectric effect \cite{Shiomi2018a} were recently observed, and in addition to its interesting transport properties, the occurrences of giant magneto-resistance effects and quantum oscillations would suggest an important role of the magnetic order of the Eu sublattice \cite{May2014,Masuda2016,Masuda2018}.
Furthermore, the Eu moments were also suggested to have an apparent interplay with the Mn moments \cite{May2014}, which might explain why the orientation of the Eu moments in EuMnBi$_2$ is reported to be different from that in EuZnBi$_2$ \cite{Masuda2016}, and the ordering temperature of Mn moments in EuMnBi$_2$ is much higher than that of SrMnBi$_2$ with the same crystal structure.
Given the large energy scale of exchange interactions, it would be very hard to tune the magnetic order of Mn ions via applied magnetic fields, unless using extremely high fields.
Nevertheless, it was found that the magnetic order of Eu ions in EuMnBi$_2$ is actually field-tunable and a spin-flop transition occurs at the applied field $H = 5.3$ T along the $c$ axis \cite{May2014,Masuda2016,Masuda2018}.
This thus suggests that such an interplay between different magnetic sublattices in this class of magnetic Dirac materials could be used to tune their intrinsic magnetic structures under moderate magnetic fields, subsequently, to impact their electronic behaviors related to Dirac fermions.
In this regard, EuMnBi$_2$ provides an ideal platform to experimentally examine possible intricate interplay between multiple magnetic sublattices, and to test the scenario of possible tuning of Dirac fermion behaviors via the magnetic degrees of freedom.
Although the magnetic structure of the Eu sublattice has already been studied by resonant X-ray magnetic scattering and neutron diffraction, a comprehensive study of the magnetic structures of both Eu and Mn sublattcies as well as a possible interplay between Eu and Mn magnetism has not been reported so far.

In this work, we present detailed neutron scattering studies of magnetic structures, field-induced spin-flop transition and the interplay between Eu and Mn magnetism in the Dirac material EuMnBi$_2$.
From polarized neutron diffractions, we have confirmed the ordered magnetic moment orientation of the Mn and Eu sublattices, and the existence of the interplay between the Eu and Mn magnetic moments based on the temperature dependence measurements.
After proper correction for the Eu neutron absorption, we have further determined the magnetic structure and ordered magnetic moment size for both the Eu and Mn sublattices by using hot-neutron single crystal diffraction.
For the spin-flop states, we have studied the field dependence of the magnetic structure of the Eu sublattice and determined the evolution process of the Eu moment direction with the applied field along the $c$ axis.
Moreover, based on a quantitative analysis of the observed spin-flop transition in our neutron diffraction study, we have determined the exchange interaction and magnetic anisotropy parameters for the Eu sublattice.
We thus propose an anisotropic XXZ spin Hamiltonian model, that includes a dominant isotropic antiferromagnetic exchange interaction with a small planar exchange anisotropy as well as a small uniaxial single-ion anisotropy, for the Eu sublattice in EuMnBi$_2$.

\section{\label{sec:experiment}Experimental Details}

\begin{figure}
\centering
\includegraphics[width=8cm]{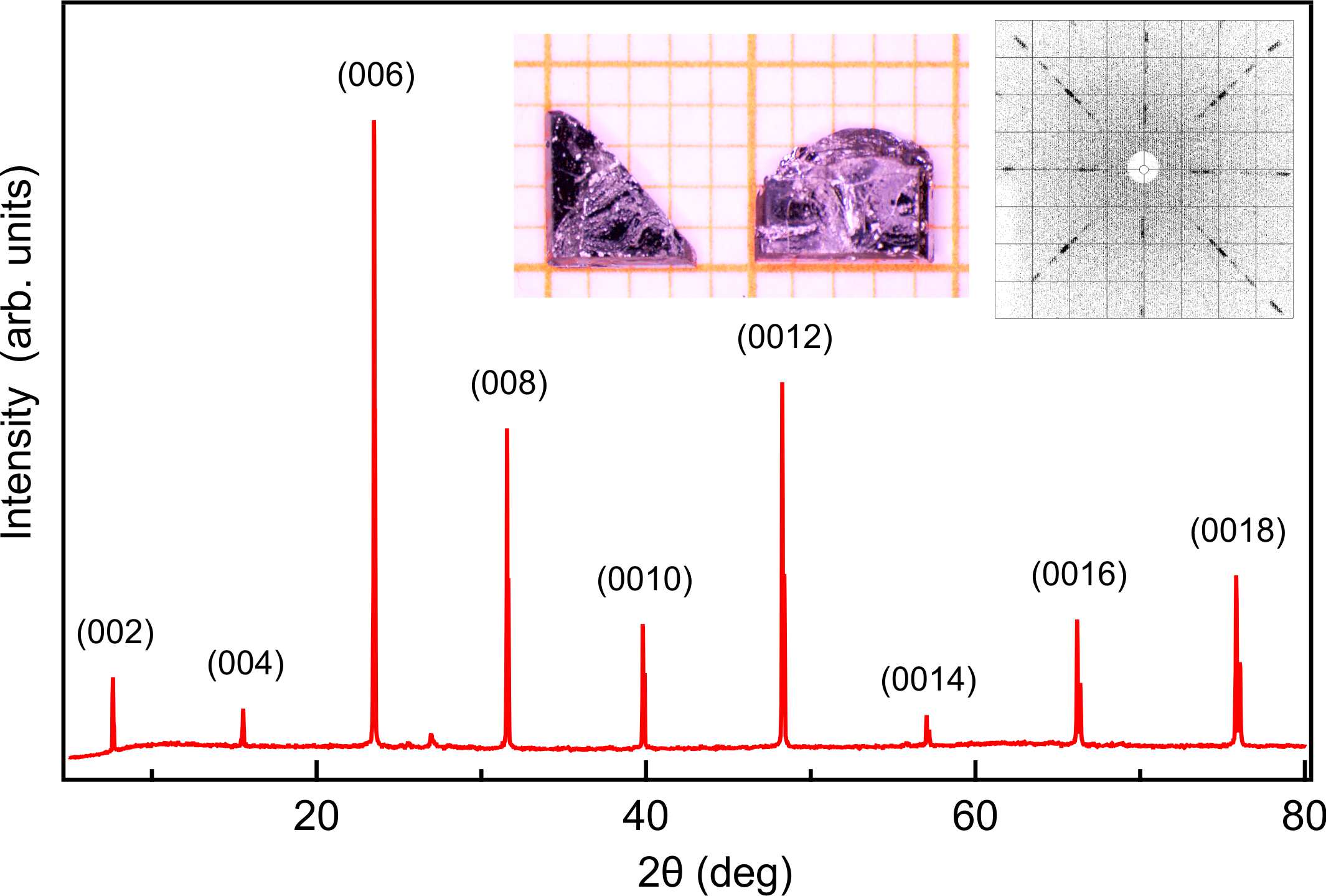}
\hspace{0.5cm}
\caption{\label{fig:Figs_XRD}
X-ray diffraction of single-crystal EuMnBi$_2$ at 300 K.
XRD pattern shows sharp (0,0,L) peaks.
The left inset is a photograph of single-crystal samples of EuMnBi$_2$, showing a typical size of $\sim$5 mm with a thickness of $\sim$1.5 mm (grid width is 1 mm) and clear rectangular natural edge;
the right inset is an X-Ray Laue pattern of the (H,K,0) reciprocal plane, and a 4-fold symmetry can be clearly seen.
}
\end{figure}

Single crystals of EuMnBi$_2$ were grown by the flux method using bismuth as self-flux.
The starting materials of Eu, Mn, and Bi were mixed in an Ar-filled glove box at a molar ratio of Eu : Mn : Bi = 1 : 1 : 10.
The mixture was placed in an alumina crucible, which was then sealed in an evacuated quartz tube.
The tube was heated up to 1000 $^{\circ}$C over 10 h and then dwelt for 20 h.
Afterwards, the tube was slowly cooled down to 600 $^{\circ}$C with a cooling speed of 2.5 $^{\circ}$C/h followed by centrifuging to separate  crystals from the Bi flux.
Shiny plate-like crystals with a typical dimension of $5\times5\times1$ mm were obtained.

Single-crystal X-ray diffraction (XRD) was performed at room temperature with an incident wavelength of 1.54 \AA\ ($Cu$-$K_\alpha$) on a Bruker D2 Phaser X-ray diffractometer.
The neutron scattering data presented in this paper were collected at the Heinz Maier-Leibnitz Zentrum(MLZ) in Garching, Germany and the ILL in Grenoble, France.
The single-crystal neutron diffraction experiment was performed at the hot-neutron 4-circle diffractometer HEIDI \cite{Meven2015} (with incident wave length $\lambda_i$ = 0.795 \AA), the polarized neutron diffraction measurement was carried out on the cold-neutron polarized spectrometer DNS \cite{Schweika2001,Su2015} (with $\lambda_i$ = 4.2 \AA), and the field dependence study was carried out at the lifting-counter thermal-neutron diffractometer D23 with a 12 T vertical-field magnet (and $\lambda_i$ = 1.2735 \AA).
By combining a wide range of polarized and non-polarized neutron diffraction techniques, the temperature, neutron polarization (non spin-flip and spin-flip) and magnetic-field dependences of both nuclear and magnetic reflections could be thoroughly investigated.
In particular, given the significantly reduced neutron absorption and extinction effects with hot neutrons, it became possible to obtain high-quality and reliable structure factors for the refinements of both nuclear and magnetic structures in materials containing strong neutron-absorbing elements such as Eu in EuMnBi$_2$.

A few selected single crystals were also used for measuring the specific heat capacity and magnetic properties by PPMS and SQUID (from Quantum Design).
Magnetic susceptibility was measured from 2 K to 350 K in various applied magnetic fields with both Zero-Field-Cooling (ZFC) and Field-Cooling (FC) conditions.
The isothermal magnetization ($M-H$) curves were measured in a sweeping field from $-$50 to 50 kOe at 2 K and 300 K, respectively.

\section{\label{sec:Results}Results}

\subsection{\label{sec:crystal}X-Ray Diffraction and Magnetic Properties}

\begin{figure}
\centering
\includegraphics[width=8cm]{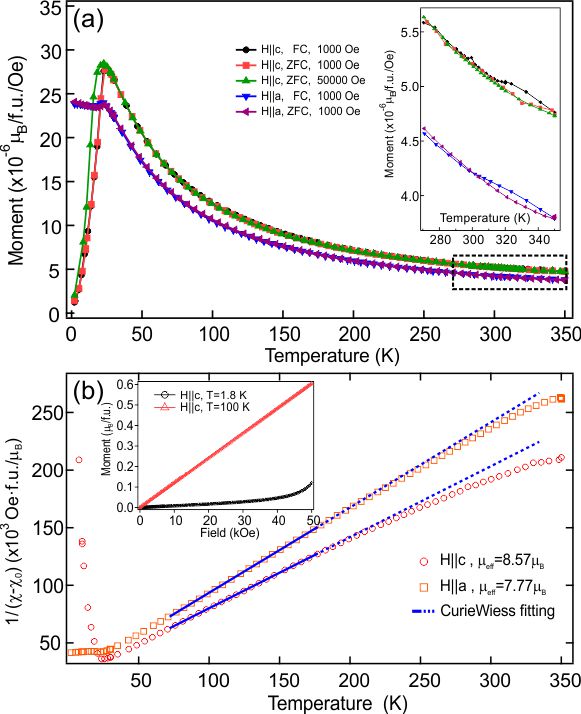}
\caption{\label{fig:Figs_ZFCFC_MH}
(a) ZFC/FC magnetization curves measured under applied magnetic fields along the $a$ and $c$ axes.
Inset shows the zoom-in plot of the high temperature range.
(b) The inverse susceptibility of the ZFC curves with Curie-Weiss fitting around the selected temperature range.
Dashed lines are the extension of the linear fitting.
}
\end{figure}

The crystalline quality and structure were checked by XRD and X-Ray Laue (Fig.\ref{fig:Figs_XRD}).
EuMnBi$_2$ shares the same structure as SrMnBi$_2$ with space group $I4/mmm$ (No. 139).
The lattice parameter $c$ = 22.614(5) \AA\ at room temperature extracted from XRD (in Fig.\ref{fig:Figs_XRD}) is quite consistent with the previous results \cite{May2014}.
It is worth noting that the small peak around 2$\theta\sim27^\circ$ in Fig.\ref{fig:Figs_XRD} is due to the residual bismuth flux on the surface.

The temperature dependence of magnetic properties (shown in Fig.\ref{fig:Figs_ZFCFC_MH}) was measured along both the $a$ and $c$ axes respectively.
As shown in the magnetic susceptibility data with applied fields $H$ = 0.1 and 5 T in Fig.\ref{fig:Figs_ZFCFC_MH}(a), the onset of AFM ordering of Eu magnetic moments can be clearly observed at around 22 K.
Below the ordering temperature, the susceptibility shows clear magnetic anisotropy, suggesting that Eu moments are more inclined along the $c$ axis.
There is no clear indication for the onset of the AFM ordering of Mn moments in the susceptibility. Nonetheless, as shown in the inset of Fig.\ref{fig:Figs_ZFCFC_MH}(a), a FC/ZFC bifurcation point could still be seen at about 337 K, which could be regarded as the signature of the AFM ordering of Mn moments.
The magnetization curves show typical Curie-Weiss behavior above $T_N^{Eu}$ = 22 K, as shown in Fig.\ref{fig:Figs_ZFCFC_MH}(b).
In order to obtain a relatively accurate effective moment size of Eu, an appropriate temperature range (70 K - 180 K) was selected for the fitting of the susceptibility.
The data were fitted to $\chi=\chi_0+\frac{C}{T-T_c}$, where $C$ is the Curie constant, $T_c$ represents the Weiss temperature and $\chi_0$ accounts for the temperature independent contributions.
The effective moments of Eu$^{2+}$ obtained in a usual manner ($\rm \mu_{eff}=\sqrt{\frac{3 k_B\cdot C}{n\mu_B}}$) are 7.77 $\rm \mu_B$ (H$\parallel$a) and 8.57 $\rm \mu_B$ (H$\parallel$c) respectively, which are quite reasonable given that the theoretical value is 7.94 $\rm \mu_B$ for Eu$^{2+}$ ($4f^7$).
The obtained effective moments along the $c$ axis (easy axis here) is a little bit larger than the theoretical one, and similar results were also seen in the previously reported works (e.g. EuMnSb$_2$: 8.0 $\rm \mu_B$ \cite{Yi2017}; EuMnBi$_2$: 8.1 $\rm \mu_B$ \cite{May2014}).
One possible explanation for this is a non-negligible Curie-Weiss contribution from the Mn moments since they may not be completely saturated in this temperature regime.

The inset in Fig.\ref{fig:Figs_ZFCFC_MH}(b) shows field dependence of the magnetic moment in EuMnBi$_2$ above and below the AFM transition temperature of Eu.
The magnetization at 1.8 K shows an accelerating upward change around $\sim$4.8 T, which is consistent with the spin-flop transition of EuMnBi$_2$ \cite{May2014,Masuda2016} under applied fields along the $c$ axis.

\subsection{\label{sec:heatCapacity}Specific Heat Capacity}

\begin{figure}
\centering
\includegraphics[width=8cm]{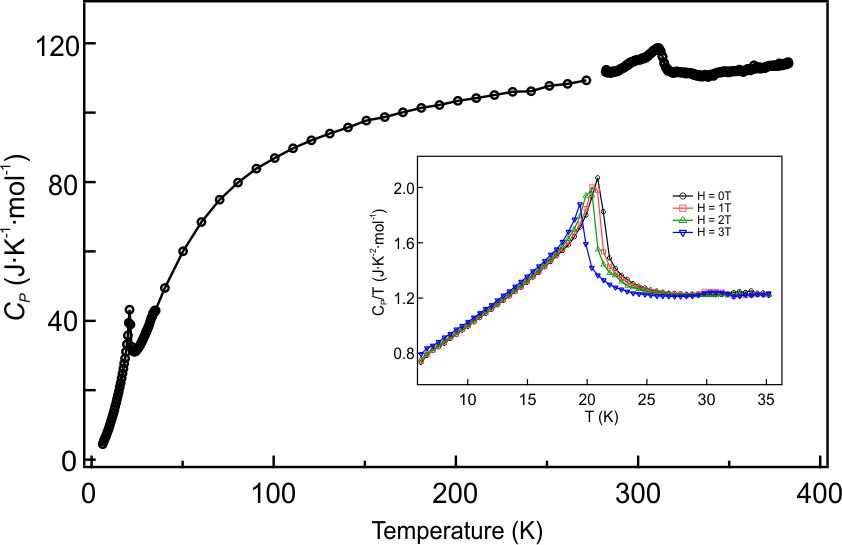}
\caption{\label{fig:HC}
The specific heat capacity ($C_p$ vs $T$ curve) of EuMnBi$_2$ single crystal.
The inset shows the field dependence of the anomaly peak, which indicate a AFM phase transition of the Eu moments.
}
\end{figure}

The specific heat capacity $C_p$ of EuMnBi$_2$ measured over a range of 2 K to 350 K shows two clear anomalies, as shown in Fig.\ref{fig:HC}.
A distinct and strong anomaly is observed near 22 K, which is believed to be associated with the ordering of the Eu magnetic moments.
As expected, the application of a magnetic field has a slight influence on this heat capacity anomaly.
The peak position was shifted to lower temperatures by increasing applied field.
Another anomaly in $C_p$ associated with the AFM ordering of the Mn moments is the observed maximum near 315 K.
This anomaly is quite broad in temperature, as previously reported \cite{May2014}.
In SrMnBi$_2$, that anomaly associated with the Mn ordering has been reported to be around 290 K and be also rather weak and broad \cite{Wang2011,May2014}.
This kind of broad peak may indicate the presence of possible short-range ordering of the Mn moments near the magnetic ordering temperature.
Compared to SrMnBi$_2$, the $\rm N\acute{e}el$ temperature of the Mn moments ordering in EuMnBi$_2$ is much higher, which is presumably due to the enhanced exchange interaction between the Mn moments.

\subsection{\label{sec:PolarizedNeutron}Polarized Neutron Diffraction}

\begin{figure}
\centering
\includegraphics[width=8cm]{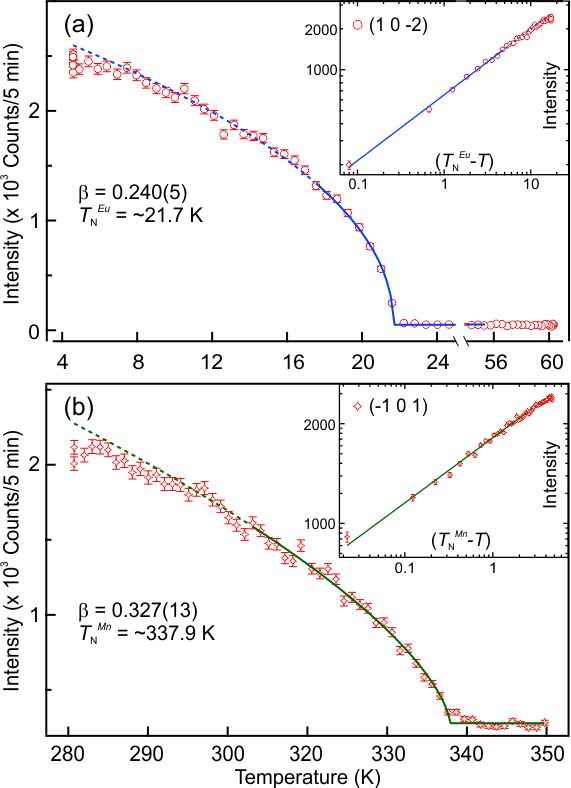}
\caption{\label{fig:Figs_DNSMT}
(a) Temperature dependence of the (1,0,-2) magnetic Bragg peak intensity of Eu.
(b) Temperature dependence of the (-1,0,1) magnetic Bragg peak intensity of Mn.
The empty circles and stars with error bar are experimental data.
The solid lines are the fittings of the experimental data by the formula $I = I_0 +A(1-T/T_N)^{2\beta}$.
The insets are the corresponding log scale plots of the peak intensity versus the reduced temperature $T_N-T$ where $T_N$ is the critical temperature for Eu and Mn respectively.
The slop represents the power parameter $2\beta$.
}
\end{figure}

\begin{figure}
\centering
\includegraphics[width=8cm]{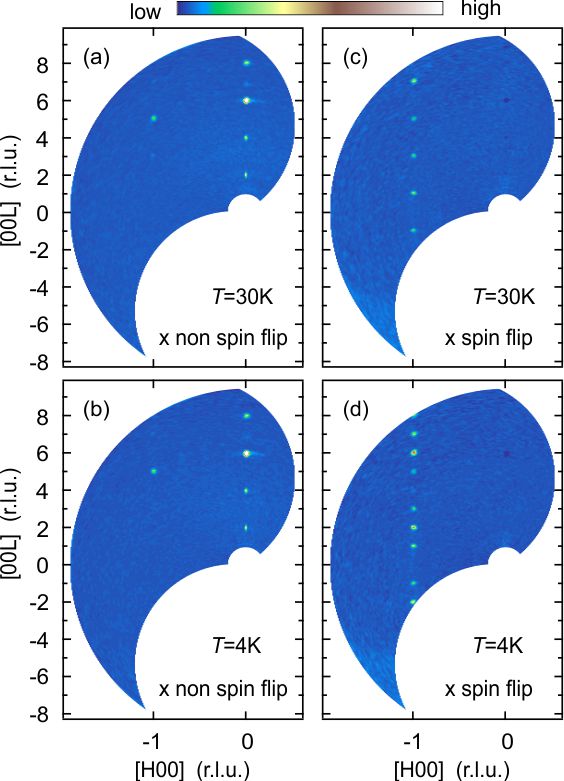}
\caption{\label{fig:Figs_DNSPlot}
Polarized neutron diffraction patterns of single crystal EuMnBi$_2$.
(a),(b) nuclear reflections in the (H,0,L) plane at 30 K and 4 K.
(c) magnetic reflections of Mn magnetic sublattice at 30 K ($T_N^{Eu}<$30 K$<T_N^{Mn}$).
(d) magnetic reflections of both Mn and Eu sublattice at 4 K (4 K$<T_N^{Eu}<T_N^{Mn}$).
}
\end{figure}

Polarized neutron scattering was performed at DNS with (H,0,L) as the horizontal scattering plane, Fig.\ref{fig:Figs_DNSMT} shows the temperature dependence of the intensities for two selected magnetic reflections (1,0,-2) and (-1,0,1) at the $x$ spin flip channel.
It needs to be mentioned that the $x$ polarization direction is along the average direction of the scattering vectors \textbf{Q} for all the detectors in the detector bank at DNS,
and the corresponding $y$ polarization direction is perpendicular to $x$ in the horizontal scattering plane, thus the $z$ polarization direction is vertical, i.e. perpendicular to both $x$ and $y$.
Two magnetic phase transitions can be seen clearly in different temperature ranges.
To obtain a reliable transition temperature and critical exponents, the temperature dependence curves were fitted with the power law equation $I = I_0 +A(1-T/T_N)^{2\beta}$ in a range of about $\pm10\%$ of $T_N$ near those transitions.
The fitted transition temperatures are $\sim$22 K and $\sim$337 K, which are in a good agreement with our heat capacity and magnetization results.
The fitted critical exponents are $\beta$ = 0.240(5) for Eu and $\beta$ = 0.327(13) for Mn, resulting as the linear slopes in the inserts of Fig.\ref{fig:Figs_DNSMT}(a) and (b).
The critical exponent of Mn is close to the classical three-dimensional Ising model ($\beta$ = 0.326).
As for Eu, the critical exponent value is just located between two-dimensional ($\beta$ = 0.125) and three-dimensional Ising model.
However, the power-law refinement holds over an unusually wide temperature range for Eu, down to $\sim$7 K until the intensity tends to saturate, well outside the usual critical region.

To gain further information of the magnetic moment orientations, two-dimensional (2D) Q-scans in the (H,0,L) planes of reciprocal space at the spin and non-spin flip modes along different neutron polarization directions were performed at 30 K and 4 K, respectively.
As shown in Fig.\ref{fig:Figs_DNSPlot}(a) and (b),  the diffraction pattern of $x$ non-spin flip stays the same and no clear temperature dependence is observed, which indicates that all reflections in $x$ non-spin flip are only related to the nuclear structure.
The scattering signal in $x$ spin flip shown in Fig.\ref{fig:Figs_DNSPlot}(c) and (d) has purely magnetic contributions.
At 30 K, only the Mn moments are ordered, therefore Fig.\ref{fig:Figs_DNSPlot}(c) basically shows pure magnetic reflections of Mn moments; but at 4 K, both Eu and Mn moments are ordered and they both contribute to the magnetic diffraction intensities, so new reflections appear in Fig.\ref{fig:Figs_DNSPlot}(d) associated with the magnetic scattering of Eu moments and it also indicates that the magnetic sublattices of Eu and Mn have different magnetic structures.
For $z$ polarization (shown in Fig.\ref{fig:fig_polar} in Appendix A), the 2D Q-scans are basically the same as those in the $x$ direction, which suggest that the magnetic moments have a net projection component in horizontal $XY$ scattering plane which is (H,0,L).
The sign of polarized neutron can only be flipped when there exists a non-zero component of magnetic moments perpendicular to the polarization of the neutron beam $\textbf{P}$ and the scattering wave vector $\textbf{Q}$.
By comparing the nuclear and magnetic diffraction patterns, the absence of magnetic (0,0,L) reflections and occurrence of (-1,0,L) reflections is an indication for all the moments along the $c$ axis, which means that there is no projected magnetic moment of Eu or Mn in the $ab$ plane,
otherwise any magnetic moment perpendicular to Q = (0,0,L) would contribute to the $x$ and $z$ spin flip channels like the (0,0,2n) nuclear reflections in Fig.\ref{fig:Figs_DNSPlot}(a),(b) which are not observed in Fig.\ref{fig:Figs_DNSPlot}(c),(d).

Last but not least, all the magnetic reflections lie on the positions of the Brillouin zone center.
This indicates that the magnetic unit cell is the same as the nuclear unit cell, so the magnetic propagation vectors $\textbf{k}$ could be (0,0,0) or (\textit{h,k,l}), where \textit{h, k, l} are integers.
Based on their different temperature dependences, the magnetic propagation vectors of Mn and Eu moments are identified as (0,0,0) and (0,0,1) respectively.

As we know, it is usually very difficult to tune the Mn moments directly by simply applying magnetic field, to overcome the exchange interaction between Mn moments, it may need an extreme high magnetic field due to its high AFM phase transition temperature ($T_N^{Mn}$ = $\sim$337 K).
In EuMnBi$_2$ system, if the exchange interaction between the Eu and Mn moments is strong enough, it would be possible to mediate the magnetic moments of Mn indirectly by tuning the Eu moments with a moderate applied magnetic field.
So, it would be interesting and worth to study the exchange interaction between Eu and Mn magnetic sublattices by applying magnetic fields or partially replace some Mn atoms with non-magnetic ions in the future.

\subsection{\label{sec:HotNeutron}Single Crystal Diffraction with Hot Neutrons}

Having determined the magnetic propagation vectors as well as the moment directions of Eu and Mn sublattices via polarized neutron diffraction, hot-neutron single crystal diffraction measurements were performed to comprehensively determine both the crystalline and magnetic structures of EuMnBi$_2$ at HEIDI.
Since EuMnBi$_2$ has two magnetic phase transitions at $T_N^{Eu}\sim$22 K for Eu$^{2+}$ and $T_N^{Mn}\sim$337 K for Mn$^{2+}$, we measured about 1700 nuclear and magnetic reflections allowed by the symmetry of space group $I4/mmm$ at 3 K and 300 K respectively.
Given the strong neutron absorption of Eu in this material, a finite element analysis method was used for neutron absorption correction.
A few reasonable approximations were used for the convenience of calculation:
the absorption of one reflection is based on its integrated intensities and center omega angle setup instead of each scanning point intensity in its whole rocking curve;
The effective neutron beam flux incident on the sample are approximately treated as a constant for all reflections (further details in Appendix.\ref{app:Absorption}).
After a proper neutron absorption correction, the corrected structure factor data were refined using Jana2006 \cite{Petricek2014a}, and the irreducible representations of possible magnetic structure models were analyzed by \emph{MAXMAGN} \cite{Perez-Mato2015} from the \emph{Bilbao Crystallographic Server} \cite{Aroyo2011,Aroyo2006,Aroyo2006a}.
The nuclear structure used here was based on the previously reported works \cite{May2014} and the parameters established from our own XRD results.

\begin{figure}
\centering
\includegraphics[width=8cm]{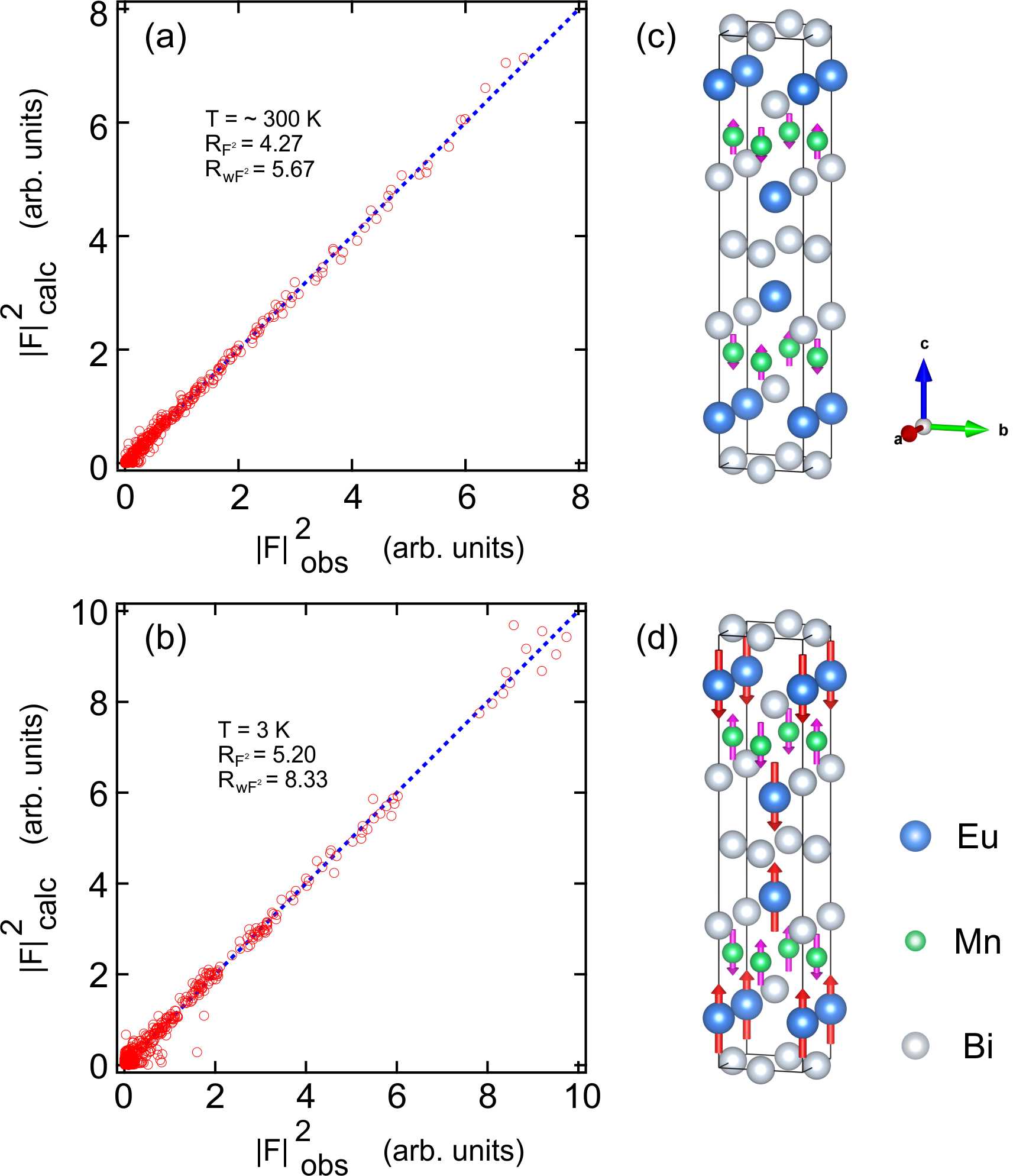}
\caption{\label{fig:Figs_HeidiRefine}
(a),(b) Integrated intensities of the Bragg reflections collected at room temperature 300 K and low temperature 3 K are plotted against the calculated values, respectively.
(c),(d) are the corresponding magnetic structure models generated by VESTA \cite{Momma2011}.
}
\end{figure}

At 300 K, only Mn$^{2+}$ moments are ordered and the propagation vector was determined to be $\textbf{k}_{Mn}$ = (0,0,0) for the Mn magnetic sublattice.
There are 12 possible maximal magnetic space groups for the parent space group $I4/mmm$ (No. 139) with the propagation vector $\textbf{k}$ = (0,0,0), and only 6 subgroups ($I4/mm'm'$, $I4'/m'm'm$, $Im'm'm$, $Im'mm$, $Fm'm'm$, $Fm'mm$) which allow non-zero magnetic moments.
Since the orientation of the Mn$^{2+}$ moments was confirmed by polarized neutron scattering and its AFM magnetic properties were also confirmed by its magnetization basically, there is only one subgroup $I4'/m'm'm$ (AFM) possible for the magnetic structure of Mn$^{2+}$.
The integrated intensities of 1375 nuclear reflections (673 unique) were collected and then could be refined very well by combining the nuclear structure and G-type (magnetic space group: $I4'/m'm'm$) AFM structure of the Mn$^{2+}$ moments.
As shown in Fig.\ref{fig:Figs_HeidiRefine}(a), the calculated intensities are quite linear with the observed intensities, the weighted R-factor of the refinement is 5.67$\%$.
Fig.\ref{fig:Figs_HeidiRefine}(c) shows the corresponding magnetic structure of the Mn moments, and the refined ordered moment size for Mn$^{2+}$ is 2.1(1) $\rm \mu_B$ at $T$ = 300 K (Table.\ref{tab:atom}).

\begin{table}[htbp]
    \caption[EuMnBi$_2$ refinement results]{\label{tab:atom}
    Refined results for the nuclear and magnetic structures of EuMnBi$_2$ at 300 K and 3 K.
    $k$ in the table represents the magnetic propagation vector.
    All magnetic structures are based on the nuclear space group $I4/mmm$.
    }
    \begin{center}
       \begin{tabular}{ccccccccc}
       \hline
       \hline
       \multicolumn{9}{c}{\textrm{T = 300 K}}\\
       \hline
       &Atom&Site&$x$&$y$&$z$&U$_{iso}$&$\textbf{k}$& \\
       \hline
       &Eu&$4e$&0&0&0.11482(8)&0.01040(41)&-& \\
       &Mn&$4d$&0&0.5&0.25&0.01391(64)&(0,0,0)& \\
       &Bi(1)&$4c$&0&0.5&0&0.01135(23)&-& \\
       &Bi(2)&$4e$&0&0&0.32855(5)&0.01135(23)&-& \\
       \hline
       &\multicolumn{8}{l}{\textrm{$a$ = 4.535(5) \AA, $c$ = 22.6110(5) \AA}}\\
       &\multicolumn{8}{l}{\textrm{$R_{F^2}$ = 4.27, $R_{wF^2}$ = 5.67}}\\
       &\multicolumn{8}{l}{\textrm{$M_{Mn}$ = 2.1(1) $\rm \mu_B$}}\\
       \hline
       \hline
       \multicolumn{9}{c}{\textrm{T = 3 K}}\\
       \hline
       &Atom&Site&$x$&$y$&$z$&U$_{iso}$&$\textbf{k}$& \\
       \hline
       &Eu&$4e$&0&0&0.1149(7)&0.00198(37)&(0,0,1)& \\
       &Mn&$4d$&0&0.5&0.25&0.00323(54)&(0,0,0)& \\
       &Bi(1)&$4c$&0&0.5&0&0.00152(24)&-& \\
       &Bi(2)&$4e$&0&0&0.3289(33)&0.00152(24)&-& \\
       \hline
       &\multicolumn{8}{l}{\textrm{$a$ = 4.512(3) \AA, $c$ = 22.23(13) \AA}}\\
       &\multicolumn{8}{l}{\textrm{$R_{F^2}$ = 5.20, $R_{wF^2}$ = 8.33}}\\
       &\multicolumn{8}{l}{\textrm{$M_{Mn}$ = 4.1(1) $\rm \mu_B$, $M_{Eu}$ = 7.7(1) $\rm \mu_B$}}\\
       \hline
       \hline
\end{tabular}
    \end{center}
\end{table}

A similar analysis was performed for the data taken at 3 K, where the Eu$^2+$ moments are ordered in addition to the same magnetic order of Mn.
The additional magnetic reflections in Fig.\ref{fig:Figs_DNSPlot}(d) indicate that the Eu$^{2+}$ magnetic order has a different magnetic propagation vector from that of Mn$^{2+}$, $\textbf{k}_{Eu}$ = (0,0,1).
By doing the same magnetic symmetry analysis as Mn$^{2+}$, there are also 6 maximal subgroups ($P_I4/mnc$, $P_I4/nnc$, $C_Amca$, $C_Amcm$, $P_Immn$, $P_Innm$) which allow non-zero magnetic moments in the total 12 possible maximal magnetic space groups for the parent space group $I4/mmm$ (No. 139) with the propagation vector $\textbf{k}=(0,0,1)$.
As for the Mn$^{2+}$ the Eu$^{2+}$ moments are also by polarization analysis found to be oriented along the $c$ axis, allowing only two magnetic space groups $P_I4/mnc$ and $P_I4/nnc$.
The two possible structures were refined separately using 1717 reflections (622 unique).
We found that the refinement result is significantly better and more reliable by using magnetic space group $P_I4/nnc$.
$P_I4/mnc$ is excluded for its unreasonable refined values for both the Eu$^{2+}$ and Mn$^{2+}$ moments size.
As Fig.\ref{fig:Figs_HeidiRefine}(b) shows, the integrated intensities $I_{obs}$ of almost all of the reflections have a nice linear behavior with the calculated intensities $I_{calc}$, the refined weighted R-factor is 8.33$\%$ which is still acceptable by considering the additional errors induced during the absorption correction process.
Fig.\ref{fig:Figs_HeidiRefine}(d) shows the corresponding magnetic structure of Eu and Mn sublattices which is consistent with the previously reported X-ray results \cite{Masuda2016}, and the final refined ordered magnetic moment at 3 K is 4.1(1) $\rm \mu_B$ for Mn$^{2+}$ and 7.7(1) $\rm \mu_B$ for Eu$^{2+}$.
All the refined parameters are shown in Table.\ref{tab:atom}, the lattice parameters at 3 K are a little bit smaller than that at 300 K.
As expected, the moment size of Eu is close to the theoretical saturated value of the isolated atoms $M_{Eu^{2+}}=g_Jm_J\mu_B=7\mu_B$, ($m_J$ = 7/2, $g_J$ = 2) and the average value extracted from the effective moment size determined in magnetization measurements ($M_{Eu^{2+}}=\sqrt{\frac{J}{J+1}}(\frac{\mu_{eff}^{\parallel a}+\mu_{eff}^{\parallel c}}{2})\approx8.2\mu_B$) indicating the electrons responsible for moments of Eu$^{2+}$ are quite local.
As for Mn, the moment size is about 20\% smaller than the full moment of the isolated Mn atoms, which may be caused by the hybridization between the localized $3d$ electrons of Mn$^{2+}$ and itinerant $6p$ electrons from the valence band of Bi.

\subsection{\label{sec:Tdep_Bdep}Spin-Flop Transition and Magnetic Anisotropy}

\begin{figure}
\centering
\includegraphics[width=8cm]{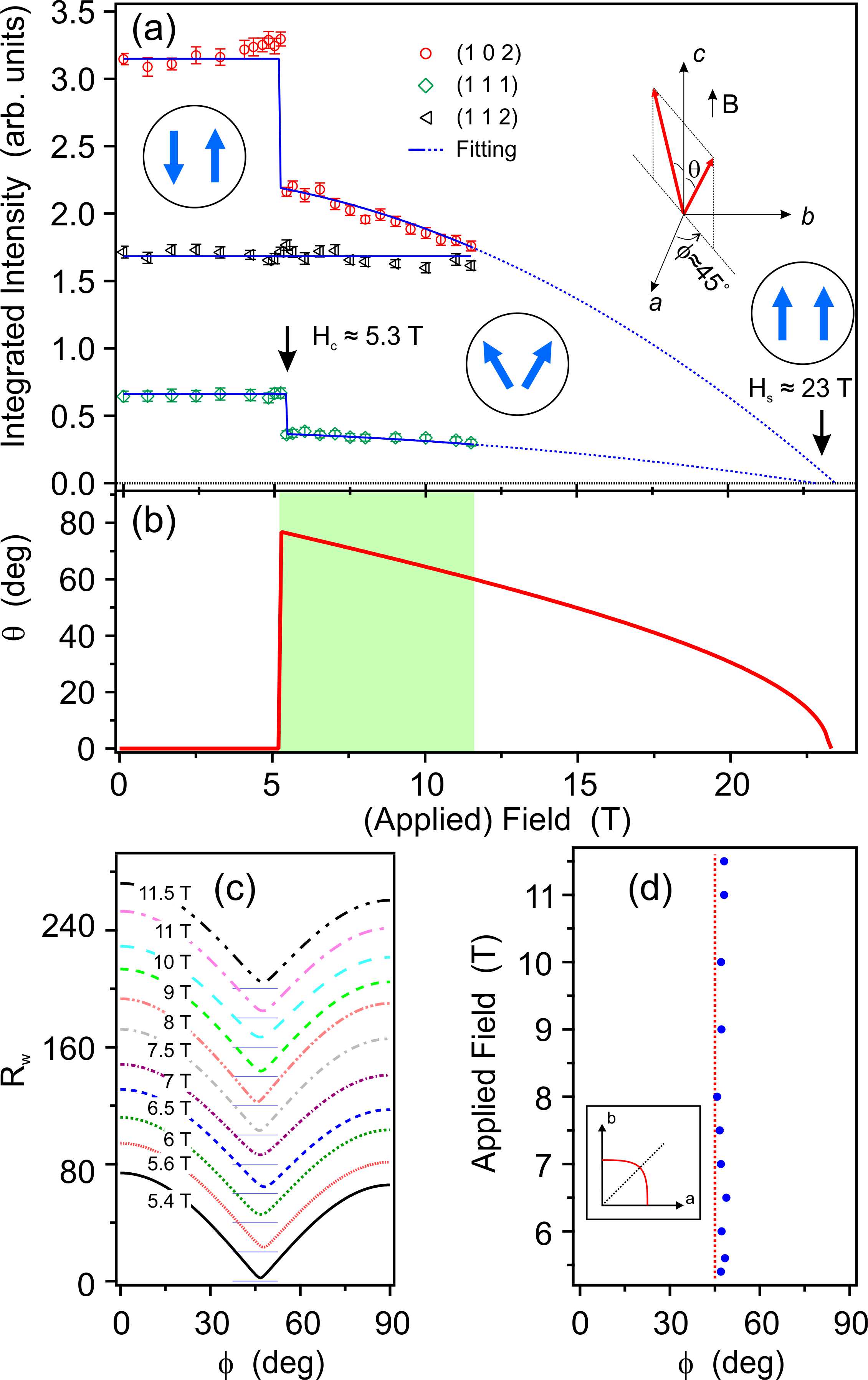}
\caption{\label{fig:Figs_spinflop}
(a) Field dependence of the integrated intensities of the selected nuclear and magnetic (contributed by Eu) reflections taken at 1.5 K.
Blue lines are the fitting results.
The insets are the schematic plots of the Eu magnetic moment directions in AFM (left), spin-flop (middle) and spin-flip (right) phase, respectively.
Blue arrows inside circles represent the spin configuration in different phases.
$H_c$ denotes the spin-flop critical field, and $H_s$ denotes the spin-flip (saturation) critical field.
(b) Calculated evolution of the corresponding tilting angle $\theta$ of Eu magnetic moments in the spin-flop process.
Light green zone corresponds to the measured field range in (a).
(c) Calculated weighted factor as a function of the in-plane azimuth angle $\phi$ of Eu magnetic moments at various applied fields.
Line profiles are shifted along the vertical axis with a step of 20 from the one at 5.4 T.
(d) shows the best $\phi$ angles with the smallest $R_w$ factor at different applied fields.
Red dash line is a reference line for $\phi = 45^\circ$.
The corresponding inset in (d) is a schematic diagram of the spatial shape for the in-plane anisotropy.
}
\end{figure}

Having determined the antiferromagnetic structures of both Eu and Mn moments comprehensively, a field dependent neutron diffraction experiment was performed at D23 with a 12 T vertical-field magnet, with the aim to shed light on the nature of the spin-flop transition and magnetic anisotropy in this compound.
Comparing to the previously reported study of the spin-flop transition via magnetization measurements \cite{Masuda2016}, neutron scattering has an irreplaceable advantage for being able to access to possible field dependent spin reorientation process of each of the Eu and Mn magnetic sublattice separately because of their distinct magnetic propagation wavevectors, and it can thus give more direct information about the field-driven evolution of the magnetic structures as well as possible interplay between Eu and Mn magnetism.
As shown in Fig.\ref{fig:Figs_spinflop}(a), a sharp intensity drop indicating a spin-flop phase transition was observed at about 5.3 T for both (1,0,2) and (1,1,1) magnetic reflections of the Eu sublattice, however the nuclear reflection (1,1,2) basically shows no field dependence.
This field-driven spin reorientation phenomenon, also known as the spin-flop transition, has already been observed and investigated in a number of different classes of antiferromagnets during the past several decades (including two-sublattice uniaxial \cite{Poulis1951,Ubbink1953,Gorter1953,Shapira1968,Poulis1954,Blazey1968,Blazey1971,Shapira1970,Foner1963,Rives1967,Becerra1988}, multi-sublattice \cite{Tsukada2001}, noncentrosymetric \cite{Zheludev1997a,Lumsden2001,Zheludev1997} antiferromagnets etc.),
and a variety of phenomenological models \cite{Blazey1968,Blazey1971,Zheludev1997,Bogdanov2002,Bogdanov2007} have been proposed to solve the spin configuration for the spin-flop transition.
A common approach is to list all possible free-energy terms in a magnetic system (Eq.\ref{eq:energy}) and minimize the free energy of this system to satisfy the equilibrium condition of the spin-flop process.
For the two-sublattice uniaxial collinear antiferromagnets, such as EuMnBi$_2$ here in this paper, the molar free energy of $N$ antiferromagnetically coupled spins $S$ at $T=0$ K can be given by \cite{Blazey1968,Blazey1971,Stephen2001,Iwasaki2018,Li2016c}:
\begin{eqnarray}\label{eq:energy}
\begin{aligned}
E=&\frac{1}{2}(g\mu_{B}SN)\Bigg[J\cos(\theta_1-\theta_2)-\frac{1}{2}K_u(\cos^2\theta_1+\cos^2\theta_2)\\
&+K_{e}\cos\theta_1\cos\theta_2-H_{\perp}(\sin\theta_1+\sin\theta_2)\\
&-H_{\parallel }(\cos\theta_1+\cos\theta_2)\Bigg],
\end{aligned}
\end{eqnarray}
where $\theta_1$ and $\theta_2$ represent the angles for the sublattice magnetizations of Eu deviated from the easy-axis (i.e. $c$-axis) directions.
The first term is the exchange energy, and the second and third terms are the magnetic anisotropy energies, whereas $J$ is the antiferromagnetical exchange interaction between the two sublattices of Eu, and $K_u$ and $K_{e}$ denote the uniaxial sinlge-ion magnetic anisotropy and exchange interaction anisotropy, respectively.
The last two terms are the Zeeman energies, where $H_{\perp}$ and $H_{\parallel }$ are the components of the applied field perpendicular and parallel to the easy axis of the magnetization.
To figure out the relation between the orientation of Eu moments and the applied field in the spin-flop phase, the following equilibrium condition
\begin{eqnarray}\label{eq:condition}
\begin{aligned}
\frac{\partial E}{\partial \theta_1}=\frac{\partial E}{\partial \theta_2}=0
\end{aligned}
\end{eqnarray}
can be utilized.
Derived from Eq.(\ref{eq:energy}) and Eq.(\ref{eq:condition}), we can have
\begin{eqnarray}\label{eq:H_perp}
\begin{aligned}
H_{\perp}\sin(\theta_1-\theta_2)=&J(\sin\theta_1+\sin\theta_2)\sin(\theta_1-\theta_2)\\
&+K_u(\cos\theta_2-\cos\theta_1)\sin\theta_1\sin\theta_2\\
&+K_e(\cos\theta_2-\cos\theta_1)\sin\theta_1\sin\theta_2
\end{aligned}
\end{eqnarray}
and
\begin{eqnarray}\label{eq:H_parallel}
\begin{aligned}
H_{\parallel}\sin(\theta_1-\theta_2)=&J(\cos\theta_1+\cos\theta_2)\sin(\theta_1-\theta_2)\\
&+K_u(\sin\theta_2-\sin\theta_1)\cos\theta_1\cos\theta_2\\
&-K_e(\cos^2\theta_1\sin\theta_2-\cos^2\theta_2\sin\theta_1).
\end{aligned}
\end{eqnarray}
Here we consider two special situations: the applied field $H$ is perpendicular or parallel to the easy axis.
First, for the perpendicular case, $\theta_1=\pi-\theta_2=\theta$, we have
\begin{eqnarray}\label{eq:H_perp_1}
\begin{aligned}
H=H_{\perp}=(2J+K_u+K_e)\sin\theta
\end{aligned}
\end{eqnarray}
and particularly when $\theta=90^\circ$, the spin-flip critical field of the saturation for the applied field perpendicular to $c$ axis will be $H_s(\perp)=2J+K_u+K_e$.
Second, for the applied field $H$ parallel to the easy axis and $\theta_1=-\theta_2=\theta$, the relation between the magnetization directions and the applied field in the spin-flop state can be extracted as
\begin{eqnarray}\label{eq:H_parallel_1}
\begin{aligned}
H=H_{\parallel}=(2J-K_u+K_e)\cos\theta
\end{aligned}
\end{eqnarray}
where $\theta=0^\circ$ is the saturation condition and naturally we will have the spin-flip (saturation) critical field for the applied field along the $c$ axis $H_s(\parallel)=2J-K_u+K_e$.
Since the magnetic diffraction intensities of the reflections (1,0,2) and (1,1,1) are sensitive to $\theta$, namely proportional to the squared in-plane AFM component (i.e. $|\textit{M}\sin\theta|^2$) for the Eu sublattice in the spin-flop phase, the intensity can be simply expressed as $I\sim 1-\frac{H^2}{(2J-K_u+K_e)^2}$.
Thus $2J-K_u+K_e$ can be easily extracted from the fitting of the field dependence curves of the reflections (1,0,2) and (1,1,1), subsequently the critical tilting angle $\theta$ of the Eu moments at the spin-flop phase transition can also be determined, as shown in Fig.\ref{fig:Figs_spinflop}(b), which is about 76.8$^\circ$.
Furthermore, both the fitted results for the reflections (1,0,2) and (1,1,1) show the saturated magnetic field $H_s(\parallel)$ at around 23 T, which is quite consistent with the previously reported magnetization in ref \cite{Masuda2016} as well as the theoretical calculation in ref \cite{Iwasaki2018} that also successfully reproduced the half-integer quantum Hall effect.

In addition, the critical field of the spin-flop transition can in principle also be calculated from the exchange and anisotropy constants.
When the applied field is parallel to the easy axis, the molar free energy at $T=0$ K for the antiferromagnetic phase and spin-flop phase can be express as
\begin{eqnarray}\label{eq:Energy_sf}
\begin{aligned}
E=
\begin{cases}
-(J+K_u+K_e),  H<H_c\\
J\cos2\theta-(K_u-K_e)\cos^2\theta-2H\cos\theta ,  H\ge  H_c
\end{cases}
\end{aligned}
\end{eqnarray}
where $H_c$ is the spin-flop critical field and $\frac{1}{2}(g\mu_BSN)$ is omitted for simplicity.
Assuming the adiabatic approximation during the spin-flop phase transition,
we can set the energy of the antiferromagnetic phase equal to that of the spin-flop phase given by Eq.(\ref{eq:Energy_sf}) when $H=H_c$.
From Eq.(\ref{eq:Energy_sf}) and Eq.(\ref{eq:H_parallel_1}), one can naturally obtain the spin-flop critical field and the critical angle $\theta$ as
\begin{subequations}\label{eq:Critical}
\begin{align}
H_c&=\sqrt{(2J-K_u+K_e)(K_u+K_e)}\label{eq:Critical_a}\\
\cos\theta&=\sqrt{\frac{K_u+K_e}{2J-K_u+K_e}}=\frac{H_c}{H_s(\parallel)}\label{eq:Critical_b}.
\end{align}
\end{subequations}

From all above equations, the exchange interaction $J$ and anisotropy parameters all can be expressed in terms of the critical fields (which can be directly measured):
\begin{subequations}\label{eq:ExC}
\begin{align}
J&=\frac{1}{2}\bigg[H_{s}(\perp)-\frac{H_c^2}{H_s(\parallel)} \bigg] \label{ExC_a}\\
K_u&=\frac{1}{2}\bigg[H_s(\perp)-H_s(\parallel) \bigg] \label{ExC_b}\\
K_e&=\frac{H_c^2}{H_s(\parallel)}-\frac{1}{2}\bigg[H_s(\perp)-H_s(\parallel) \bigg] \label{ExC_c},
\end{align}
\end{subequations}
where $H_c=5.3$ T and $H_s(\parallel)=23$ T can be extracted from our fitting results.
While the value of $H_s(\perp)$has not been confirmed in any experiments directly, we can still give an estimation according to the previously proposed relations near $T_N$ in ref \cite{Shapira1970}:
\begin{equation}
\begin{aligned}
T_N-T=\frac{g^2\mu_B^2(2S^2+2S+1)H_s(\perp)^2}{120k_B^2T_N},{}
\end{aligned}
\end{equation}
subsequently, $H_s(\perp)$ is calculated as 30.4 T with $T_N\approx22$ K, $S=\frac{7}{2}$ for Eu ions and the actual neutron experimental temperature $T=1.5$ K.
Besides, we can also give out a reference value of $H_s(\perp)$ from the low-field isothermal magnetization data \cite{ffz_mag} at low temperature by linear extension to the saturation condition (i.e. for $M_s=7\mu_B$ per Eu, $H_s(\perp)$ will be estimated as 29.3 T).
Hence, with the known critical fields, one obtains the exchange interaction $J=14.04$ T (0.81 meV), uniaxial magnetic anisotropy $K_u=3.15$ T (0.18 meV), exchange interaction anisotropy $K_e=-1.93$ T (-0.11 meV) at $T=1.5$ K.
Actually, a standard formula for the spin-flop critical field as a function of uniaxial magnetic anisotropy energy is known as \cite{Nagamiya1955,Poulis1954}:
\begin{equation}\label{eq:Ku}
\begin{aligned}
H_c=\sqrt{\frac{2K_u}{\chi_{\perp}-\chi_{\parallel}}},
\end{aligned}
\end{equation}
where $\chi_{\perp}$ and $\chi_{\parallel}$ are the susceptibilities in a small applied magnetic field at $T=0$ K for $H\perp c$ and $H\parallel c$ respectively.
Combining the above Eq.(\ref{eq:Ku}) and the susceptibilities at $T=2$ K in Fig.\ref{fig:Figs_ZFCFC_MH}(a), the uniaxial magnetic anisotropy parameter can be obtained, $K_u=3.35$ T (0.19 meV).
As expected, the values of $K_u$ that we calculated by using two methods are quite close, which indicates that the model we used above is suitable for the spin-flop transition of the Eu sublattice in this uniaxial antiferromagnet EuMnBi$_2$.
The positive $K_u$ implies that a single-ion easy axis is along the $c$ axis.
With all the known exchange interaction and anisotropy parameters, the spin Hamiltonian of a two-sublattice $(i,j)$ antiferromagnet \cite{Becerra1988,Rohrer1969} for Eu can be approximately written in the so-called XXZ model as:
\begin{equation}\label{eq:Hamiltonian}
\begin{aligned}
\mathbf{H}_{Eu} &=\sum_{i<j}\left [ J\left ( S_i^x S_j^x+S_i^y S_j^y \right )  +\left ( J+K_e \right ) S_i^z S_j^z \right ] \\
&- K_u \sum_i S_i^z S_i^z-g\mu_BH\sum_i S_i^z.
\end{aligned}
\end{equation}
The ratio between the out-of-plane and the in-plane components of the exchange interaction $r=(J+K_e)/J$ can be used as an indicator to distinguish from the various classical spin models (i.e. $r=1$ for Heisenberg model, $r=0$ for XY model, $r=\infty$ for Ising model).
Given that the exchange anisotropy $K_e$ is only $\sim-13\%$ of $J$ which makes $r=0.87$, it can thus be strongly suggested that the Eu magnetism should be described by a dominant Heisenberg exchange interaction with a small planar exchange anisotropy.
Nevertheless, given that $K_u> \left | K_e \right | $, the ordered moment direction of Eu along the $c$ axis is largely dictated by $K_u$.
Hence the classical 3D/2D Ising model is likely not suitable for describing the magnetic interactions of Eu here in EuMnBi$_2$.
This result may explain why the fitted critical exponents $\beta$ for Eu has a strong deviation from that of the 3D or 2D Ising models.
Given the difficulties for a potential inelastic neutron scattering measurement on spin-wave excitations due to strong neutron absorption of Eu, our estimation of the magnetic exchange interaction as well as magnetic anisotropy parameters based on a quantitative analysis of the spin-flop transition clearly provides very valuable microscopic understanding of the magnetism in this compound.

\begin{figure}
\centering
\includegraphics[width=8cm]{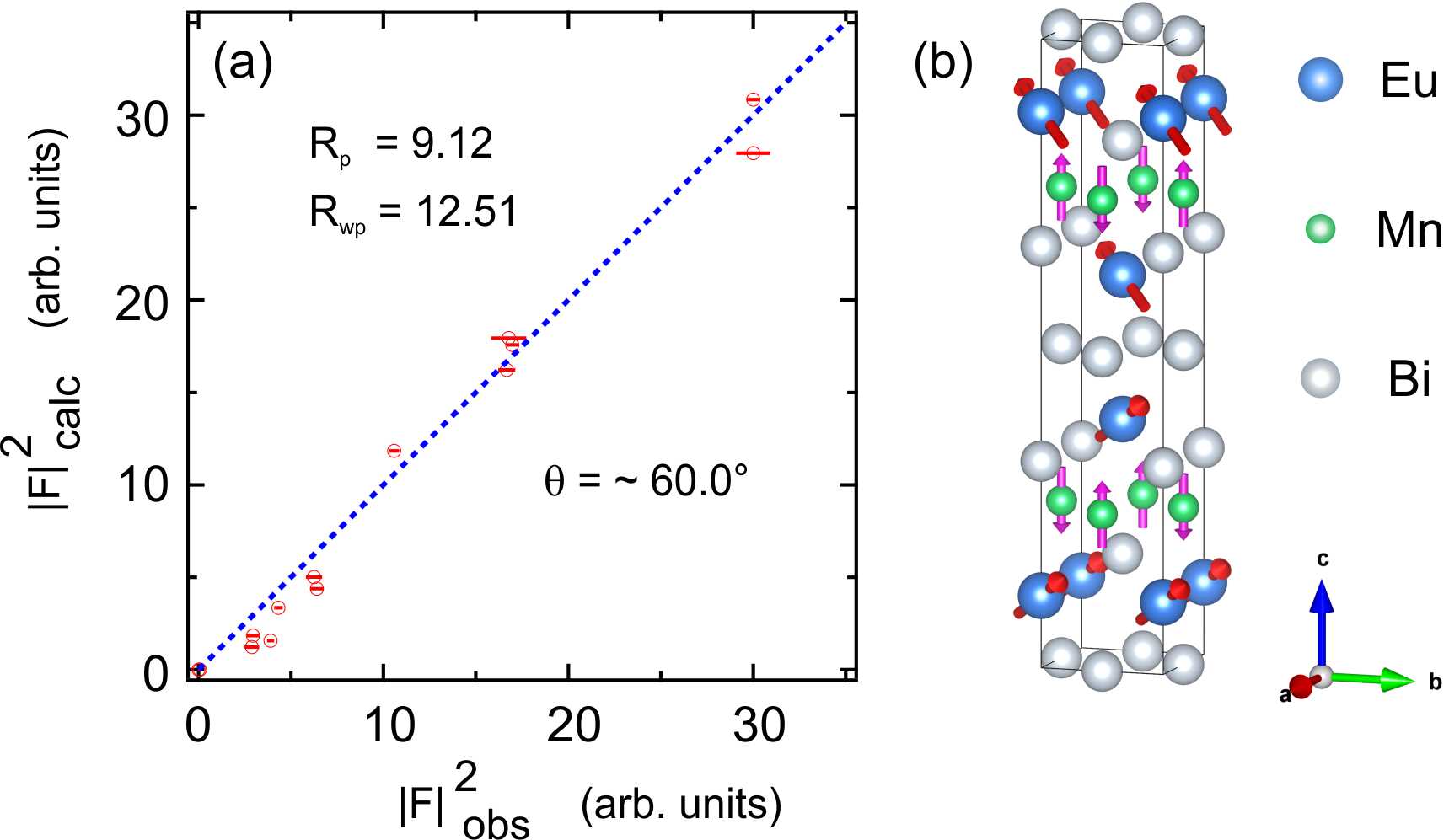}
\caption{\label{fig:Figs_D23Refine}
(a) Comparison of the observed and calculated squared structure factors of neutron diffraction data taken at 1.5 K with 11.5 T magnetic field along $c$ axis.
(b) The corresponding magnetic structure models under 11 T field, the magnetic moment direction of Eu atoms are tilted by about 60$^{\circ}$ away from $c$ axis with in-plane AFM component along the [1,1,0] direction.
}
\end{figure}

Now we go one more step further to study the spin configuration in the spin-flop sates as well as in-plane anisotropy.
Performing a series of neutron diffraction experiments at various applied fields would be a direct and effective method to figure out the in-plane preferred directions for the magnetic moment of Eu in its spin-flop states.
Because the magnetic structure factor is proportional to the component of the magnetic moment that is perpendicular to the scattering wave vector \textbf{Q}, so the intensity of magnetic diffractions changes as the moment direction changes.
By analyzing the intensity change of non-equivalent magnetic reflections at the spin-flop transition, the orientated directions of in-plane magnetic components can be confirmed.
With appropriate numerical calculations and also taking magnetic domains into consideration, we get the best in-plane azimuth angles $\phi$ for the applied fields, as shown in Fig.\ref{fig:Figs_spinflop}(c) and (d).
The parameter $R_w$ is the weighted profile factor for angle $\phi$ refinement to check how good it will fit, which here is defined as
\begin{eqnarray}
R_w=100\cdot \left[\frac{\sum_{i=1}^{n} w_{i}\cdot \left|\frac{I_{exp,i}^{SF}}{I_{exp,i}^{AFM}} -\frac{I_{calc,i}^{SF}}{I_{calc,i}^{AFM}}\right|^{2}}{\sum_{i=1}^{n} w_{i}\cdot (\frac{I_{exp,i}^{SF}}{I_{exp,i}^{AFM}})^{2}}\right]^{1/2}
\end{eqnarray}
in which, only the ratios of intensities between the spin-flop and the zero-field AFM states matter.
We found that the best fitted azimuth angles $\phi$ in the $ab$ plane are very close to 45$^\circ$, as shown in Fig.\ref{fig:Figs_spinflop}(d), which indicates that the in-plane AFM components of Eu magnetic moments are basically along the <1,1,0> directions, as shown in the schematic plot in Fig.\ref{fig:Figs_spinflop}(a).
Since the preferred in-plane orientation is already known, the spatial shape of the anisotropy tensor can be easily imagined, that is, the component of the anisotropy tensor along the diagonal $ab$ direction (i.e. [1,1,0]) is larger than that along the $a$ or $b$ axis, as shown in the inset of Fig.\ref{fig:Figs_spinflop}(d), and the out-of-plane component is strongest of all.

Meanwhile, we also collected 161 reflections in a reasonable Q range in a magnetic field of 11.5 T (11 non-equivalent reflections for Q $<$ 3.1 \AA$^{-1}$).
The observed squared structure factors versus the calculated ones are plotted in Fig.\ref{fig:Figs_D23Refine}(a), and the corresponding $\left | F \right |^2_{calc}$ were calculated based on the structure parameters in Table.\ref{tab:atom} and the magnetic structure models in Fig.\ref{fig:Figs_D23Refine}(b).
The tilting angle $\theta$ is refined as $\sim 60.0^\circ$ which is well consistent with our fitted field dependence, resulting in $\theta=60.4^\circ$ at 11.5 T in Fig.\ref{fig:Figs_spinflop}(b).
For now, the magnetic structures of the Eu sublattices in EuMnBi$_2$ including their evolutions in the field along the $c$ axis are comprehensively determined.

\begin{figure}
\centering
\includegraphics[width=8cm]{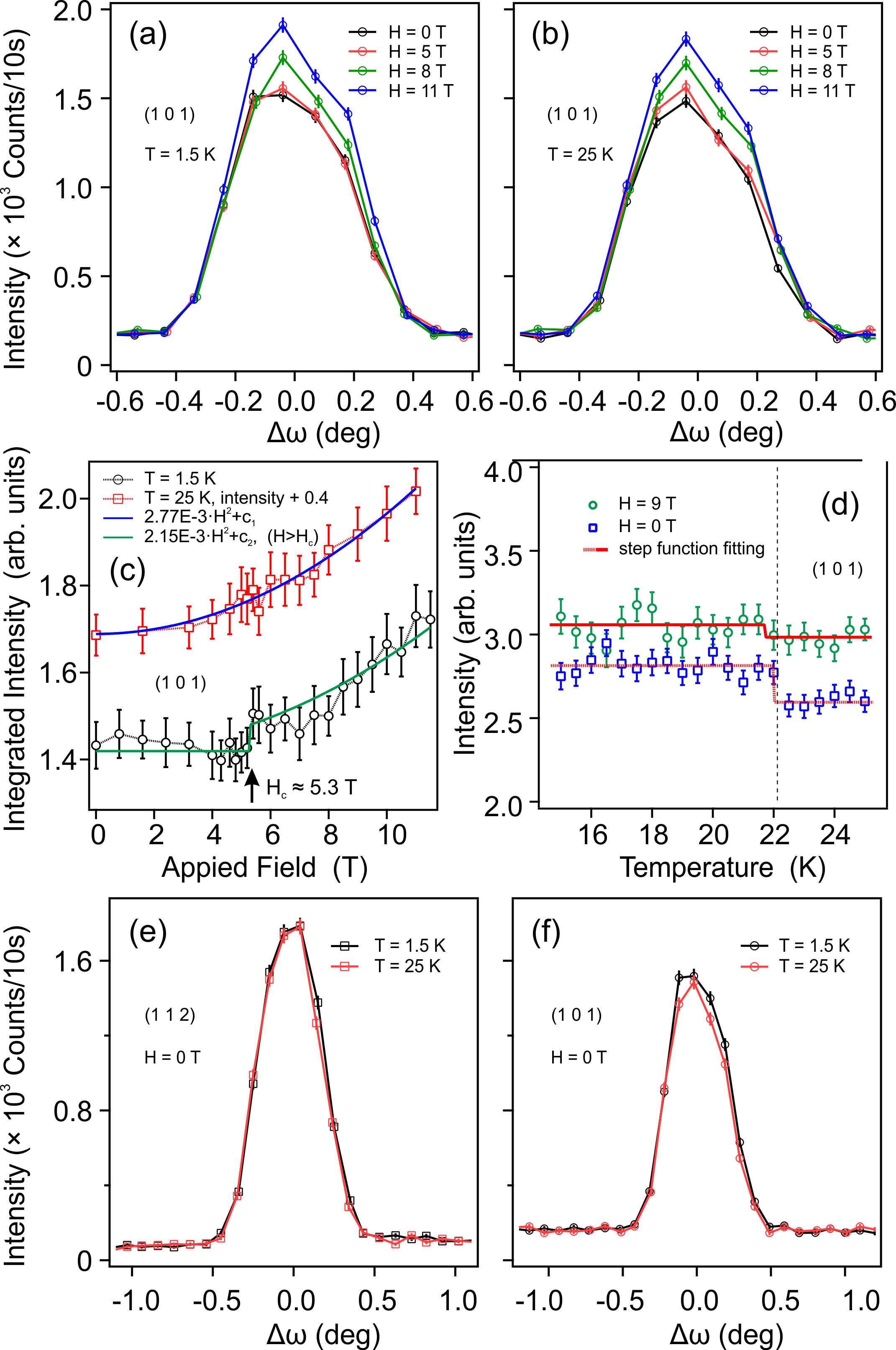}
\caption{\label{fig:Figs_D23_Hdep}
Field and temperature dependences of the selected magnetic reflections.
(a) and (b) Rocking curve scans of the magnetic reflection (1,0,1) under various applied fields along the $c$ axis at 2 K and 25 K respectively.
(c) the extracted field dependence of integrated intensities of (1,0,1) at 1.5 K and 25 K.
The data at 25 K is shifted up by 0.4 for an easy comparison with that at 1.5 K.
Solid blue line is the quadratic function fitting for 25 K and solid green line is a piecewise fitting of a quadratic function and a constant for 1.5 K.
(d) Temperature dependence of the magnetic reflection (1,0,1) with $H$ = 0 T and 9 T.
Red solid lines are the fittings of step function.
A clear kink is observed at around 22 K for magnetic reflection (1,0,1) under zero field.
Dash line shows the ordered temperature $T_N^{Eu}$.
Rocking curve scans of the nuclear reflection (e) (1,1,2) and magnetic reflection (f) (1,0,1) at 2 K and 25 K under zero field.
}
\end{figure}

\subsection{\label{sec:Coupling}Coupling of Eu and Mn Magnetism}

To reveal possible interplay between Eu and Mn magnetism, we now turn to the temperature and field dependence of two representative magnetic reflections (1,0,1) and (1,0,2), which are attributed to the AFM ordering of Mn and Eu respectively.
Detailed field dependence experiments of magnetic reflection (1,0,1) at 1.5 K and 25 K were performed.
Some selected rocking curve scans are plotted in Fig.\ref{fig:Figs_D23_Hdep}(a) and (b).
For both 1.5 K and 25 K, the intensities are found to be enhanced a little by an increase of the magnetic field.
This could be due to a very small mis-alignment between the $c$ axis of the sample and the direction of the applied vertical field, and a non-zero in-plane component of the applied field may induce a canted states \cite{Bogdanov2007},
thus making the magnetic moment component perpendicular to $\textbf{Q}_{(\pm1,0,1)}$ linearly
increased/decreased by the field, similar behavior can also be observed on the magnetic reflection (1,0,2) in Fig.\ref{fig:Figs_spinflop}(a) just before the spin-flop transition occurs.
A clear kink is observed at $H$ = 5.3 T from the extracted field dependence of the integrated intensities of (1,0,1) for T = 1.5 K.
The integrated intensity basically stays in the same level for fields $H$ $<$ 5.3 T, as shown in Fig.\ref{fig:Figs_D23_Hdep}(c).
That the AFM order of Mn responds to the occurrence of the spin-flop phase transition of Eu moments at B$_c$ strongly suggests the existence of interplay between Eu and Mn magnetism.
On the other hand, no clear anomaly is seen in the data at 25 K which can be well fitted by the quadratic curve.
The fitting results also show that the coefficient of the quadratic term at 25 K is obviously much larger than that at 1.5 K.
This implies that the Mn moment strongly prefers to be oriented along the $c$ axis as long as the Eu moment does the same.
Such a $c$ axis preferred magnetic anisotropy likely results from a strong coupling between Eu and Mn moments.
Furthermore, the temperature dependence of the magnetic reflection (1,0,1) is also monitored during the cooling process near $T_N^{Eu}$ with or without applied field, as shown in Fig.\ref{fig:Figs_D23_Hdep}(d).
At zero field, a sudden jump of intensity happens at T = 22 K, i.e. exactly at the magnetic phase transition of Eu; in a field of $H$ = 9 T, the intensity jump is not as clear as that at zero field.
This suggests the interaction between Eu and Mn moments is significantly weakened when the Eu magnetic sublattice enters into the spin-flop phase.
It is necessary to mention that such kind of intensity increase in Fig.\ref{fig:Figs_D23_Hdep}(d) could also be caused by the extinction release during the phase transition.
As shown in Fig.\ref{fig:Figs_D23_Hdep}(e) and (f), the intensity stays basically the same for the nuclear reflection (1,1,2) but increases a little bit for the magnetic reflection (1,0,1), suggesting that this is an intrinsic intensity increase instead of the effect from extinction release.

\begin{figure}
\centering
\includegraphics[width=8cm]{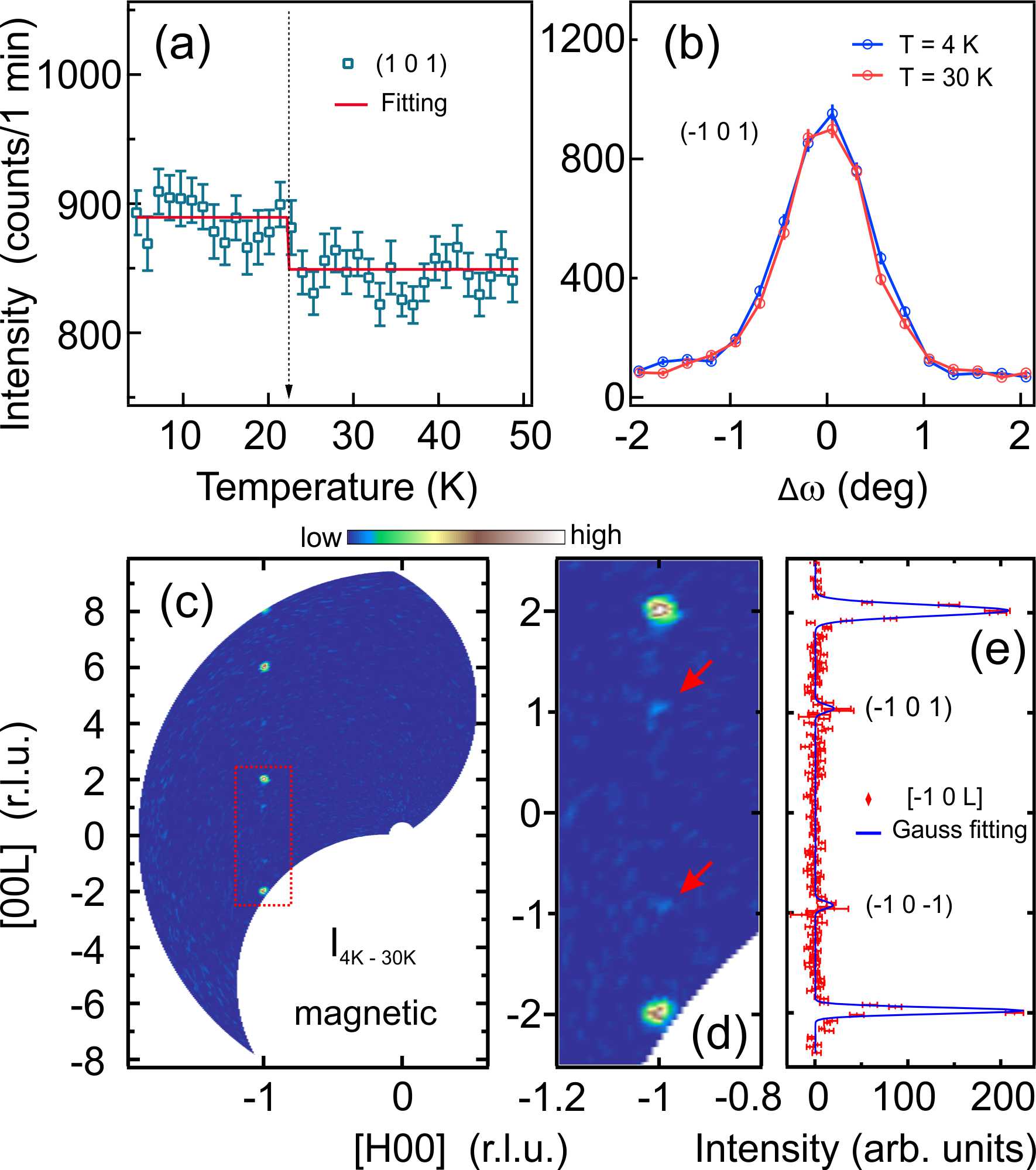}
\caption{\label{fig:Figs_P101Tdep_DNS}
Temperature dependence of polarized neutron diffraction at DNS.
(a) intensities of the magnetic reflection (-1,0,1) of Mn as a function of temperature near T$_N^{Eu}$.
Solid red line is a fitting of step function, dash line with arrow shows the AFM phase transition temperature of Eu.
(b) rocking curve scans of the magnetic reflection (-1,0,1) measured at 4 K and 30 K.
(c) the difference of polarized neutron diffraction patterns between 4 K and 30 K in the (H,0,L) scattering plane.
(d) zoom-in plot of the red rectangle part in (c).
red arrows show the extra intensity of magnetic reflections (1,0,1) and (-1,0,1).
(e) the corresponding line profile extracted from (d) along the [0,0,L] direction at H = -1.
Solid blue line is the multi-peak gauss fitting.
}
\end{figure}

A similar zero field temperature dependence experiment was also performed with polarized neutrons by using another EuMnBi$_2$ sample at DNS.
As shown in Fig.\ref{fig:Figs_P101Tdep_DNS}(a), a small but finite increase of the intensity of the Mn magnetic refection (1,0,1) can clearly be observed at T$_N^{Eu}=\sim$22 K.
Such an increase is also visible from a comparison of the rocking curve scans at 4 K and 30 K, as shown in Fig.10(b).
With a large Q-range mapping in the x spin flip channel, the difference diffraction pattern in the (H,0,L) plane between 4 K and 30 K is shown in Fig.\ref{fig:Figs_P101Tdep_DNS}(c), from which all the magnetic intensity enhancement can be easily seen.
Except the Eu magnetic reflections (-1,0,$\pm$2), there are some extra intensities on (-1,0,$\pm$1) as shown in the zoom-in plot in Fig.\ref{fig:Figs_P101Tdep_DNS}(d), and in the corresponding line cut (Fig.\ref{fig:Figs_P101Tdep_DNS}(e)).
Therefore, both field dependence and temperature dependence of magnetic reflections suggest a strong interplay between the two magnetic sublattices in EuMnBi$_2$.

\section{\label{sec:Discussion and Conclusion}Discussion and Conclusion}

We note that the propagation vectors of Mn and Eu sublattices in EuMnBi$_2$ are $\textbf{k}=(0,0,0)$ and (0,0,1) respectively.
At low temperature, the refined magnetic moment size for Mn ions is smaller than the usual one, while on the contrary, the moment size of Eu ions is a little bit larger than the theoretical value for isolated ions.
This could be a result of strong coupling between $3d$, $4f$ and itinerant electrons.
The field and temperature dependence experiments do show strong evidence for the interplay between Eu and Mn sublattices, and the strength of the interplay could be affected by the applied field.
This coupling between Eu and Mn sublattices could be the result of the change of the magnetic anisotropy caused by the ordering of Eu moments.
The magnetic ordering of Eu would enhance the magnetic anisotropy, which would in turn increase the tendency of the Mn moments being orientated along the $c$ axis.
On the other side, a weakened magnetic anisotropy in the spin-flop phase of Eu is expected to also weaken the Eu-Mn coupling strength, as demonstrated in Fig.\ref{fig:Figs_D23_Hdep}(c) and (d).
This thus suggests that the coupling of Eu and Mn magnetism strongly depends on the magnetic structure of the Eu sublattice in this system.
Given that, the magnetic structure of the Eu sublattice can be tuned by an applied field, this may bring new possibilities to continuously tune the interaction between rare-earth and transition metal magnetic ions by an external magnetic field instead of chemical doping \cite{Jin2019,Shang2013}.
So it is also worth to study the Eu-Mn coupling in the future with various field directions.
Fields especially along the [1,0,0] and [1,1,0] directions, unlike the applied field along the $c$ axis, will much easier align the magnetic moments of the Eu ions and cant them into the $ab$ plane, which may help to tilt the Mn moments subsequently.

It is known that the magnetic order of the Eu sublattice shows a remarkable impact on the Dirac fermions in this layered antiferromagnet as demonstrated by Shubnikov–de Haas (SdH) oscillation measurement and first-principles calculations \cite{Masuda2018}.
Coincidently, signatures of spin-fermion coupling between the magnetic Mn layer and Dirac fermions of the Bi layer were just reported in a similar system YbMnBi$_2$ \cite{Sapkota2020}.
Therefore, there is a good reason to believe that not only the Eu sublattice but also the Mn sublattice could play an important role on the Dirac band structures in EuMnBi$_2$.
Such an intricate interplay of $3d$, $4f$ and itinerant electrons may be used to realize novel correlated Dirac fermion states in a solid, which can offer a promising approach to emerging topological spintronics.

In summary, the magnetic phase transitions of the Mn and Eu sublattices of EuMnBi$_2$ were studied by magnetization, heat capacity and neutron scattering, the transition temperature for Eu and Mn are confirmed as $\sim$22 K and $\sim$337 K from the temperature dependence of corresponding magnetic reflections;
also, the detailed AFM structures of the Mn and Eu sublattices were directly investigated by using complementary polarized and unpolarized single-crystal neutron diffraction, and all the magnetic moments are found aligned along the $c$ axis at zero field.
At 300 K, the magnetic moment size is estimated as $\sim$2.1 $\rm \mu_B$ for the Mn ions; at 3 K, the ordered moment sizes are $\sim$4.1 $\rm \mu_B$ for the Mn ions and $\sim$7.7 $\rm \mu_B$ for the Eu ions.
Furthermore, the spin-flop process of the Eu sublattice, the corresponding magnetic structure and its evolution in the spin-flop phase were microscopically investigated in detail by neutron diffraction.
By constructing the molar free energy of this antiferromagnetic system and combining the equilibrium condition, the exchange interaction $J$ and anisotropy parameters $K_u$, $K_e$ are extracted from the fitted critical fields.
We found that $J\gg K_e$, $K_u$ showing the isotropic antiferromagnetic exchange interaction dominates in the spin Hamiltonian of the Eu sublattice.
A Heisenberg model modified with a small exchange anisotropy term as perturbation (namely the XXZ model) should be sufficient to describe the magnetic exchange interactions for the Eu sublattice.
It has also been determined from the refinement that spins tilt up to the $c$ axis along the <1,1,0> directions upon increasing applied fields in spin-flop states, which gives us more information about the in-plane magnetic anisotropy, as the schematic inset shows qualitatively in Fig.\ref{fig:Figs_spinflop}(d).
Furthermore, by measuring field and temperature dependence of the selected magnetic reflections, the existence of the interplay between Eu and Mn sublattices was revealed.
For future studies, the interplay between the localized magnetism and itinerant electrons in this class of Dirac fermion systems are highly desired, since EuMnBi$_2$ belongs to a large AMnPn$_2$ family of compounds, which is attracting strong interest due to the potential in spintronic applications like other magnetic topological materials.

Note added:
Just before the submission of our manuscript, we have noticed another recent study on the magnetic structure of the Eu moments in the spin-flop state of EuMnBi$_2$ \cite{Masuda2020}.
While the results from both studies are largely consistent with each other, we found certain discrepancies between them mainly concerning the magnetic structure of the Eu sublattice in the spin-flop state, namely, the in-plane component of Eu magnetic moments in our work is orientated along the [1,1,0] direction, instead of [1,0,0] as reported in ref \cite{Masuda2020}, and the reported moment tilting angle from the $c$ axis is also slightly different, especially in the vicinity of the spin-flop transition.
Due to subtle differences in both experimental approach and data analysis between our work and the study in ref \cite{Masuda2020}, we will not attempt to speculate possible causes for those discrepancies here.
Nevertheless, we believe that there are several advantages in our experiments that would make our conclusion very reliable.
First, we measured the field dependence of the magnetic structure in the spin-flop state under the magnetic field up to 11.5 T, not just in the vicinity of the spin-flop transition.
Second, two non-equivalent magnetic diffractions instead of only one as used in ref \cite{Masuda2020} were used to determine the spin reorientation of the Eu moments in the spin-flop state in our study.
Third, both polarized and nonpolarized neutrons diffraction methods are combined to determine the magnetic structures and the ordered moment size, in addition, the proposed spin-flop magnetic structure of the Eu moments is further confirmed by our refinement of magnetic structural factors measured in the spin-flop state.

\section*{ACKNOWLEDGMENTS}

This work is supported by the HGF–OCPC Postdoctoral Program.
We would like to acknowledge Susanne Mayr for assistance with the orientation of the crystal and Wentao Jin, Hao Deng, Erxi Feng and Navid Qureshi for helpful discussions.
This work is based on the experiments performed at DNS, HEIDI and D23 instruments. DNS is operated by J$\rm \ddot{u}$lich Centre for Neutron Science (JCNS) at the Heinz Maier-Leibnitz Zentrum (MLZ), Garching, Germany.
HEIDI is operated jointly by RWTH Aachen University and JCNS within the JARA alliance.
D23 is operated by the Institut Laue-Langevin (ILL) and CEA, Grenoble, France.

\appendix

\section{\label{app:Polarization}Magnetic scattering cross section for polarized neutrons}

\begin{table}[htbp]
    \caption[x, y, z polarization information nutshell]{\label{tab:porlariztion}
    The decomposed components of magnetization and nuclear scattering that contribute to the intensities of the different polarization channels are listed blow.
    }
    \begin{center}
       \begin{tabular}{c|c|c}
       \hline
       Polarization&Spin flip&non spin flip \\
       \hline
       $\textbf{P}$ $\parallel$ x $\parallel$ $\textbf{Q}$&$M^{\perp Q}_y$+$M^{\perp Q}_z$&nuclear \\
       \hline
       $\textbf{P}$ $\parallel$ y $\perp$ $\textbf{Q}$&$M^{\perp Q}_z$&nuclear+$M^{\perp Q}_y$ \\
       \hline
       $\textbf{P}$ $\parallel$ z $\perp$ $\textbf{Q}$&$M^{\perp Q}_y$&nuclear+$M^{\perp Q}_z$ \\
       \hline
       \end{tabular}
    \end{center}
\end{table}

\begin{figure}
\centering
\includegraphics[width=8cm]{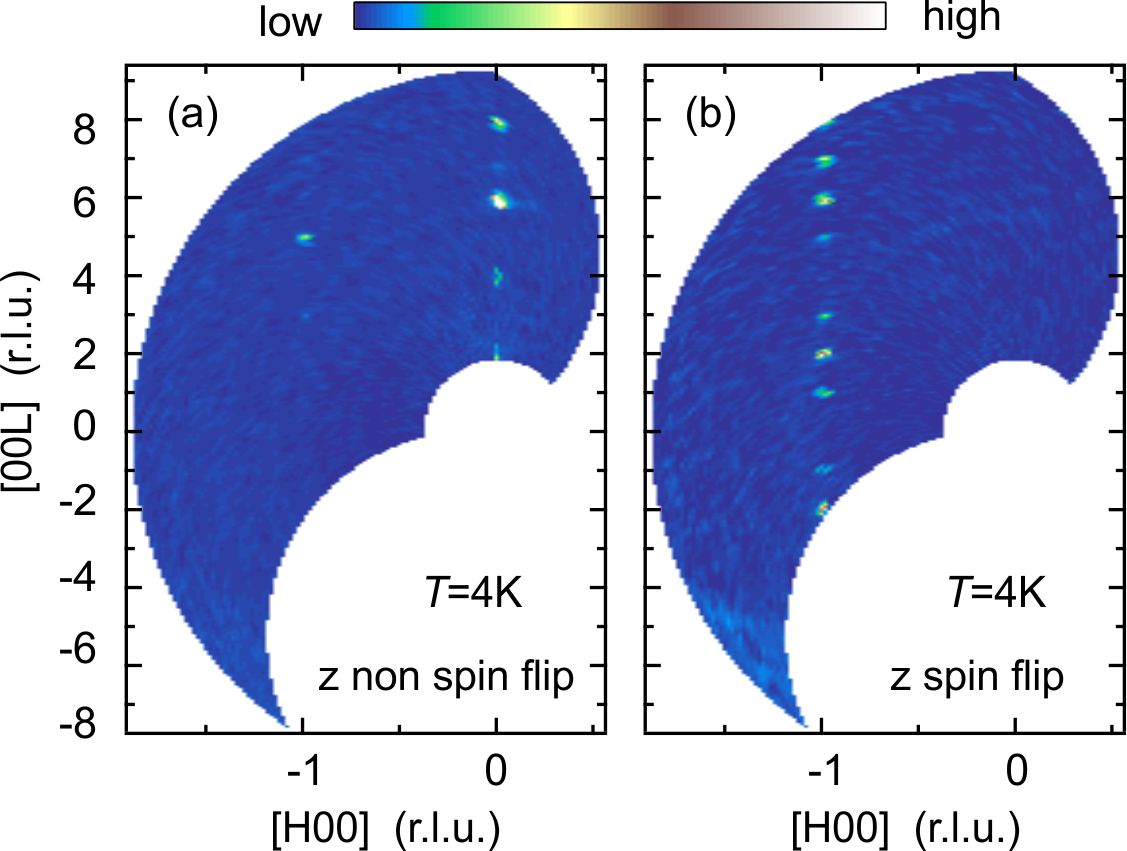}
\caption{\label{fig:fig_polar}
Polarized neutron diffraction patterns of single crystal EuMnBi$_2$ in the (H,0,L) plane at 4 K, (a) $z$ non spin flip channel, (b) $z$ spin flip channel.
}
\end{figure}

\begin{figure}
\centering
\includegraphics[width=8cm]{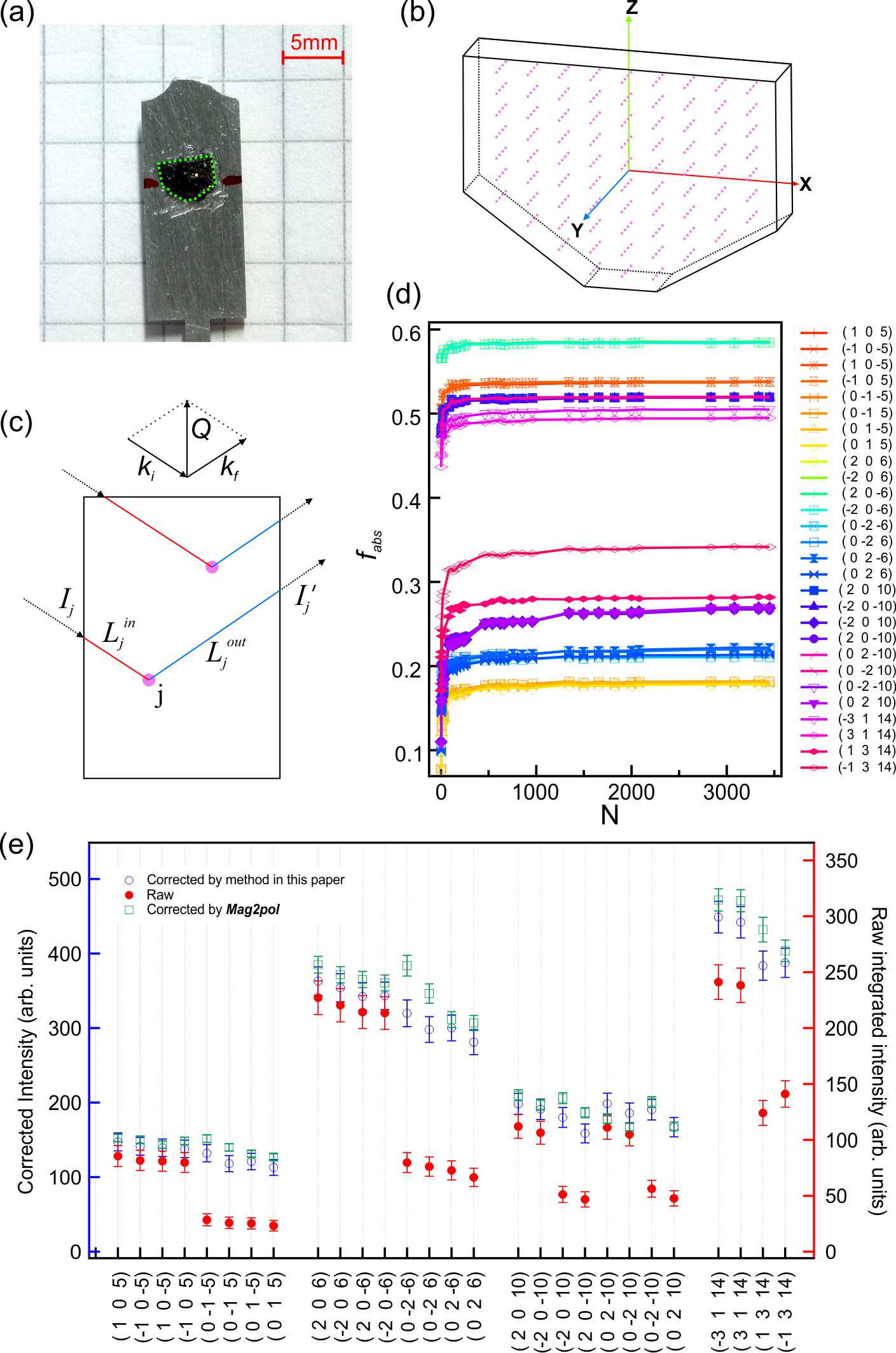}
\caption{\label{fig:fig_Abs}
(a) Picture of single crystal sample on holder for HEIDI.
(b) A 3D model with approximate same shape and size as our measured sample in (a).
The purple dots inside the 3D model represent the positions to calculate the neutron paths.
(c) A schematic drawing for showing the path of incident and scattered neutron beam at different positions inside the sample.
(d) The plot of absorption factors $f_{abs}$ for some reflections versus the number $N$ of selected scattering points inside sample.
$f_{abs}$ quickly get saturated when $N$ increase.
(e) comparison between raw and corrected intensities of some selected equivalent reflections.
}
\end{figure}

The full magnetic scattering cross section for polarized neutrons is quite complicated and the reader is referred to Ref \cite{Schweika2001} for more details.
For simplicity, we here only give a qualitative description to help to understand the polarization analysis.
As for the magnetic scattering of a given $\textbf{Q}$, it leads to
\begin{eqnarray}
\frac{d\sigma}{d\Omega}_{mag}\sim \left |\left \langle f\left | \mathbf{\hat{\sigma}_n}\cdot \mathbf{M_{\perp Q}} \right | i \right \rangle\right |^2
\end{eqnarray}
in which, $\mathbf{\hat{\sigma}_n}$ are the Pauli spin matrices for the neutron spin, $\mathbf{M_{\perp Q}}$ represents the component of the magnetization that is perpendicular to the scattering vector $\textbf{Q}$.
Here, we use the common convention for the specific orthogonal setting with $x$ parallel to $\textbf{Q}$, and $y$ and $z$ perpendicular to $\textbf{Q}$, in the horizontal scattering plane and perpendicular to this plane.
Only non zero magnetization components both perpendicular to the $\textbf{Q}$ and the polarization vector $\textbf{P}$ can have contributions in the spin flip channel.
Hence, we summarize the possible magnetization components responsible for the scattering intensity in Table.\ref{tab:porlariztion} for $x$, $y$, $z$ polarization in the spin and non spin flip channel.
The background and the incoherent scattering are not included here.
As a supplement to Fig.\ref{fig:Figs_DNSPlot}, the $z$ polarization data is shown in Fig.\ref{fig:fig_polar}.

\section{\label{app:Absorption}Neutron absorption correction}

Since the absorption cross section of Eu is $\sim$4530 barn which is by far larger than that of Mn and Bi, we only corrected the absorption of Eu.
As shown in Fig.\ref{fig:fig_Abs}(a), the sample we measured is flake like with irregular shape.
Therefore, the dimension parameters were measured and a 3D model (Fig.\ref{fig:fig_Abs}(b)) was established for the absorption calculation.
According to Beer–Lambert's Law, we have
\begin{eqnarray}
I'=I\cdot exp(-n_s\cdot\sigma_{abs}\cdot L),
\end{eqnarray}
where $n_s$ is the density of scattering units in the sample, $\sigma_{abs}$ is the absorption cross section, $L$ is the length of the neutron absorption path inside the sample.
In order to simplify the calculation, a representation of isolated points were selected and treated as scattering positions inside the sample, as shown in Fig.\ref{fig:fig_Abs}(b).
Assuming the neutron beam flux is uniform and time independent, then for a certain scattering condition as shown in Fig.\ref{fig:fig_Abs}(c), the scattering intensity will be as following:
\begin{eqnarray}
I'_{total}
&\sim& \frac{I_0}{N}\cdot\sum_{j=1}^{N}e^{-n_s\cdot\sigma_{abs}\cdot (L_{j}^{in}+L_{j}^{out})},
\end{eqnarray}
in which $I_0$ is the total incident intensity on the sample, which should be principally related to the sample shape, sample rotation center and beam uniformity depends on the diffraction conditions of Q = (h,k,l).
By considering the small size of sample and the inevitable divergence of the neutron beam, $I_0$ could be further approximated as being a constant for all diffraction centers, so the absorption factor can be simply expressed as
\begin{eqnarray}
f_{abs}
&=& \frac{1}{N}\sum_{j=1}^{N}e^{-\mu\cdot L_{j}}.
\end{eqnarray}
Principally, the calculated $f_{abs}$ will converge to a certain value using a reasonable limited number of scattering points $N$.

In Fig.\ref{fig:fig_Abs}(d), for some selected reflections, the absorption factors quickly start to get saturated as $N$ increases.
In this paper, we use $N=3500$ to make sure that all the absorption factors are convergent and in considerable credibility.
Raw integrated intensities and corrected intensities of some selected strong reflections are plotted in Fig.\ref{fig:fig_Abs}(e), the variance becomes significantly better for equivalent reflections after proper correction.
In addition, we noticed that there is a new and much more efficient software $\textit{Mag2pol}$ \cite{Qureshi2019} made by Navid Qureshi, which can also do the neutron absorption correction, and gives similar correction results as our method, shown in Fig.\ref{fig:fig_Abs}(e).

\nocite{*}

\bibliography{ref}

\begin{thebibliography}{76}%
\makeatletter
\providecommand \@ifxundefined [1]{%
 \@ifx{#1\undefined}
}%
\providecommand \@ifnum [1]{%
 \ifnum #1\expandafter \@firstoftwo
 \else \expandafter \@secondoftwo
 \fi
}%
\providecommand \@ifx [1]{%
 \ifx #1\expandafter \@firstoftwo
 \else \expandafter \@secondoftwo
 \fi
}%
\providecommand \natexlab [1]{#1}%
\providecommand \enquote  [1]{``#1''}%
\providecommand \bibnamefont  [1]{#1}%
\providecommand \bibfnamefont [1]{#1}%
\providecommand \citenamefont [1]{#1}%
\providecommand \href@noop [0]{\@secondoftwo}%
\providecommand \href [0]{\begingroup \@sanitize@url \@href}%
\providecommand \@href[1]{\@@startlink{#1}\@@href}%
\providecommand \@@href[1]{\endgroup#1\@@endlink}%
\providecommand \@sanitize@url [0]{\catcode `\\12\catcode `\$12\catcode
  `\&12\catcode `\#12\catcode `\^12\catcode `\_12\catcode `\%12\relax}%
\providecommand \@@startlink[1]{}%
\providecommand \@@endlink[0]{}%
\providecommand \url  [0]{\begingroup\@sanitize@url \@url }%
\providecommand \@url [1]{\endgroup\@href {#1}{\urlprefix }}%
\providecommand \urlprefix  [0]{URL }%
\providecommand \Eprint [0]{\href }%
\providecommand \doibase [0]{http://dx.doi.org/}%
\providecommand \selectlanguage [0]{\@gobble}%
\providecommand \bibinfo  [0]{\@secondoftwo}%
\providecommand \bibfield  [0]{\@secondoftwo}%
\providecommand \translation [1]{[#1]}%
\providecommand \BibitemOpen [0]{}%
\providecommand \bibitemStop [0]{}%
\providecommand \bibitemNoStop [0]{.\EOS\space}%
\providecommand \EOS [0]{\spacefactor3000\relax}%
\providecommand \BibitemShut  [1]{\csname bibitem#1\endcsname}%
\let\auto@bib@innerbib\@empty
\bibitem [{\citenamefont {Armitage}\ \emph {et~al.}(2018)\citenamefont
  {Armitage}, \citenamefont {Mele},\ and\ \citenamefont
  {Vishwanath}}]{Armitage2018}%
  \BibitemOpen
  \bibfield  {author} {\bibinfo {author} {\bibfnamefont {N.~P.}\ \bibnamefont
  {Armitage}}, \bibinfo {author} {\bibfnamefont {E.~J.}\ \bibnamefont {Mele}},
  \ and\ \bibinfo {author} {\bibfnamefont {A.}~\bibnamefont {Vishwanath}},\
  }\href {\doibase 10.1103/RevModPhys.90.015001} {\bibfield  {journal}
  {\bibinfo  {journal} {Rev. Mod. Phys.}\ }\textbf {\bibinfo {volume} {90}},\
  \bibinfo {pages} {015001} (\bibinfo {year} {2018})}\BibitemShut {NoStop}%
\bibitem [{\citenamefont {Hasan}\ and\ \citenamefont {Kane}(2010)}]{Hasan2010}%
  \BibitemOpen
  \bibfield  {author} {\bibinfo {author} {\bibfnamefont {M.~Z.}\ \bibnamefont
  {Hasan}}\ and\ \bibinfo {author} {\bibfnamefont {C.~L.}\ \bibnamefont
  {Kane}},\ }\href {\doibase 10.1103/RevModPhys.82.3045} {\bibfield  {journal}
  {\bibinfo  {journal} {Rev. Mod. Phys.}\ }\textbf {\bibinfo {volume} {82}},\
  \bibinfo {pages} {3045} (\bibinfo {year} {2010})}\BibitemShut {NoStop}%
\bibitem [{\citenamefont {Pesin}\ and\ \citenamefont
  {Balents}(2010)}]{Pesin2010}%
  \BibitemOpen
  \bibfield  {author} {\bibinfo {author} {\bibfnamefont {D.}~\bibnamefont
  {Pesin}}\ and\ \bibinfo {author} {\bibfnamefont {L.}~\bibnamefont
  {Balents}},\ }\href {\doibase 10.1038/nphys1606} {\bibfield  {journal}
  {\bibinfo  {journal} {Nat. Phys.}\ }\textbf {\bibinfo {volume} {6}},\
  \bibinfo {pages} {376} (\bibinfo {year} {2010})}\BibitemShut {NoStop}%
\bibitem [{\citenamefont {Rau}\ \emph {et~al.}(2016)\citenamefont {Rau},
  \citenamefont {Lee},\ and\ \citenamefont {Kee}}]{Rau2016}%
  \BibitemOpen
  \bibfield  {author} {\bibinfo {author} {\bibfnamefont {J.~G.}\ \bibnamefont
  {Rau}}, \bibinfo {author} {\bibfnamefont {E.~K.-H.}\ \bibnamefont {Lee}}, \
  and\ \bibinfo {author} {\bibfnamefont {H.-Y.}\ \bibnamefont {Kee}},\ }\href
  {\doibase 10.1146/annurev-conmatphys-031115-011319} {\bibfield  {journal}
  {\bibinfo  {journal} {Annu. Rev. Condens. Matter Phys.}\ }\textbf {\bibinfo
  {volume} {7}},\ \bibinfo {pages} {195} (\bibinfo {year} {2016})}\BibitemShut
  {NoStop}%
\bibitem [{\citenamefont {Sapkota}\ \emph {et~al.}(2020)\citenamefont
  {Sapkota}, \citenamefont {Classen}, \citenamefont {Stone}, \citenamefont
  {Savici}, \citenamefont {Garlea}, \citenamefont {Wang}, \citenamefont
  {Tranquada}, \citenamefont {Petrovic},\ and\ \citenamefont
  {Zaliznyak}}]{Sapkota2020}%
  \BibitemOpen
  \bibfield  {author} {\bibinfo {author} {\bibfnamefont {A.}~\bibnamefont
  {Sapkota}}, \bibinfo {author} {\bibfnamefont {L.}~\bibnamefont {Classen}},
  \bibinfo {author} {\bibfnamefont {M.~B.}\ \bibnamefont {Stone}}, \bibinfo
  {author} {\bibfnamefont {A.~T.}\ \bibnamefont {Savici}}, \bibinfo {author}
  {\bibfnamefont {V.~O.}\ \bibnamefont {Garlea}}, \bibinfo {author}
  {\bibfnamefont {A.}~\bibnamefont {Wang}}, \bibinfo {author} {\bibfnamefont
  {J.~M.}\ \bibnamefont {Tranquada}}, \bibinfo {author} {\bibfnamefont
  {C.}~\bibnamefont {Petrovic}}, \ and\ \bibinfo {author} {\bibfnamefont
  {I.~A.}\ \bibnamefont {Zaliznyak}},\ }\href {\doibase
  10.1103/PhysRevB.101.041111} {\bibfield  {journal} {\bibinfo  {journal}
  {Phys. Rev. B}\ }\textbf {\bibinfo {volume} {101}},\ \bibinfo {pages}
  {041111(R)} (\bibinfo {year} {2020})}\BibitemShut {NoStop}%
\bibitem [{\citenamefont {Masuda}\ \emph {et~al.}(2016)\citenamefont {Masuda},
  \citenamefont {Sakai}, \citenamefont {Tokunaga}, \citenamefont {Yamasaki},
  \citenamefont {Miyake}, \citenamefont {Shiogai}, \citenamefont {Nakamura},
  \citenamefont {Awaji}, \citenamefont {Tsukazaki}, \citenamefont {Nakao},
  \citenamefont {Murakami}, \citenamefont {Arima}, \citenamefont {Tokura},\
  and\ \citenamefont {Ishiwata}}]{Masuda2016}%
  \BibitemOpen
  \bibfield  {author} {\bibinfo {author} {\bibfnamefont {H.}~\bibnamefont
  {Masuda}}, \bibinfo {author} {\bibfnamefont {H.}~\bibnamefont {Sakai}},
  \bibinfo {author} {\bibfnamefont {M.}~\bibnamefont {Tokunaga}}, \bibinfo
  {author} {\bibfnamefont {Y.}~\bibnamefont {Yamasaki}}, \bibinfo {author}
  {\bibfnamefont {A.}~\bibnamefont {Miyake}}, \bibinfo {author} {\bibfnamefont
  {J.}~\bibnamefont {Shiogai}}, \bibinfo {author} {\bibfnamefont
  {S.}~\bibnamefont {Nakamura}}, \bibinfo {author} {\bibfnamefont
  {S.}~\bibnamefont {Awaji}}, \bibinfo {author} {\bibfnamefont
  {A.}~\bibnamefont {Tsukazaki}}, \bibinfo {author} {\bibfnamefont
  {H.}~\bibnamefont {Nakao}}, \bibinfo {author} {\bibfnamefont
  {Y.}~\bibnamefont {Murakami}}, \bibinfo {author} {\bibfnamefont {T.-h.}\
  \bibnamefont {Arima}}, \bibinfo {author} {\bibfnamefont {Y.}~\bibnamefont
  {Tokura}}, \ and\ \bibinfo {author} {\bibfnamefont {S.}~\bibnamefont
  {Ishiwata}},\ }\href {\doibase 10.1126/sciadv.1501117} {\bibfield  {journal}
  {\bibinfo  {journal} {Sci. Adv.}\ }\textbf {\bibinfo {volume} {2}},\ \bibinfo
  {pages} {e1501117} (\bibinfo {year} {2016})}\BibitemShut {NoStop}%
\bibitem [{\citenamefont {{\v{S}}mejkal}\ \emph {et~al.}(2017)\citenamefont
  {{\v{S}}mejkal}, \citenamefont {Jungwirth},\ and\ \citenamefont
  {Sinova}}]{Smejkal2017}%
  \BibitemOpen
  \bibfield  {author} {\bibinfo {author} {\bibfnamefont {L.}~\bibnamefont
  {{\v{S}}mejkal}}, \bibinfo {author} {\bibfnamefont {T.}~\bibnamefont
  {Jungwirth}}, \ and\ \bibinfo {author} {\bibfnamefont {J.}~\bibnamefont
  {Sinova}},\ }\href {\doibase 10.1002/pssr.201700044} {\bibfield  {journal}
  {\bibinfo  {journal} {Phys. status solidi - Rapid Res. Lett.}\ }\textbf
  {\bibinfo {volume} {11}},\ \bibinfo {pages} {1700044} (\bibinfo {year}
  {2017})}\BibitemShut {NoStop}%
\bibitem [{\citenamefont {{\v{S}}mejkal}\ \emph {et~al.}(2018)\citenamefont
  {{\v{S}}mejkal}, \citenamefont {Mokrousov}, \citenamefont {Yan},\ and\
  \citenamefont {MacDonald}}]{Smejkal2018}%
  \BibitemOpen
  \bibfield  {author} {\bibinfo {author} {\bibfnamefont {L.}~\bibnamefont
  {{\v{S}}mejkal}}, \bibinfo {author} {\bibfnamefont {Y.}~\bibnamefont
  {Mokrousov}}, \bibinfo {author} {\bibfnamefont {B.}~\bibnamefont {Yan}}, \
  and\ \bibinfo {author} {\bibfnamefont {A.~H.}\ \bibnamefont {MacDonald}},\
  }\href {\doibase 10.1038/s41567-018-0064-5} {\bibfield  {journal} {\bibinfo
  {journal} {Nat. Phys.}\ }\textbf {\bibinfo {volume} {14}},\ \bibinfo {pages}
  {242} (\bibinfo {year} {2018})}\BibitemShut {NoStop}%
\bibitem [{\citenamefont {Guo}\ \emph {et~al.}(2014)\citenamefont {Guo},
  \citenamefont {Princep}, \citenamefont {Zhang}, \citenamefont {Manuel},
  \citenamefont {Khalyavin}, \citenamefont {Mazin}, \citenamefont {Shi},\ and\
  \citenamefont {Boothroyd}}]{Guo2014}%
  \BibitemOpen
  \bibfield  {author} {\bibinfo {author} {\bibfnamefont {Y.~F.}\ \bibnamefont
  {Guo}}, \bibinfo {author} {\bibfnamefont {A.~J.}\ \bibnamefont {Princep}},
  \bibinfo {author} {\bibfnamefont {X.}~\bibnamefont {Zhang}}, \bibinfo
  {author} {\bibfnamefont {P.}~\bibnamefont {Manuel}}, \bibinfo {author}
  {\bibfnamefont {D.}~\bibnamefont {Khalyavin}}, \bibinfo {author}
  {\bibfnamefont {I.~I.}\ \bibnamefont {Mazin}}, \bibinfo {author}
  {\bibfnamefont {Y.~G.}\ \bibnamefont {Shi}}, \ and\ \bibinfo {author}
  {\bibfnamefont {A.~T.}\ \bibnamefont {Boothroyd}},\ }\href {\doibase
  10.1103/PhysRevB.90.075120} {\bibfield  {journal} {\bibinfo  {journal} {Phys.
  Rev. B}\ }\textbf {\bibinfo {volume} {90}},\ \bibinfo {pages} {075120}
  (\bibinfo {year} {2014})}\BibitemShut {NoStop}%
\bibitem [{\citenamefont {Zhang}\ \emph {et~al.}(2016)\citenamefont {Zhang},
  \citenamefont {Liu}, \citenamefont {Yi}, \citenamefont {Zhao}, \citenamefont
  {Xia}, \citenamefont {Ji}, \citenamefont {Shi}, \citenamefont {Yu},
  \citenamefont {Wang}, \citenamefont {Chen},\ and\ \citenamefont
  {Zhang}}]{Zhang2016}%
  \BibitemOpen
  \bibfield  {author} {\bibinfo {author} {\bibfnamefont {A.}~\bibnamefont
  {Zhang}}, \bibinfo {author} {\bibfnamefont {C.}~\bibnamefont {Liu}}, \bibinfo
  {author} {\bibfnamefont {C.}~\bibnamefont {Yi}}, \bibinfo {author}
  {\bibfnamefont {G.}~\bibnamefont {Zhao}}, \bibinfo {author} {\bibfnamefont
  {T.-l.}\ \bibnamefont {Xia}}, \bibinfo {author} {\bibfnamefont
  {J.}~\bibnamefont {Ji}}, \bibinfo {author} {\bibfnamefont {Y.}~\bibnamefont
  {Shi}}, \bibinfo {author} {\bibfnamefont {R.}~\bibnamefont {Yu}}, \bibinfo
  {author} {\bibfnamefont {X.}~\bibnamefont {Wang}}, \bibinfo {author}
  {\bibfnamefont {C.}~\bibnamefont {Chen}}, \ and\ \bibinfo {author}
  {\bibfnamefont {Q.}~\bibnamefont {Zhang}},\ }\href {\doibase
  10.1038/ncomms13833} {\bibfield  {journal} {\bibinfo  {journal} {Nat.
  Commun.}\ }\textbf {\bibinfo {volume} {7}},\ \bibinfo {pages} {13833}
  (\bibinfo {year} {2016})}\BibitemShut {NoStop}%
\bibitem [{\citenamefont {Zhang}\ \emph
  {et~al.}(2019{\natexlab{a}})\citenamefont {Zhang}, \citenamefont {Okamoto},
  \citenamefont {Stone}, \citenamefont {Liu}, \citenamefont {Zhu},
  \citenamefont {DiTusa}, \citenamefont {Mao},\ and\ \citenamefont
  {Tennant}}]{Zhang2019}%
  \BibitemOpen
  \bibfield  {author} {\bibinfo {author} {\bibfnamefont {Q.}~\bibnamefont
  {Zhang}}, \bibinfo {author} {\bibfnamefont {S.}~\bibnamefont {Okamoto}},
  \bibinfo {author} {\bibfnamefont {M.~B.}\ \bibnamefont {Stone}}, \bibinfo
  {author} {\bibfnamefont {J.}~\bibnamefont {Liu}}, \bibinfo {author}
  {\bibfnamefont {Y.}~\bibnamefont {Zhu}}, \bibinfo {author} {\bibfnamefont
  {J.}~\bibnamefont {DiTusa}}, \bibinfo {author} {\bibfnamefont
  {Z.}~\bibnamefont {Mao}}, \ and\ \bibinfo {author} {\bibfnamefont {D.~A.}\
  \bibnamefont {Tennant}},\ }\href {\doibase 10.1103/PhysRevB.100.205105}
  {\bibfield  {journal} {\bibinfo  {journal} {Phys. Rev. B}\ }\textbf {\bibinfo
  {volume} {100}},\ \bibinfo {pages} {205105} (\bibinfo {year}
  {2019}{\natexlab{a}})}\BibitemShut {NoStop}%
\bibitem [{\citenamefont {Xu}\ \emph {et~al.}(2018)\citenamefont {Xu},
  \citenamefont {Liu}, \citenamefont {Shi}, \citenamefont {Muechler},
  \citenamefont {Gayles}, \citenamefont {Felser},\ and\ \citenamefont
  {Sun}}]{Xu2018a}%
  \BibitemOpen
  \bibfield  {author} {\bibinfo {author} {\bibfnamefont {Q.}~\bibnamefont
  {Xu}}, \bibinfo {author} {\bibfnamefont {E.}~\bibnamefont {Liu}}, \bibinfo
  {author} {\bibfnamefont {W.}~\bibnamefont {Shi}}, \bibinfo {author}
  {\bibfnamefont {L.}~\bibnamefont {Muechler}}, \bibinfo {author}
  {\bibfnamefont {J.}~\bibnamefont {Gayles}}, \bibinfo {author} {\bibfnamefont
  {C.}~\bibnamefont {Felser}}, \ and\ \bibinfo {author} {\bibfnamefont
  {Y.}~\bibnamefont {Sun}},\ }\href {\doibase 10.1103/PhysRevB.97.235416}
  {\bibfield  {journal} {\bibinfo  {journal} {Phys. Rev. B}\ }\textbf {\bibinfo
  {volume} {97}},\ \bibinfo {pages} {235416} (\bibinfo {year}
  {2018})}\BibitemShut {NoStop}%
\bibitem [{\citenamefont {Liu}\ \emph {et~al.}(2018)\citenamefont {Liu},
  \citenamefont {Sun}, \citenamefont {Kumar}, \citenamefont {Muechler},
  \citenamefont {Sun}, \citenamefont {Jiao}, \citenamefont {Yang},
  \citenamefont {Liu}, \citenamefont {Liang}, \citenamefont {Xu}, \citenamefont
  {Kroder}, \citenamefont {S{\"{u}}{\ss}}, \citenamefont {Borrmann},
  \citenamefont {Shekhar}, \citenamefont {Wang}, \citenamefont {Xi},
  \citenamefont {Wang}, \citenamefont {Schnelle}, \citenamefont {Wirth},
  \citenamefont {Chen}, \citenamefont {Goennenwein},\ and\ \citenamefont
  {Felser}}]{Liu2018d}%
  \BibitemOpen
  \bibfield  {author} {\bibinfo {author} {\bibfnamefont {E.}~\bibnamefont
  {Liu}}, \bibinfo {author} {\bibfnamefont {Y.}~\bibnamefont {Sun}}, \bibinfo
  {author} {\bibfnamefont {N.}~\bibnamefont {Kumar}}, \bibinfo {author}
  {\bibfnamefont {L.}~\bibnamefont {Muechler}}, \bibinfo {author}
  {\bibfnamefont {A.}~\bibnamefont {Sun}}, \bibinfo {author} {\bibfnamefont
  {L.}~\bibnamefont {Jiao}}, \bibinfo {author} {\bibfnamefont {S.-Y.}\
  \bibnamefont {Yang}}, \bibinfo {author} {\bibfnamefont {D.}~\bibnamefont
  {Liu}}, \bibinfo {author} {\bibfnamefont {A.}~\bibnamefont {Liang}}, \bibinfo
  {author} {\bibfnamefont {Q.}~\bibnamefont {Xu}}, \bibinfo {author}
  {\bibfnamefont {J.}~\bibnamefont {Kroder}}, \bibinfo {author} {\bibfnamefont
  {V.}~\bibnamefont {S{\"{u}}{\ss}}}, \bibinfo {author} {\bibfnamefont
  {H.}~\bibnamefont {Borrmann}}, \bibinfo {author} {\bibfnamefont
  {C.}~\bibnamefont {Shekhar}}, \bibinfo {author} {\bibfnamefont
  {Z.}~\bibnamefont {Wang}}, \bibinfo {author} {\bibfnamefont {C.}~\bibnamefont
  {Xi}}, \bibinfo {author} {\bibfnamefont {W.}~\bibnamefont {Wang}}, \bibinfo
  {author} {\bibfnamefont {W.}~\bibnamefont {Schnelle}}, \bibinfo {author}
  {\bibfnamefont {S.}~\bibnamefont {Wirth}}, \bibinfo {author} {\bibfnamefont
  {Y.}~\bibnamefont {Chen}}, \bibinfo {author} {\bibfnamefont {S.~T.~B.}\
  \bibnamefont {Goennenwein}}, \ and\ \bibinfo {author} {\bibfnamefont
  {C.}~\bibnamefont {Felser}},\ }\href {\doibase 10.1038/s41567-018-0234-5}
  {\bibfield  {journal} {\bibinfo  {journal} {Nat. Phys.}\ }\textbf {\bibinfo
  {volume} {14}},\ \bibinfo {pages} {1125} (\bibinfo {year}
  {2018})}\BibitemShut {NoStop}%
\bibitem [{\citenamefont {Li}\ \emph {et~al.}(2019)\citenamefont {Li},
  \citenamefont {Li}, \citenamefont {Du}, \citenamefont {Wang}, \citenamefont
  {Gu}, \citenamefont {Zhang}, \citenamefont {He}, \citenamefont {Duan},\ and\
  \citenamefont {Xu}}]{Li2019c}%
  \BibitemOpen
  \bibfield  {author} {\bibinfo {author} {\bibfnamefont {J.}~\bibnamefont
  {Li}}, \bibinfo {author} {\bibfnamefont {Y.}~\bibnamefont {Li}}, \bibinfo
  {author} {\bibfnamefont {S.}~\bibnamefont {Du}}, \bibinfo {author}
  {\bibfnamefont {Z.}~\bibnamefont {Wang}}, \bibinfo {author} {\bibfnamefont
  {B.-L.}\ \bibnamefont {Gu}}, \bibinfo {author} {\bibfnamefont {S.-C.}\
  \bibnamefont {Zhang}}, \bibinfo {author} {\bibfnamefont {K.}~\bibnamefont
  {He}}, \bibinfo {author} {\bibfnamefont {W.}~\bibnamefont {Duan}}, \ and\
  \bibinfo {author} {\bibfnamefont {Y.}~\bibnamefont {Xu}},\ }\href {\doibase
  10.1126/sciadv.aaw5685} {\bibfield  {journal} {\bibinfo  {journal} {Sci.
  Adv.}\ }\textbf {\bibinfo {volume} {5}},\ \bibinfo {pages} {eaaw5685}
  (\bibinfo {year} {2019})}\BibitemShut {NoStop}%
\bibitem [{\citenamefont {Zhang}\ \emph
  {et~al.}(2019{\natexlab{b}})\citenamefont {Zhang}, \citenamefont {Shi},
  \citenamefont {Zhu}, \citenamefont {Xing}, \citenamefont {Zhang},\ and\
  \citenamefont {Wang}}]{Zhang2019d}%
  \BibitemOpen
  \bibfield  {author} {\bibinfo {author} {\bibfnamefont {D.}~\bibnamefont
  {Zhang}}, \bibinfo {author} {\bibfnamefont {M.}~\bibnamefont {Shi}}, \bibinfo
  {author} {\bibfnamefont {T.}~\bibnamefont {Zhu}}, \bibinfo {author}
  {\bibfnamefont {D.}~\bibnamefont {Xing}}, \bibinfo {author} {\bibfnamefont
  {H.}~\bibnamefont {Zhang}}, \ and\ \bibinfo {author} {\bibfnamefont
  {J.}~\bibnamefont {Wang}},\ }\href {\doibase 10.1103/PhysRevLett.122.206401}
  {\bibfield  {journal} {\bibinfo  {journal} {Phys. Rev. Lett.}\ }\textbf
  {\bibinfo {volume} {122}},\ \bibinfo {pages} {206401} (\bibinfo {year}
  {2019}{\natexlab{b}})}\BibitemShut {NoStop}%
\bibitem [{\citenamefont {Gong}\ \emph {et~al.}(2019)\citenamefont {Gong},
  \citenamefont {Guo}, \citenamefont {Li}, \citenamefont {Zhu}, \citenamefont
  {Liao}, \citenamefont {Liu}, \citenamefont {Zhang}, \citenamefont {Gu},
  \citenamefont {Tang}, \citenamefont {Feng}, \citenamefont {Zhang},
  \citenamefont {Li}, \citenamefont {Song}, \citenamefont {Wang}, \citenamefont
  {Yu}, \citenamefont {Chen}, \citenamefont {Wang}, \citenamefont {Yao},
  \citenamefont {Duan}, \citenamefont {Xu}, \citenamefont {Zhang},
  \citenamefont {Ma}, \citenamefont {Xue},\ and\ \citenamefont
  {He}}]{Gong2019}%
  \BibitemOpen
  \bibfield  {author} {\bibinfo {author} {\bibfnamefont {Y.}~\bibnamefont
  {Gong}}, \bibinfo {author} {\bibfnamefont {J.}~\bibnamefont {Guo}}, \bibinfo
  {author} {\bibfnamefont {J.}~\bibnamefont {Li}}, \bibinfo {author}
  {\bibfnamefont {K.}~\bibnamefont {Zhu}}, \bibinfo {author} {\bibfnamefont
  {M.}~\bibnamefont {Liao}}, \bibinfo {author} {\bibfnamefont {X.}~\bibnamefont
  {Liu}}, \bibinfo {author} {\bibfnamefont {Q.}~\bibnamefont {Zhang}}, \bibinfo
  {author} {\bibfnamefont {L.}~\bibnamefont {Gu}}, \bibinfo {author}
  {\bibfnamefont {L.}~\bibnamefont {Tang}}, \bibinfo {author} {\bibfnamefont
  {X.}~\bibnamefont {Feng}}, \bibinfo {author} {\bibfnamefont {D.}~\bibnamefont
  {Zhang}}, \bibinfo {author} {\bibfnamefont {W.}~\bibnamefont {Li}}, \bibinfo
  {author} {\bibfnamefont {C.}~\bibnamefont {Song}}, \bibinfo {author}
  {\bibfnamefont {L.}~\bibnamefont {Wang}}, \bibinfo {author} {\bibfnamefont
  {P.}~\bibnamefont {Yu}}, \bibinfo {author} {\bibfnamefont {X.}~\bibnamefont
  {Chen}}, \bibinfo {author} {\bibfnamefont {Y.}~\bibnamefont {Wang}}, \bibinfo
  {author} {\bibfnamefont {H.}~\bibnamefont {Yao}}, \bibinfo {author}
  {\bibfnamefont {W.}~\bibnamefont {Duan}}, \bibinfo {author} {\bibfnamefont
  {Y.}~\bibnamefont {Xu}}, \bibinfo {author} {\bibfnamefont {S.-C.}\
  \bibnamefont {Zhang}}, \bibinfo {author} {\bibfnamefont {X.}~\bibnamefont
  {Ma}}, \bibinfo {author} {\bibfnamefont {Q.-K.}\ \bibnamefont {Xue}}, \ and\
  \bibinfo {author} {\bibfnamefont {K.}~\bibnamefont {He}},\ }\href {\doibase
  10.1088/0256-307X/36/7/076801} {\bibfield  {journal} {\bibinfo  {journal}
  {Chin. Phys. Lett.}\ }\textbf {\bibinfo {volume} {36}},\ \bibinfo {pages}
  {076801} (\bibinfo {year} {2019})}\BibitemShut {NoStop}%
\bibitem [{\citenamefont {Park}\ \emph {et~al.}(2011)\citenamefont {Park},
  \citenamefont {Lee}, \citenamefont {Wolff-Fabris}, \citenamefont {Koh},
  \citenamefont {Eom}, \citenamefont {Kim}, \citenamefont {Farhan},
  \citenamefont {Jo}, \citenamefont {Kim}, \citenamefont {Shim},\ and\
  \citenamefont {Kim}}]{Park2011}%
  \BibitemOpen
  \bibfield  {author} {\bibinfo {author} {\bibfnamefont {J.}~\bibnamefont
  {Park}}, \bibinfo {author} {\bibfnamefont {G.}~\bibnamefont {Lee}}, \bibinfo
  {author} {\bibfnamefont {F.}~\bibnamefont {Wolff-Fabris}}, \bibinfo {author}
  {\bibfnamefont {Y.~Y.}\ \bibnamefont {Koh}}, \bibinfo {author} {\bibfnamefont
  {M.~J.}\ \bibnamefont {Eom}}, \bibinfo {author} {\bibfnamefont {Y.~K.}\
  \bibnamefont {Kim}}, \bibinfo {author} {\bibfnamefont {M.~A.}\ \bibnamefont
  {Farhan}}, \bibinfo {author} {\bibfnamefont {Y.~J.}\ \bibnamefont {Jo}},
  \bibinfo {author} {\bibfnamefont {C.}~\bibnamefont {Kim}}, \bibinfo {author}
  {\bibfnamefont {J.~H.}\ \bibnamefont {Shim}}, \ and\ \bibinfo {author}
  {\bibfnamefont {J.~S.}\ \bibnamefont {Kim}},\ }\href {\doibase
  10.1103/PhysRevLett.107.126402} {\bibfield  {journal} {\bibinfo  {journal}
  {Phys. Rev. Lett.}\ }\textbf {\bibinfo {volume} {107}},\ \bibinfo {pages}
  {126402} (\bibinfo {year} {2011})}\BibitemShut {NoStop}%
\bibitem [{\citenamefont {Lee}\ \emph {et~al.}(2013)\citenamefont {Lee},
  \citenamefont {Farhan}, \citenamefont {Kim},\ and\ \citenamefont
  {Shim}}]{Lee2013}%
  \BibitemOpen
  \bibfield  {author} {\bibinfo {author} {\bibfnamefont {G.}~\bibnamefont
  {Lee}}, \bibinfo {author} {\bibfnamefont {M.~A.}\ \bibnamefont {Farhan}},
  \bibinfo {author} {\bibfnamefont {J.~S.}\ \bibnamefont {Kim}}, \ and\
  \bibinfo {author} {\bibfnamefont {J.~H.}\ \bibnamefont {Shim}},\ }\href
  {\doibase 10.1103/PhysRevB.87.245104} {\bibfield  {journal} {\bibinfo
  {journal} {Phys. Rev. B}\ }\textbf {\bibinfo {volume} {87}},\ \bibinfo
  {pages} {245104} (\bibinfo {year} {2013})}\BibitemShut {NoStop}%
\bibitem [{\citenamefont {Farhan}\ \emph {et~al.}(2014)\citenamefont {Farhan},
  \citenamefont {Lee},\ and\ \citenamefont {Shim}}]{Farhan2014}%
  \BibitemOpen
  \bibfield  {author} {\bibinfo {author} {\bibfnamefont {M.~A.}\ \bibnamefont
  {Farhan}}, \bibinfo {author} {\bibfnamefont {G.}~\bibnamefont {Lee}}, \ and\
  \bibinfo {author} {\bibfnamefont {J.~H.}\ \bibnamefont {Shim}},\ }\href
  {\doibase 10.1088/0953-8984/26/4/042201} {\bibfield  {journal} {\bibinfo
  {journal} {J. Phys. Condens. Matter}\ }\textbf {\bibinfo {volume} {26}},\
  \bibinfo {pages} {042201} (\bibinfo {year} {2014})}\BibitemShut {NoStop}%
\bibitem [{\citenamefont {Wang}\ \emph {et~al.}(2011)\citenamefont {Wang},
  \citenamefont {Zhao}, \citenamefont {Yin}, \citenamefont {Kotliar},
  \citenamefont {Kim}, \citenamefont {Aronson},\ and\ \citenamefont
  {Morosan}}]{Wang2011}%
  \BibitemOpen
  \bibfield  {author} {\bibinfo {author} {\bibfnamefont {J.~K.}\ \bibnamefont
  {Wang}}, \bibinfo {author} {\bibfnamefont {L.~L.}\ \bibnamefont {Zhao}},
  \bibinfo {author} {\bibfnamefont {Q.}~\bibnamefont {Yin}}, \bibinfo {author}
  {\bibfnamefont {G.}~\bibnamefont {Kotliar}}, \bibinfo {author} {\bibfnamefont
  {M.~S.}\ \bibnamefont {Kim}}, \bibinfo {author} {\bibfnamefont {M.~C.}\
  \bibnamefont {Aronson}}, \ and\ \bibinfo {author} {\bibfnamefont
  {E.}~\bibnamefont {Morosan}},\ }\href {\doibase 10.1103/PhysRevB.84.064428}
  {\bibfield  {journal} {\bibinfo  {journal} {Phys. Rev. B}\ }\textbf {\bibinfo
  {volume} {84}},\ \bibinfo {pages} {064428} (\bibinfo {year}
  {2011})}\BibitemShut {NoStop}%
\bibitem [{\citenamefont {Wang}\ \emph
  {et~al.}(2012{\natexlab{a}})\citenamefont {Wang}, \citenamefont {Graf},
  \citenamefont {Wang}, \citenamefont {Lei}, \citenamefont {Tozer},\ and\
  \citenamefont {Petrovic}}]{Wang2012}%
  \BibitemOpen
  \bibfield  {author} {\bibinfo {author} {\bibfnamefont {K.}~\bibnamefont
  {Wang}}, \bibinfo {author} {\bibfnamefont {D.}~\bibnamefont {Graf}}, \bibinfo
  {author} {\bibfnamefont {L.}~\bibnamefont {Wang}}, \bibinfo {author}
  {\bibfnamefont {H.}~\bibnamefont {Lei}}, \bibinfo {author} {\bibfnamefont
  {S.~W.}\ \bibnamefont {Tozer}}, \ and\ \bibinfo {author} {\bibfnamefont
  {C.}~\bibnamefont {Petrovic}},\ }\href {\doibase 10.1103/PhysRevB.85.041101}
  {\bibfield  {journal} {\bibinfo  {journal} {Phys. Rev. B}\ }\textbf {\bibinfo
  {volume} {85}},\ \bibinfo {pages} {041101(R)} (\bibinfo {year}
  {2012}{\natexlab{a}})}\BibitemShut {NoStop}%
\bibitem [{\citenamefont {Borisenko}\ \emph {et~al.}(2019)\citenamefont
  {Borisenko}, \citenamefont {Evtushinsky}, \citenamefont {Gibson},
  \citenamefont {Yaresko}, \citenamefont {Koepernik}, \citenamefont {Kim},
  \citenamefont {Ali}, \citenamefont {van~den Brink}, \citenamefont {Hoesch},
  \citenamefont {Fedorov}, \citenamefont {Haubold}, \citenamefont
  {Kushnirenko}, \citenamefont {Soldatov}, \citenamefont {Sch{\"{a}}fer},\ and\
  \citenamefont {Cava}}]{Borisenko2015a}%
  \BibitemOpen
  \bibfield  {author} {\bibinfo {author} {\bibfnamefont {S.}~\bibnamefont
  {Borisenko}}, \bibinfo {author} {\bibfnamefont {D.}~\bibnamefont
  {Evtushinsky}}, \bibinfo {author} {\bibfnamefont {Q.}~\bibnamefont {Gibson}},
  \bibinfo {author} {\bibfnamefont {A.}~\bibnamefont {Yaresko}}, \bibinfo
  {author} {\bibfnamefont {K.}~\bibnamefont {Koepernik}}, \bibinfo {author}
  {\bibfnamefont {T.}~\bibnamefont {Kim}}, \bibinfo {author} {\bibfnamefont
  {M.}~\bibnamefont {Ali}}, \bibinfo {author} {\bibfnamefont {J.}~\bibnamefont
  {van~den Brink}}, \bibinfo {author} {\bibfnamefont {M.}~\bibnamefont
  {Hoesch}}, \bibinfo {author} {\bibfnamefont {A.}~\bibnamefont {Fedorov}},
  \bibinfo {author} {\bibfnamefont {E.}~\bibnamefont {Haubold}}, \bibinfo
  {author} {\bibfnamefont {Y.}~\bibnamefont {Kushnirenko}}, \bibinfo {author}
  {\bibfnamefont {I.}~\bibnamefont {Soldatov}}, \bibinfo {author}
  {\bibfnamefont {R.}~\bibnamefont {Sch{\"{a}}fer}}, \ and\ \bibinfo {author}
  {\bibfnamefont {R.~J.}\ \bibnamefont {Cava}},\ }\href {\doibase
  10.1038/s41467-019-11393-5} {\bibfield  {journal} {\bibinfo  {journal} {Nat.
  Commun}\ }\textbf {\bibinfo {volume} {10}},\ \bibinfo {pages} {3424}
  (\bibinfo {year} {2019})}\BibitemShut {NoStop}%
\bibitem [{\citenamefont {He}\ \emph {et~al.}(2012)\citenamefont {He},
  \citenamefont {Wang},\ and\ \citenamefont {Chen}}]{He2012}%
  \BibitemOpen
  \bibfield  {author} {\bibinfo {author} {\bibfnamefont {J.~B.}\ \bibnamefont
  {He}}, \bibinfo {author} {\bibfnamefont {D.~M.}\ \bibnamefont {Wang}}, \ and\
  \bibinfo {author} {\bibfnamefont {G.~F.}\ \bibnamefont {Chen}},\ }\href
  {\doibase 10.1063/1.3694760} {\bibfield  {journal} {\bibinfo  {journal}
  {Appl. Phys. Lett.}\ }\textbf {\bibinfo {volume} {100}},\ \bibinfo {pages}
  {112405} (\bibinfo {year} {2012})}\BibitemShut {NoStop}%
\bibitem [{\citenamefont {Wang}\ \emph
  {et~al.}(2012{\natexlab{b}})\citenamefont {Wang}, \citenamefont {Wang},\ and\
  \citenamefont {Petrovic}}]{Wang2012b}%
  \BibitemOpen
  \bibfield  {author} {\bibinfo {author} {\bibfnamefont {K.}~\bibnamefont
  {Wang}}, \bibinfo {author} {\bibfnamefont {L.}~\bibnamefont {Wang}}, \ and\
  \bibinfo {author} {\bibfnamefont {C.}~\bibnamefont {Petrovic}},\ }\href
  {\doibase 10.1063/1.3695155} {\bibfield  {journal} {\bibinfo  {journal}
  {Appl. Phys. Lett.}\ }\textbf {\bibinfo {volume} {100}},\ \bibinfo {pages}
  {112111} (\bibinfo {year} {2012}{\natexlab{b}})}\BibitemShut {NoStop}%
\bibitem [{\citenamefont {Feng}\ \emph {et~al.}(2015)\citenamefont {Feng},
  \citenamefont {Wang}, \citenamefont {Chen}, \citenamefont {Shi},
  \citenamefont {Xie}, \citenamefont {Yi}, \citenamefont {Liang}, \citenamefont
  {He}, \citenamefont {He}, \citenamefont {Peng}, \citenamefont {Liu},
  \citenamefont {Liu}, \citenamefont {Zhao}, \citenamefont {Liu}, \citenamefont
  {Dong}, \citenamefont {Zhang}, \citenamefont {Chen}, \citenamefont {Xu},
  \citenamefont {Dai}, \citenamefont {Fang},\ and\ \citenamefont
  {Zhou}}]{Feng2015}%
  \BibitemOpen
  \bibfield  {author} {\bibinfo {author} {\bibfnamefont {Y.}~\bibnamefont
  {Feng}}, \bibinfo {author} {\bibfnamefont {Z.}~\bibnamefont {Wang}}, \bibinfo
  {author} {\bibfnamefont {C.}~\bibnamefont {Chen}}, \bibinfo {author}
  {\bibfnamefont {Y.}~\bibnamefont {Shi}}, \bibinfo {author} {\bibfnamefont
  {Z.}~\bibnamefont {Xie}}, \bibinfo {author} {\bibfnamefont {H.}~\bibnamefont
  {Yi}}, \bibinfo {author} {\bibfnamefont {A.}~\bibnamefont {Liang}}, \bibinfo
  {author} {\bibfnamefont {S.}~\bibnamefont {He}}, \bibinfo {author}
  {\bibfnamefont {J.}~\bibnamefont {He}}, \bibinfo {author} {\bibfnamefont
  {Y.}~\bibnamefont {Peng}}, \bibinfo {author} {\bibfnamefont {X.}~\bibnamefont
  {Liu}}, \bibinfo {author} {\bibfnamefont {Y.}~\bibnamefont {Liu}}, \bibinfo
  {author} {\bibfnamefont {L.}~\bibnamefont {Zhao}}, \bibinfo {author}
  {\bibfnamefont {G.}~\bibnamefont {Liu}}, \bibinfo {author} {\bibfnamefont
  {X.}~\bibnamefont {Dong}}, \bibinfo {author} {\bibfnamefont {J.}~\bibnamefont
  {Zhang}}, \bibinfo {author} {\bibfnamefont {C.}~\bibnamefont {Chen}},
  \bibinfo {author} {\bibfnamefont {Z.}~\bibnamefont {Xu}}, \bibinfo {author}
  {\bibfnamefont {X.}~\bibnamefont {Dai}}, \bibinfo {author} {\bibfnamefont
  {Z.}~\bibnamefont {Fang}}, \ and\ \bibinfo {author} {\bibfnamefont {X.~J.}\
  \bibnamefont {Zhou}},\ }\href {\doibase 10.1038/srep05385} {\bibfield
  {journal} {\bibinfo  {journal} {Sci. Rep.}\ }\textbf {\bibinfo {volume}
  {4}},\ \bibinfo {pages} {5385} (\bibinfo {year} {2015})}\BibitemShut
  {NoStop}%
\bibitem [{\citenamefont {Li}\ \emph {et~al.}(2016)\citenamefont {Li},
  \citenamefont {Wang}, \citenamefont {Graf}, \citenamefont {Wang},
  \citenamefont {Wang},\ and\ \citenamefont {Petrovic}}]{Li2016}%
  \BibitemOpen
  \bibfield  {author} {\bibinfo {author} {\bibfnamefont {L.}~\bibnamefont
  {Li}}, \bibinfo {author} {\bibfnamefont {K.}~\bibnamefont {Wang}}, \bibinfo
  {author} {\bibfnamefont {D.}~\bibnamefont {Graf}}, \bibinfo {author}
  {\bibfnamefont {L.}~\bibnamefont {Wang}}, \bibinfo {author} {\bibfnamefont
  {A.}~\bibnamefont {Wang}}, \ and\ \bibinfo {author} {\bibfnamefont
  {C.}~\bibnamefont {Petrovic}},\ }\href {\doibase 10.1103/PhysRevB.93.115141}
  {\bibfield  {journal} {\bibinfo  {journal} {Phys. Rev. B}\ }\textbf {\bibinfo
  {volume} {93}},\ \bibinfo {pages} {115141} (\bibinfo {year}
  {2016})}\BibitemShut {NoStop}%
\bibitem [{\citenamefont {Wang}\ \emph {et~al.}(2016)\citenamefont {Wang},
  \citenamefont {Yu},\ and\ \citenamefont {Xia}}]{Wang2016c}%
  \BibitemOpen
  \bibfield  {author} {\bibinfo {author} {\bibfnamefont {Y.-Y.}\ \bibnamefont
  {Wang}}, \bibinfo {author} {\bibfnamefont {Q.-H.}\ \bibnamefont {Yu}}, \ and\
  \bibinfo {author} {\bibfnamefont {T.-L.}\ \bibnamefont {Xia}},\ }\href
  {\doibase 10.1088/1674-1056/25/10/107503} {\bibfield  {journal} {\bibinfo
  {journal} {Chin. Phys. B}\ }\textbf {\bibinfo {volume} {25}},\ \bibinfo
  {pages} {107503} (\bibinfo {year} {2016})}\BibitemShut {NoStop}%
\bibitem [{\citenamefont {Liu}\ \emph {et~al.}(2017{\natexlab{a}})\citenamefont
  {Liu}, \citenamefont {Hu}, \citenamefont {Graf}, \citenamefont {Zou},
  \citenamefont {Zhu}, \citenamefont {Shi}, \citenamefont {Che}, \citenamefont
  {Radmanesh}, \citenamefont {Lau}, \citenamefont {Spinu}, \citenamefont {Cao},
  \citenamefont {Ke},\ and\ \citenamefont {Mao}}]{Liu2017b}%
  \BibitemOpen
  \bibfield  {author} {\bibinfo {author} {\bibfnamefont {J.~Y.}\ \bibnamefont
  {Liu}}, \bibinfo {author} {\bibfnamefont {J.}~\bibnamefont {Hu}}, \bibinfo
  {author} {\bibfnamefont {D.}~\bibnamefont {Graf}}, \bibinfo {author}
  {\bibfnamefont {T.}~\bibnamefont {Zou}}, \bibinfo {author} {\bibfnamefont
  {M.}~\bibnamefont {Zhu}}, \bibinfo {author} {\bibfnamefont {Y.}~\bibnamefont
  {Shi}}, \bibinfo {author} {\bibfnamefont {S.}~\bibnamefont {Che}}, \bibinfo
  {author} {\bibfnamefont {S.~M.~A.}\ \bibnamefont {Radmanesh}}, \bibinfo
  {author} {\bibfnamefont {C.~N.}\ \bibnamefont {Lau}}, \bibinfo {author}
  {\bibfnamefont {L.}~\bibnamefont {Spinu}}, \bibinfo {author} {\bibfnamefont
  {H.~B.}\ \bibnamefont {Cao}}, \bibinfo {author} {\bibfnamefont
  {X.}~\bibnamefont {Ke}}, \ and\ \bibinfo {author} {\bibfnamefont {Z.~Q.}\
  \bibnamefont {Mao}},\ }\href {\doibase 10.1038/s41467-017-00673-7} {\bibfield
   {journal} {\bibinfo  {journal} {Nat. Commun.}\ }\textbf {\bibinfo {volume}
  {8}},\ \bibinfo {pages} {646} (\bibinfo {year}
  {2017}{\natexlab{a}})}\BibitemShut {NoStop}%
\bibitem [{\citenamefont {Park}\ \emph {et~al.}(2016)\citenamefont {Park},
  \citenamefont {Sandilands}, \citenamefont {You}, \citenamefont {Ji},
  \citenamefont {Sohn}, \citenamefont {Han}, \citenamefont {Moon},
  \citenamefont {Kim}, \citenamefont {Shim}, \citenamefont {Kim},\ and\
  \citenamefont {Noh}}]{Park2016}%
  \BibitemOpen
  \bibfield  {author} {\bibinfo {author} {\bibfnamefont {H.~J.}\ \bibnamefont
  {Park}}, \bibinfo {author} {\bibfnamefont {L.~J.}\ \bibnamefont
  {Sandilands}}, \bibinfo {author} {\bibfnamefont {J.~S.}\ \bibnamefont {You}},
  \bibinfo {author} {\bibfnamefont {H.~S.}\ \bibnamefont {Ji}}, \bibinfo
  {author} {\bibfnamefont {C.~H.}\ \bibnamefont {Sohn}}, \bibinfo {author}
  {\bibfnamefont {J.~W.}\ \bibnamefont {Han}}, \bibinfo {author} {\bibfnamefont
  {S.~J.}\ \bibnamefont {Moon}}, \bibinfo {author} {\bibfnamefont {K.~W.}\
  \bibnamefont {Kim}}, \bibinfo {author} {\bibfnamefont {J.~H.}\ \bibnamefont
  {Shim}}, \bibinfo {author} {\bibfnamefont {J.~S.}\ \bibnamefont {Kim}}, \
  and\ \bibinfo {author} {\bibfnamefont {T.~W.}\ \bibnamefont {Noh}},\ }\href
  {\doibase 10.1103/PhysRevB.93.205122} {\bibfield  {journal} {\bibinfo
  {journal} {Phys. Rev. B}\ }\textbf {\bibinfo {volume} {93}},\ \bibinfo
  {pages} {205122} (\bibinfo {year} {2016})}\BibitemShut {NoStop}%
\bibitem [{\citenamefont {Liu}\ \emph {et~al.}(2017{\natexlab{b}})\citenamefont
  {Liu}, \citenamefont {Hu}, \citenamefont {Zhang}, \citenamefont {Graf},
  \citenamefont {Cao}, \citenamefont {Radmanesh}, \citenamefont {Adams},
  \citenamefont {Zhu}, \citenamefont {Cheng}, \citenamefont {Liu},
  \citenamefont {Phelan}, \citenamefont {Wei}, \citenamefont {Jaime},
  \citenamefont {Balakirev}, \citenamefont {Tennant}, \citenamefont {DiTusa},
  \citenamefont {Chiorescu}, \citenamefont {Spinu},\ and\ \citenamefont
  {Mao}}]{Liu2017}%
  \BibitemOpen
  \bibfield  {author} {\bibinfo {author} {\bibfnamefont {J.~Y.}\ \bibnamefont
  {Liu}}, \bibinfo {author} {\bibfnamefont {J.}~\bibnamefont {Hu}}, \bibinfo
  {author} {\bibfnamefont {Q.}~\bibnamefont {Zhang}}, \bibinfo {author}
  {\bibfnamefont {D.}~\bibnamefont {Graf}}, \bibinfo {author} {\bibfnamefont
  {H.~B.}\ \bibnamefont {Cao}}, \bibinfo {author} {\bibfnamefont {S.~M.~A.}\
  \bibnamefont {Radmanesh}}, \bibinfo {author} {\bibfnamefont {D.~J.}\
  \bibnamefont {Adams}}, \bibinfo {author} {\bibfnamefont {Y.~L.}\ \bibnamefont
  {Zhu}}, \bibinfo {author} {\bibfnamefont {G.~F.}\ \bibnamefont {Cheng}},
  \bibinfo {author} {\bibfnamefont {X.}~\bibnamefont {Liu}}, \bibinfo {author}
  {\bibfnamefont {W.~A.}\ \bibnamefont {Phelan}}, \bibinfo {author}
  {\bibfnamefont {J.}~\bibnamefont {Wei}}, \bibinfo {author} {\bibfnamefont
  {M.}~\bibnamefont {Jaime}}, \bibinfo {author} {\bibfnamefont
  {F.}~\bibnamefont {Balakirev}}, \bibinfo {author} {\bibfnamefont {D.~A.}\
  \bibnamefont {Tennant}}, \bibinfo {author} {\bibfnamefont {J.~F.}\
  \bibnamefont {DiTusa}}, \bibinfo {author} {\bibfnamefont {I.}~\bibnamefont
  {Chiorescu}}, \bibinfo {author} {\bibfnamefont {L.}~\bibnamefont {Spinu}}, \
  and\ \bibinfo {author} {\bibfnamefont {Z.~Q.}\ \bibnamefont {Mao}},\ }\href
  {\doibase 10.1038/nmat4953} {\bibfield  {journal} {\bibinfo  {journal} {Nat.
  Mater.}\ }\textbf {\bibinfo {volume} {16}},\ \bibinfo {pages} {905} (\bibinfo
  {year} {2017}{\natexlab{b}})}\BibitemShut {NoStop}%
\bibitem [{\citenamefont {Liu}\ \emph {et~al.}(2016)\citenamefont {Liu},
  \citenamefont {Hu}, \citenamefont {Cao}, \citenamefont {Zhu}, \citenamefont
  {Chuang}, \citenamefont {Graf}, \citenamefont {Adams}, \citenamefont
  {Radmanesh}, \citenamefont {Spinu}, \citenamefont {Chiorescu},\ and\
  \citenamefont {Mao}}]{Liu2016}%
  \BibitemOpen
  \bibfield  {author} {\bibinfo {author} {\bibfnamefont {J.}~\bibnamefont
  {Liu}}, \bibinfo {author} {\bibfnamefont {J.}~\bibnamefont {Hu}}, \bibinfo
  {author} {\bibfnamefont {H.}~\bibnamefont {Cao}}, \bibinfo {author}
  {\bibfnamefont {Y.}~\bibnamefont {Zhu}}, \bibinfo {author} {\bibfnamefont
  {A.}~\bibnamefont {Chuang}}, \bibinfo {author} {\bibfnamefont
  {D.}~\bibnamefont {Graf}}, \bibinfo {author} {\bibfnamefont {D.~J.}\
  \bibnamefont {Adams}}, \bibinfo {author} {\bibfnamefont {S.~M.~A.}\
  \bibnamefont {Radmanesh}}, \bibinfo {author} {\bibfnamefont {L.}~\bibnamefont
  {Spinu}}, \bibinfo {author} {\bibfnamefont {I.}~\bibnamefont {Chiorescu}}, \
  and\ \bibinfo {author} {\bibfnamefont {Z.}~\bibnamefont {Mao}},\ }\href
  {\doibase 10.1038/srep30525} {\bibfield  {journal} {\bibinfo  {journal} {Sci.
  Rep.}\ }\textbf {\bibinfo {volume} {6}},\ \bibinfo {pages} {30525} (\bibinfo
  {year} {2016})}\BibitemShut {NoStop}%
\bibitem [{\citenamefont {Huang}\ \emph {et~al.}(2017)\citenamefont {Huang},
  \citenamefont {Kim}, \citenamefont {Shelton}, \citenamefont {Plummer},\ and\
  \citenamefont {Jin}}]{Huang2017}%
  \BibitemOpen
  \bibfield  {author} {\bibinfo {author} {\bibfnamefont {S.}~\bibnamefont
  {Huang}}, \bibinfo {author} {\bibfnamefont {J.}~\bibnamefont {Kim}}, \bibinfo
  {author} {\bibfnamefont {W.~A.}\ \bibnamefont {Shelton}}, \bibinfo {author}
  {\bibfnamefont {E.~W.}\ \bibnamefont {Plummer}}, \ and\ \bibinfo {author}
  {\bibfnamefont {R.}~\bibnamefont {Jin}},\ }\href {\doibase
  10.1073/pnas.1706657114} {\bibfield  {journal} {\bibinfo  {journal} {Proc.
  Natl. Acad. Sci.}\ }\textbf {\bibinfo {volume} {114}},\ \bibinfo {pages}
  {6256} (\bibinfo {year} {2017})}\BibitemShut {NoStop}%
\bibitem [{\citenamefont {Zhu}\ \emph {et~al.}(2019)\citenamefont {Zhu},
  \citenamefont {Mao}, \citenamefont {Xu}, \citenamefont {Du}, \citenamefont
  {Chen}, \citenamefont {Yang}, \citenamefont {Chen}, \citenamefont {Cao},
  \citenamefont {Wang},\ and\ \citenamefont {Fang}}]{Zhu2019}%
  \BibitemOpen
  \bibfield  {author} {\bibinfo {author} {\bibfnamefont {Q.}~\bibnamefont
  {Zhu}}, \bibinfo {author} {\bibfnamefont {Q.}~\bibnamefont {Mao}}, \bibinfo
  {author} {\bibfnamefont {B.}~\bibnamefont {Xu}}, \bibinfo {author}
  {\bibfnamefont {J.}~\bibnamefont {Du}}, \bibinfo {author} {\bibfnamefont
  {M.}~\bibnamefont {Chen}}, \bibinfo {author} {\bibfnamefont {J.}~\bibnamefont
  {Yang}}, \bibinfo {author} {\bibfnamefont {B.}~\bibnamefont {Chen}}, \bibinfo
  {author} {\bibfnamefont {C.}~\bibnamefont {Cao}}, \bibinfo {author}
  {\bibfnamefont {H.}~\bibnamefont {Wang}}, \ and\ \bibinfo {author}
  {\bibfnamefont {M.}~\bibnamefont {Fang}},\ }\href {\doibase
  10.1088/1361-648X/ab0482} {\bibfield  {journal} {\bibinfo  {journal} {J.
  Phys. Condens. Matter}\ }\textbf {\bibinfo {volume} {31}},\ \bibinfo {pages}
  {185701} (\bibinfo {year} {2019})}\BibitemShut {NoStop}%
\bibitem [{\citenamefont {Kealhofer}\ \emph {et~al.}(2018)\citenamefont
  {Kealhofer}, \citenamefont {Jang}, \citenamefont {Griffin}, \citenamefont
  {John}, \citenamefont {Benavides}, \citenamefont {Doyle}, \citenamefont
  {Helm}, \citenamefont {Moll}, \citenamefont {Neaton}, \citenamefont {Chan},
  \citenamefont {Denlinger},\ and\ \citenamefont {Analytis}}]{Kealhofer2018}%
  \BibitemOpen
  \bibfield  {author} {\bibinfo {author} {\bibfnamefont {R.}~\bibnamefont
  {Kealhofer}}, \bibinfo {author} {\bibfnamefont {S.}~\bibnamefont {Jang}},
  \bibinfo {author} {\bibfnamefont {S.~M.}\ \bibnamefont {Griffin}}, \bibinfo
  {author} {\bibfnamefont {C.}~\bibnamefont {John}}, \bibinfo {author}
  {\bibfnamefont {K.~A.}\ \bibnamefont {Benavides}}, \bibinfo {author}
  {\bibfnamefont {S.}~\bibnamefont {Doyle}}, \bibinfo {author} {\bibfnamefont
  {T.}~\bibnamefont {Helm}}, \bibinfo {author} {\bibfnamefont {P.~J.~W.}\
  \bibnamefont {Moll}}, \bibinfo {author} {\bibfnamefont {J.~B.}\ \bibnamefont
  {Neaton}}, \bibinfo {author} {\bibfnamefont {J.~Y.}\ \bibnamefont {Chan}},
  \bibinfo {author} {\bibfnamefont {J.~D.}\ \bibnamefont {Denlinger}}, \ and\
  \bibinfo {author} {\bibfnamefont {J.~G.}\ \bibnamefont {Analytis}},\ }\href
  {\doibase 10.1103/PhysRevB.97.045109} {\bibfield  {journal} {\bibinfo
  {journal} {Phys. Rev. B}\ }\textbf {\bibinfo {volume} {97}},\ \bibinfo
  {pages} {45109} (\bibinfo {year} {2018})}\BibitemShut {NoStop}%
\bibitem [{\citenamefont {Masuda}\ \emph {et~al.}(2018)\citenamefont {Masuda},
  \citenamefont {Sakai}, \citenamefont {Tokunaga}, \citenamefont {Ochi},
  \citenamefont {Takahashi}, \citenamefont {Akiba}, \citenamefont {Miyake},
  \citenamefont {Kuroki}, \citenamefont {Tokura},\ and\ \citenamefont
  {Ishiwata}}]{Masuda2018}%
  \BibitemOpen
  \bibfield  {author} {\bibinfo {author} {\bibfnamefont {H.}~\bibnamefont
  {Masuda}}, \bibinfo {author} {\bibfnamefont {H.}~\bibnamefont {Sakai}},
  \bibinfo {author} {\bibfnamefont {M.}~\bibnamefont {Tokunaga}}, \bibinfo
  {author} {\bibfnamefont {M.}~\bibnamefont {Ochi}}, \bibinfo {author}
  {\bibfnamefont {H.}~\bibnamefont {Takahashi}}, \bibinfo {author}
  {\bibfnamefont {K.}~\bibnamefont {Akiba}}, \bibinfo {author} {\bibfnamefont
  {A.}~\bibnamefont {Miyake}}, \bibinfo {author} {\bibfnamefont
  {K.}~\bibnamefont {Kuroki}}, \bibinfo {author} {\bibfnamefont
  {Y.}~\bibnamefont {Tokura}}, \ and\ \bibinfo {author} {\bibfnamefont
  {S.}~\bibnamefont {Ishiwata}},\ }\href {\doibase 10.1103/PhysRevB.98.161108}
  {\bibfield  {journal} {\bibinfo  {journal} {Phys. Rev. B}\ }\textbf {\bibinfo
  {volume} {98}},\ \bibinfo {pages} {161108(R)} (\bibinfo {year}
  {2018})}\BibitemShut {NoStop}%
\bibitem [{\citenamefont {Rahn}\ \emph {et~al.}(2017)\citenamefont {Rahn},
  \citenamefont {Princep}, \citenamefont {Piovano}, \citenamefont {Kulda},
  \citenamefont {Guo}, \citenamefont {Shi},\ and\ \citenamefont
  {Boothroyd}}]{Rahn2017}%
  \BibitemOpen
  \bibfield  {author} {\bibinfo {author} {\bibfnamefont {M.~C.}\ \bibnamefont
  {Rahn}}, \bibinfo {author} {\bibfnamefont {A.~J.}\ \bibnamefont {Princep}},
  \bibinfo {author} {\bibfnamefont {A.}~\bibnamefont {Piovano}}, \bibinfo
  {author} {\bibfnamefont {J.}~\bibnamefont {Kulda}}, \bibinfo {author}
  {\bibfnamefont {Y.~F.}\ \bibnamefont {Guo}}, \bibinfo {author} {\bibfnamefont
  {Y.~G.}\ \bibnamefont {Shi}}, \ and\ \bibinfo {author} {\bibfnamefont
  {A.~T.}\ \bibnamefont {Boothroyd}},\ }\href {\doibase
  10.1103/PhysRevB.95.134405} {\bibfield  {journal} {\bibinfo  {journal} {Phys.
  Rev. B}\ }\textbf {\bibinfo {volume} {95}},\ \bibinfo {pages} {134405}
  (\bibinfo {year} {2017})}\BibitemShut {NoStop}%
\bibitem [{\citenamefont {Soh}\ \emph {et~al.}(2019)\citenamefont {Soh},
  \citenamefont {Jacobsen}, \citenamefont {Ouladdiaf}, \citenamefont {Ivanov},
  \citenamefont {Piovano}, \citenamefont {Tejsner}, \citenamefont {Feng},
  \citenamefont {Wang}, \citenamefont {Su}, \citenamefont {Guo}, \citenamefont
  {Shi},\ and\ \citenamefont {Boothroyd}}]{Soh2019b}%
  \BibitemOpen
  \bibfield  {author} {\bibinfo {author} {\bibfnamefont {J.-R.}\ \bibnamefont
  {Soh}}, \bibinfo {author} {\bibfnamefont {H.}~\bibnamefont {Jacobsen}},
  \bibinfo {author} {\bibfnamefont {B.}~\bibnamefont {Ouladdiaf}}, \bibinfo
  {author} {\bibfnamefont {A.}~\bibnamefont {Ivanov}}, \bibinfo {author}
  {\bibfnamefont {A.}~\bibnamefont {Piovano}}, \bibinfo {author} {\bibfnamefont
  {T.}~\bibnamefont {Tejsner}}, \bibinfo {author} {\bibfnamefont
  {Z.}~\bibnamefont {Feng}}, \bibinfo {author} {\bibfnamefont {H.}~\bibnamefont
  {Wang}}, \bibinfo {author} {\bibfnamefont {H.}~\bibnamefont {Su}}, \bibinfo
  {author} {\bibfnamefont {Y.}~\bibnamefont {Guo}}, \bibinfo {author}
  {\bibfnamefont {Y.}~\bibnamefont {Shi}}, \ and\ \bibinfo {author}
  {\bibfnamefont {A.~T.}\ \bibnamefont {Boothroyd}},\ }\href {\doibase
  10.1103/PhysRevB.100.144431} {\bibfield  {journal} {\bibinfo  {journal}
  {Phys. Rev. B}\ }\textbf {\bibinfo {volume} {100}},\ \bibinfo {pages}
  {144431} (\bibinfo {year} {2019})}\BibitemShut {NoStop}%
\bibitem [{\citenamefont {Shiomi}\ \emph {et~al.}(2019)\citenamefont {Shiomi},
  \citenamefont {Watanabe}, \citenamefont {Masuda}, \citenamefont {Takahashi},
  \citenamefont {Yanase},\ and\ \citenamefont {Ishiwata}}]{Shiomi2018a}%
  \BibitemOpen
  \bibfield  {author} {\bibinfo {author} {\bibfnamefont {Y.}~\bibnamefont
  {Shiomi}}, \bibinfo {author} {\bibfnamefont {H.}~\bibnamefont {Watanabe}},
  \bibinfo {author} {\bibfnamefont {H.}~\bibnamefont {Masuda}}, \bibinfo
  {author} {\bibfnamefont {H.}~\bibnamefont {Takahashi}}, \bibinfo {author}
  {\bibfnamefont {Y.}~\bibnamefont {Yanase}}, \ and\ \bibinfo {author}
  {\bibfnamefont {S.}~\bibnamefont {Ishiwata}},\ }\href {\doibase
  10.1103/PhysRevLett.122.127207} {\bibfield  {journal} {\bibinfo  {journal}
  {Phys. Rev. Lett.}\ }\textbf {\bibinfo {volume} {122}},\ \bibinfo {pages}
  {127207} (\bibinfo {year} {2019})}\BibitemShut {NoStop}%
\bibitem [{\citenamefont {May}\ \emph {et~al.}(2014)\citenamefont {May},
  \citenamefont {McGuire},\ and\ \citenamefont {Sales}}]{May2014}%
  \BibitemOpen
  \bibfield  {author} {\bibinfo {author} {\bibfnamefont {A.~F.}\ \bibnamefont
  {May}}, \bibinfo {author} {\bibfnamefont {M.~A.}\ \bibnamefont {McGuire}}, \
  and\ \bibinfo {author} {\bibfnamefont {B.~C.}\ \bibnamefont {Sales}},\ }\href
  {\doibase 10.1103/PhysRevB.90.075109} {\bibfield  {journal} {\bibinfo
  {journal} {Phys. Rev. B}\ }\textbf {\bibinfo {volume} {90}},\ \bibinfo
  {pages} {075109} (\bibinfo {year} {2014})}\BibitemShut {NoStop}%
\bibitem [{\citenamefont {Meven}\ and\ \citenamefont
  {Sazonov}(2015)}]{Meven2015}%
  \BibitemOpen
  \bibfield  {author} {\bibinfo {author} {\bibfnamefont {M.}~\bibnamefont
  {Meven}}\ and\ \bibinfo {author} {\bibfnamefont {A.}~\bibnamefont
  {Sazonov}},\ }\href {\doibase 10.17815/jlsrf-1-20} {\bibfield  {journal}
  {\bibinfo  {journal} {J. large-scale Res. Facil. JLSRF}\ }\textbf {\bibinfo
  {volume} {1}},\ \bibinfo {pages} {A7} (\bibinfo {year} {2015})}\BibitemShut
  {NoStop}%
\bibitem [{\citenamefont {Schweika}\ and\ \citenamefont
  {B{\"{o}}ni}(2001)}]{Schweika2001}%
  \BibitemOpen
  \bibfield  {author} {\bibinfo {author} {\bibfnamefont {W.}~\bibnamefont
  {Schweika}}\ and\ \bibinfo {author} {\bibfnamefont {P.}~\bibnamefont
  {B{\"{o}}ni}},\ }\href {\doibase 10.1016/S0921-4526(00)00858-9} {\bibfield
  {journal} {\bibinfo  {journal} {Phys. B Condens. Matter}\ }\textbf {\bibinfo
  {volume} {297}},\ \bibinfo {pages} {155} (\bibinfo {year}
  {2001})}\BibitemShut {NoStop}%
\bibitem [{\citenamefont {Su}\ \emph {et~al.}(2015)\citenamefont {Su},
  \citenamefont {Nemkovskiy},\ and\ \citenamefont {Demirdiş}}]{Su2015}%
  \BibitemOpen
  \bibfield  {author} {\bibinfo {author} {\bibfnamefont {Y.}~\bibnamefont
  {Su}}, \bibinfo {author} {\bibfnamefont {K.}~\bibnamefont {Nemkovskiy}}, \
  and\ \bibinfo {author} {\bibfnamefont {S.}~\bibnamefont {Demirdiş}},\ }\href
  {\doibase 10.17815/jlsrf-1-33} {\bibfield  {journal} {\bibinfo  {journal} {J.
  large-scale Res. Facil. JLSRF}\ }\textbf {\bibinfo {volume} {1}},\ \bibinfo
  {pages} {A27} (\bibinfo {year} {2015})}\BibitemShut {NoStop}%
\bibitem [{\citenamefont {Yi}\ \emph {et~al.}(2017)\citenamefont {Yi},
  \citenamefont {Yang}, \citenamefont {Yang}, \citenamefont {Wang},
  \citenamefont {Matsushita}, \citenamefont {Miao}, \citenamefont {Jiao},
  \citenamefont {Cheng}, \citenamefont {Li}, \citenamefont {Yamaura},
  \citenamefont {Shi},\ and\ \citenamefont {Luo}}]{Yi2017}%
  \BibitemOpen
  \bibfield  {author} {\bibinfo {author} {\bibfnamefont {C.}~\bibnamefont
  {Yi}}, \bibinfo {author} {\bibfnamefont {S.}~\bibnamefont {Yang}}, \bibinfo
  {author} {\bibfnamefont {M.}~\bibnamefont {Yang}}, \bibinfo {author}
  {\bibfnamefont {L.}~\bibnamefont {Wang}}, \bibinfo {author} {\bibfnamefont
  {Y.}~\bibnamefont {Matsushita}}, \bibinfo {author} {\bibfnamefont
  {S.}~\bibnamefont {Miao}}, \bibinfo {author} {\bibfnamefont {Y.}~\bibnamefont
  {Jiao}}, \bibinfo {author} {\bibfnamefont {J.}~\bibnamefont {Cheng}},
  \bibinfo {author} {\bibfnamefont {Y.}~\bibnamefont {Li}}, \bibinfo {author}
  {\bibfnamefont {K.}~\bibnamefont {Yamaura}}, \bibinfo {author} {\bibfnamefont
  {Y.}~\bibnamefont {Shi}}, \ and\ \bibinfo {author} {\bibfnamefont
  {J.}~\bibnamefont {Luo}},\ }\href {\doibase 10.1103/PhysRevB.96.205103}
  {\bibfield  {journal} {\bibinfo  {journal} {Phys. Rev. B}\ }\textbf {\bibinfo
  {volume} {96}},\ \bibinfo {pages} {205103} (\bibinfo {year}
  {2017})}\BibitemShut {NoStop}%
\bibitem [{\citenamefont {Petř{\'{i}}{\v{c}}ek}\ \emph
  {et~al.}(2014)\citenamefont {Petř{\'{i}}{\v{c}}ek}, \citenamefont
  {Du{\v{s}}ek},\ and\ \citenamefont {Palatinus}}]{Petricek2014a}%
  \BibitemOpen
  \bibfield  {author} {\bibinfo {author} {\bibfnamefont {V.}~\bibnamefont
  {Petř{\'{i}}{\v{c}}ek}}, \bibinfo {author} {\bibfnamefont {M.}~\bibnamefont
  {Du{\v{s}}ek}}, \ and\ \bibinfo {author} {\bibfnamefont {L.}~\bibnamefont
  {Palatinus}},\ }\href {\doibase 10.1515/zkri-2014-1737} {\bibfield  {journal}
  {\bibinfo  {journal} {Zeitschrift f{\"{u}}r Krist. - Cryst. Mater.}\ }\textbf
  {\bibinfo {volume} {229}},\ \bibinfo {pages} {345} (\bibinfo {year}
  {2014})}\BibitemShut {NoStop}%
\bibitem [{\citenamefont {Perez-Mato}\ \emph {et~al.}(2015)\citenamefont
  {Perez-Mato}, \citenamefont {Gallego}, \citenamefont {Tasci}, \citenamefont
  {Elcoro}, \citenamefont {de~la Flor},\ and\ \citenamefont
  {Aroyo}}]{Perez-Mato2015}%
  \BibitemOpen
  \bibfield  {author} {\bibinfo {author} {\bibfnamefont {J.}~\bibnamefont
  {Perez-Mato}}, \bibinfo {author} {\bibfnamefont {S.}~\bibnamefont {Gallego}},
  \bibinfo {author} {\bibfnamefont {E.}~\bibnamefont {Tasci}}, \bibinfo
  {author} {\bibfnamefont {L.}~\bibnamefont {Elcoro}}, \bibinfo {author}
  {\bibfnamefont {G.}~\bibnamefont {de~la Flor}}, \ and\ \bibinfo {author}
  {\bibfnamefont {M.}~\bibnamefont {Aroyo}},\ }\href {\doibase
  10.1146/annurev-matsci-070214-021008} {\bibfield  {journal} {\bibinfo
  {journal} {Annu. Rev. Mater. Res.}\ }\textbf {\bibinfo {volume} {45}},\
  \bibinfo {pages} {217} (\bibinfo {year} {2015})}\BibitemShut {NoStop}%
\bibitem [{\citenamefont {Aroyo}\ \emph {et~al.}(2011)\citenamefont {Aroyo},
  \citenamefont {Perez-Mato}, \citenamefont {Orobengoa}, \citenamefont {Tasci},
  \citenamefont {{De La Flor}},\ and\ \citenamefont {Kirov}}]{Aroyo2011}%
  \BibitemOpen
  \bibfield  {author} {\bibinfo {author} {\bibfnamefont {M.~I.}\ \bibnamefont
  {Aroyo}}, \bibinfo {author} {\bibfnamefont {J.~M.}\ \bibnamefont
  {Perez-Mato}}, \bibinfo {author} {\bibfnamefont {D.}~\bibnamefont
  {Orobengoa}}, \bibinfo {author} {\bibfnamefont {E.}~\bibnamefont {Tasci}},
  \bibinfo {author} {\bibfnamefont {G.}~\bibnamefont {{De La Flor}}}, \ and\
  \bibinfo {author} {\bibfnamefont {A.}~\bibnamefont {Kirov}},\ }\href@noop {}
  {\bibfield  {journal} {\bibinfo  {journal} {Bulg. Chem. Commun.}\ }\textbf
  {\bibinfo {volume} {43}},\ \bibinfo {pages} {183} (\bibinfo {year}
  {2011})}\BibitemShut {NoStop}%
\bibitem [{\citenamefont {Aroyo}\ \emph
  {et~al.}(2006{\natexlab{a}})\citenamefont {Aroyo}, \citenamefont {Kirov},
  \citenamefont {Capillas}, \citenamefont {Perez-Mato},\ and\ \citenamefont
  {Wondratschek}}]{Aroyo2006}%
  \BibitemOpen
  \bibfield  {author} {\bibinfo {author} {\bibfnamefont {M.~I.}\ \bibnamefont
  {Aroyo}}, \bibinfo {author} {\bibfnamefont {A.}~\bibnamefont {Kirov}},
  \bibinfo {author} {\bibfnamefont {C.}~\bibnamefont {Capillas}}, \bibinfo
  {author} {\bibfnamefont {J.~M.}\ \bibnamefont {Perez-Mato}}, \ and\ \bibinfo
  {author} {\bibfnamefont {H.}~\bibnamefont {Wondratschek}},\ }\href {\doibase
  10.1107/S0108767305040286} {\bibfield  {journal} {\bibinfo  {journal} {Acta
  Crystallogr. Sect. A Found. Crystallogr.}\ }\textbf {\bibinfo {volume}
  {62}},\ \bibinfo {pages} {115} (\bibinfo {year}
  {2006}{\natexlab{a}})}\BibitemShut {NoStop}%
\bibitem [{\citenamefont {Aroyo}\ \emph
  {et~al.}(2006{\natexlab{b}})\citenamefont {Aroyo}, \citenamefont
  {Perez-Mato}, \citenamefont {Capillas}, \citenamefont {Kroumova},
  \citenamefont {Ivantchev}, \citenamefont {Madariaga}, \citenamefont {Kirov},\
  and\ \citenamefont {Wondratschek}}]{Aroyo2006a}%
  \BibitemOpen
  \bibfield  {author} {\bibinfo {author} {\bibfnamefont {M.~I.}\ \bibnamefont
  {Aroyo}}, \bibinfo {author} {\bibfnamefont {J.~M.}\ \bibnamefont
  {Perez-Mato}}, \bibinfo {author} {\bibfnamefont {C.}~\bibnamefont
  {Capillas}}, \bibinfo {author} {\bibfnamefont {E.}~\bibnamefont {Kroumova}},
  \bibinfo {author} {\bibfnamefont {S.}~\bibnamefont {Ivantchev}}, \bibinfo
  {author} {\bibfnamefont {G.}~\bibnamefont {Madariaga}}, \bibinfo {author}
  {\bibfnamefont {A.}~\bibnamefont {Kirov}}, \ and\ \bibinfo {author}
  {\bibfnamefont {H.}~\bibnamefont {Wondratschek}},\ }\href {\doibase
  10.1524/zkri.2006.221.1.15} {\bibfield  {journal} {\bibinfo  {journal}
  {Zeitschrift f{\"{u}}r Krist. - Cryst. Mater.}\ }\textbf {\bibinfo {volume}
  {221}},\ \bibinfo {pages} {15} (\bibinfo {year}
  {2006}{\natexlab{b}})}\BibitemShut {NoStop}%
\bibitem [{\citenamefont {Momma}\ and\ \citenamefont
  {Izumi}(2011)}]{Momma2011}%
  \BibitemOpen
  \bibfield  {author} {\bibinfo {author} {\bibfnamefont {K.}~\bibnamefont
  {Momma}}\ and\ \bibinfo {author} {\bibfnamefont {F.}~\bibnamefont {Izumi}},\
  }\href {\doibase 10.1107/S0021889811038970} {\bibfield  {journal} {\bibinfo
  {journal} {J. Appl. Crystallogr.}\ }\textbf {\bibinfo {volume} {44}},\
  \bibinfo {pages} {1272} (\bibinfo {year} {2011})}\BibitemShut {NoStop}%
\bibitem [{\citenamefont {Poulis}\ \emph {et~al.}(1951)\citenamefont {Poulis},
  \citenamefont {van~den Handel}, \citenamefont {Ubbink}, \citenamefont
  {Poulis},\ and\ \citenamefont {Gorter}}]{Poulis1951}%
  \BibitemOpen
  \bibfield  {author} {\bibinfo {author} {\bibfnamefont {N.~J.}\ \bibnamefont
  {Poulis}}, \bibinfo {author} {\bibfnamefont {J.}~\bibnamefont {van~den
  Handel}}, \bibinfo {author} {\bibfnamefont {J.}~\bibnamefont {Ubbink}},
  \bibinfo {author} {\bibfnamefont {J.~A.}\ \bibnamefont {Poulis}}, \ and\
  \bibinfo {author} {\bibfnamefont {C.~J.}\ \bibnamefont {Gorter}},\ }\href
  {\doibase 10.1103/PhysRev.82.552} {\bibfield  {journal} {\bibinfo  {journal}
  {Phys. Rev.}\ }\textbf {\bibinfo {volume} {82}},\ \bibinfo {pages} {552}
  (\bibinfo {year} {1951})}\BibitemShut {NoStop}%
\bibitem [{\citenamefont {Ubbink}\ \emph {et~al.}(1953)\citenamefont {Ubbink},
  \citenamefont {Poulis}, \citenamefont {Gerritsen},\ and\ \citenamefont
  {Gorter}}]{Ubbink1953}%
  \BibitemOpen
  \bibfield  {author} {\bibinfo {author} {\bibfnamefont {J.}~\bibnamefont
  {Ubbink}}, \bibinfo {author} {\bibfnamefont {J.}~\bibnamefont {Poulis}},
  \bibinfo {author} {\bibfnamefont {H.}~\bibnamefont {Gerritsen}}, \ and\
  \bibinfo {author} {\bibfnamefont {C.}~\bibnamefont {Gorter}},\ }\href
  {\doibase 10.1016/S0031-8914(53)80104-7} {\bibfield  {journal} {\bibinfo
  {journal} {Physica}\ }\textbf {\bibinfo {volume} {19}},\ \bibinfo {pages}
  {928} (\bibinfo {year} {1953})}\BibitemShut {NoStop}%
\bibitem [{\citenamefont {Gorter}(1953)}]{Gorter1953}%
  \BibitemOpen
  \bibfield  {author} {\bibinfo {author} {\bibfnamefont {C.~J.}\ \bibnamefont
  {Gorter}},\ }\href {\doibase 10.1103/RevModPhys.25.332} {\bibfield  {journal}
  {\bibinfo  {journal} {Rev. Mod. Phys.}\ }\textbf {\bibinfo {volume} {25}},\
  \bibinfo {pages} {332} (\bibinfo {year} {1953})}\BibitemShut {NoStop}%
\bibitem [{\citenamefont {Shapira}\ and\ \citenamefont
  {Zak}(1968)}]{Shapira1968}%
  \BibitemOpen
  \bibfield  {author} {\bibinfo {author} {\bibfnamefont {Y.}~\bibnamefont
  {Shapira}}\ and\ \bibinfo {author} {\bibfnamefont {J.}~\bibnamefont {Zak}},\
  }\href {\doibase 10.1103/PhysRev.170.503} {\bibfield  {journal} {\bibinfo
  {journal} {Phys. Rev.}\ }\textbf {\bibinfo {volume} {170}},\ \bibinfo {pages}
  {503} (\bibinfo {year} {1968})}\BibitemShut {NoStop}%
\bibitem [{\citenamefont {Poulis}\ and\ \citenamefont
  {Hardeman}(1954)}]{Poulis1954}%
  \BibitemOpen
  \bibfield  {author} {\bibinfo {author} {\bibfnamefont {N.}~\bibnamefont
  {Poulis}}\ and\ \bibinfo {author} {\bibfnamefont {G.}~\bibnamefont
  {Hardeman}},\ }\href {\doibase 10.1016/S0031-8914(54)80004-8} {\bibfield
  {journal} {\bibinfo  {journal} {Physica}\ }\textbf {\bibinfo {volume} {20}},\
  \bibinfo {pages} {7} (\bibinfo {year} {1954})}\BibitemShut {NoStop}%
\bibitem [{\citenamefont {Blazey}\ and\ \citenamefont
  {Rohrer}(1968)}]{Blazey1968}%
  \BibitemOpen
  \bibfield  {author} {\bibinfo {author} {\bibfnamefont {K.~W.}\ \bibnamefont
  {Blazey}}\ and\ \bibinfo {author} {\bibfnamefont {H.}~\bibnamefont
  {Rohrer}},\ }\href {\doibase 10.1103/PhysRev.173.574} {\bibfield  {journal}
  {\bibinfo  {journal} {Phys. Rev.}\ }\textbf {\bibinfo {volume} {173}},\
  \bibinfo {pages} {574} (\bibinfo {year} {1968})}\BibitemShut {NoStop}%
\bibitem [{\citenamefont {Blazey}\ \emph {et~al.}(1971)\citenamefont {Blazey},
  \citenamefont {Rohrer},\ and\ \citenamefont {Webster}}]{Blazey1971}%
  \BibitemOpen
  \bibfield  {author} {\bibinfo {author} {\bibfnamefont {K.~W.}\ \bibnamefont
  {Blazey}}, \bibinfo {author} {\bibfnamefont {H.}~\bibnamefont {Rohrer}}, \
  and\ \bibinfo {author} {\bibfnamefont {R.}~\bibnamefont {Webster}},\ }\href
  {\doibase 10.1103/PhysRevB.4.2287} {\bibfield  {journal} {\bibinfo  {journal}
  {Phys. Rev. B}\ }\textbf {\bibinfo {volume} {4}},\ \bibinfo {pages} {2287}
  (\bibinfo {year} {1971})}\BibitemShut {NoStop}%
\bibitem [{\citenamefont {Shapira}\ and\ \citenamefont
  {Foner}(1970)}]{Shapira1970}%
  \BibitemOpen
  \bibfield  {author} {\bibinfo {author} {\bibfnamefont {Y.}~\bibnamefont
  {Shapira}}\ and\ \bibinfo {author} {\bibfnamefont {S.}~\bibnamefont
  {Foner}},\ }\href {\doibase 10.1103/PhysRevB.1.3083} {\bibfield  {journal}
  {\bibinfo  {journal} {Phys. Rev. B}\ }\textbf {\bibinfo {volume} {1}},\
  \bibinfo {pages} {3083} (\bibinfo {year} {1970})}\BibitemShut {NoStop}%
\bibitem [{\citenamefont {Foner}(1963)}]{Foner1963}%
  \BibitemOpen
  \bibfield  {author} {\bibinfo {author} {\bibfnamefont {S.}~\bibnamefont
  {Foner}},\ }\href {\doibase 10.1103/PhysRev.130.183} {\bibfield  {journal}
  {\bibinfo  {journal} {Phys. Rev.}\ }\textbf {\bibinfo {volume} {130}},\
  \bibinfo {pages} {183} (\bibinfo {year} {1963})}\BibitemShut {NoStop}%
\bibitem [{\citenamefont {Rives}(1967)}]{Rives1967}%
  \BibitemOpen
  \bibfield  {author} {\bibinfo {author} {\bibfnamefont {J.~E.}\ \bibnamefont
  {Rives}},\ }\href {\doibase 10.1103/PhysRev.162.491} {\bibfield  {journal}
  {\bibinfo  {journal} {Phys. Rev.}\ }\textbf {\bibinfo {volume} {162}},\
  \bibinfo {pages} {491} (\bibinfo {year} {1967})}\BibitemShut {NoStop}%
\bibitem [{\citenamefont {Becerra}\ \emph {et~al.}(1988)\citenamefont
  {Becerra}, \citenamefont {Oliveira}, \citenamefont {Paduan-Filho},
  \citenamefont {Figueiredo},\ and\ \citenamefont {Souza}}]{Becerra1988}%
  \BibitemOpen
  \bibfield  {author} {\bibinfo {author} {\bibfnamefont {C.~C.}\ \bibnamefont
  {Becerra}}, \bibinfo {author} {\bibfnamefont {N.~F.}\ \bibnamefont
  {Oliveira}}, \bibinfo {author} {\bibfnamefont {A.}~\bibnamefont
  {Paduan-Filho}}, \bibinfo {author} {\bibfnamefont {W.}~\bibnamefont
  {Figueiredo}}, \ and\ \bibinfo {author} {\bibfnamefont {M.~V.~P.}\
  \bibnamefont {Souza}},\ }\href {\doibase 10.1103/PhysRevB.38.6887} {\bibfield
   {journal} {\bibinfo  {journal} {Phys. Rev. B}\ }\textbf {\bibinfo {volume}
  {38}},\ \bibinfo {pages} {6887} (\bibinfo {year} {1988})}\BibitemShut
  {NoStop}%
\bibitem [{\citenamefont {Tsukada}\ \emph {et~al.}(2001)\citenamefont
  {Tsukada}, \citenamefont {Takeya}, \citenamefont {Masuda},\ and\
  \citenamefont {Uchinokura}}]{Tsukada2001}%
  \BibitemOpen
  \bibfield  {author} {\bibinfo {author} {\bibfnamefont {I.}~\bibnamefont
  {Tsukada}}, \bibinfo {author} {\bibfnamefont {J.}~\bibnamefont {Takeya}},
  \bibinfo {author} {\bibfnamefont {T.}~\bibnamefont {Masuda}}, \ and\ \bibinfo
  {author} {\bibfnamefont {K.}~\bibnamefont {Uchinokura}},\ }\href {\doibase
  10.1103/PhysRevLett.87.127203} {\bibfield  {journal} {\bibinfo  {journal}
  {Phys. Rev. Lett.}\ }\textbf {\bibinfo {volume} {87}},\ \bibinfo {pages}
  {127203} (\bibinfo {year} {2001})}\BibitemShut {NoStop}%
\bibitem [{\citenamefont {Zheludev}\ \emph
  {et~al.}(1997{\natexlab{a}})\citenamefont {Zheludev}, \citenamefont {Maslov},
  \citenamefont {Shirane}, \citenamefont {Sasago}, \citenamefont {Koide},
  \citenamefont {Uchinokura}, \citenamefont {Tennant},\ and\ \citenamefont
  {Nagler}}]{Zheludev1997a}%
  \BibitemOpen
  \bibfield  {author} {\bibinfo {author} {\bibfnamefont {A.}~\bibnamefont
  {Zheludev}}, \bibinfo {author} {\bibfnamefont {S.}~\bibnamefont {Maslov}},
  \bibinfo {author} {\bibfnamefont {G.}~\bibnamefont {Shirane}}, \bibinfo
  {author} {\bibfnamefont {Y.}~\bibnamefont {Sasago}}, \bibinfo {author}
  {\bibfnamefont {N.}~\bibnamefont {Koide}}, \bibinfo {author} {\bibfnamefont
  {K.}~\bibnamefont {Uchinokura}}, \bibinfo {author} {\bibfnamefont {D.~A.}\
  \bibnamefont {Tennant}}, \ and\ \bibinfo {author} {\bibfnamefont {S.~E.}\
  \bibnamefont {Nagler}},\ }\href {\doibase 10.1103/PhysRevB.56.14006}
  {\bibfield  {journal} {\bibinfo  {journal} {Phys. Rev. B}\ }\textbf {\bibinfo
  {volume} {56}},\ \bibinfo {pages} {14006} (\bibinfo {year}
  {1997}{\natexlab{a}})}\BibitemShut {NoStop}%
\bibitem [{\citenamefont {Lumsden}\ \emph {et~al.}(2001)\citenamefont
  {Lumsden}, \citenamefont {Sales}, \citenamefont {Mandrus}, \citenamefont
  {Nagler},\ and\ \citenamefont {Thompson}}]{Lumsden2001}%
  \BibitemOpen
  \bibfield  {author} {\bibinfo {author} {\bibfnamefont {M.~D.}\ \bibnamefont
  {Lumsden}}, \bibinfo {author} {\bibfnamefont {B.~C.}\ \bibnamefont {Sales}},
  \bibinfo {author} {\bibfnamefont {D.}~\bibnamefont {Mandrus}}, \bibinfo
  {author} {\bibfnamefont {S.~E.}\ \bibnamefont {Nagler}}, \ and\ \bibinfo
  {author} {\bibfnamefont {J.~R.}\ \bibnamefont {Thompson}},\ }\href {\doibase
  10.1103/PhysRevLett.86.159} {\bibfield  {journal} {\bibinfo  {journal} {Phys.
  Rev. Lett.}\ }\textbf {\bibinfo {volume} {86}},\ \bibinfo {pages} {159}
  (\bibinfo {year} {2001})}\BibitemShut {NoStop}%
\bibitem [{\citenamefont {Zheludev}\ \emph
  {et~al.}(1997{\natexlab{b}})\citenamefont {Zheludev}, \citenamefont {Maslov},
  \citenamefont {Shirane}, \citenamefont {Sasago}, \citenamefont {Koide},\ and\
  \citenamefont {Uchinokura}}]{Zheludev1997}%
  \BibitemOpen
  \bibfield  {author} {\bibinfo {author} {\bibfnamefont {A.}~\bibnamefont
  {Zheludev}}, \bibinfo {author} {\bibfnamefont {S.}~\bibnamefont {Maslov}},
  \bibinfo {author} {\bibfnamefont {G.}~\bibnamefont {Shirane}}, \bibinfo
  {author} {\bibfnamefont {Y.}~\bibnamefont {Sasago}}, \bibinfo {author}
  {\bibfnamefont {N.}~\bibnamefont {Koide}}, \ and\ \bibinfo {author}
  {\bibfnamefont {K.}~\bibnamefont {Uchinokura}},\ }\href {\doibase
  10.1103/PhysRevLett.78.4857} {\bibfield  {journal} {\bibinfo  {journal}
  {Phys. Rev. Lett.}\ }\textbf {\bibinfo {volume} {78}},\ \bibinfo {pages}
  {4857} (\bibinfo {year} {1997}{\natexlab{b}})}\BibitemShut {NoStop}%
\bibitem [{\citenamefont {Bogdanov}\ \emph {et~al.}(2002)\citenamefont
  {Bogdanov}, \citenamefont {R{\"{o}}{\ss}ler}, \citenamefont {Wolf},\ and\
  \citenamefont {M{\"{u}}ller}}]{Bogdanov2002}%
  \BibitemOpen
  \bibfield  {author} {\bibinfo {author} {\bibfnamefont {A.~N.}\ \bibnamefont
  {Bogdanov}}, \bibinfo {author} {\bibfnamefont {U.~K.}\ \bibnamefont
  {R{\"{o}}{\ss}ler}}, \bibinfo {author} {\bibfnamefont {M.}~\bibnamefont
  {Wolf}}, \ and\ \bibinfo {author} {\bibfnamefont {K.-H.}\ \bibnamefont
  {M{\"{u}}ller}},\ }\href {\doibase 10.1103/PhysRevB.66.214410} {\bibfield
  {journal} {\bibinfo  {journal} {Phys. Rev. B}\ }\textbf {\bibinfo {volume}
  {66}},\ \bibinfo {pages} {214410} (\bibinfo {year} {2002})}\BibitemShut
  {NoStop}%
\bibitem [{\citenamefont {Bogdanov}\ \emph {et~al.}(2007)\citenamefont
  {Bogdanov}, \citenamefont {Zhuravlev},\ and\ \citenamefont
  {R{\"{o}}{\ss}ler}}]{Bogdanov2007}%
  \BibitemOpen
  \bibfield  {author} {\bibinfo {author} {\bibfnamefont {A.~N.}\ \bibnamefont
  {Bogdanov}}, \bibinfo {author} {\bibfnamefont {A.~V.}\ \bibnamefont
  {Zhuravlev}}, \ and\ \bibinfo {author} {\bibfnamefont {U.~K.}\ \bibnamefont
  {R{\"{o}}{\ss}ler}},\ }\href {\doibase 10.1103/PhysRevB.75.094425} {\bibfield
   {journal} {\bibinfo  {journal} {Phys. Rev. B}\ }\textbf {\bibinfo {volume}
  {75}},\ \bibinfo {pages} {094425} (\bibinfo {year} {2007})}\BibitemShut
  {NoStop}%
\bibitem [{\citenamefont {Blundell}(2001)}]{Stephen2001}%
  \BibitemOpen
  \bibfield  {author} {\bibinfo {author} {\bibfnamefont {S.}~\bibnamefont
  {Blundell}},\ }\href@noop {} {\emph {\bibinfo {title} {{Magnetism in
  Condensed Matter}}}}\ (\bibinfo  {publisher} {Oxford University Press},\
  \bibinfo {address} {Oxford ; New York},\ \bibinfo {year} {2001})\BibitemShut
  {NoStop}%
\bibitem [{\citenamefont {Iwasaki}\ and\ \citenamefont
  {Morinari}(2018)}]{Iwasaki2018}%
  \BibitemOpen
  \bibfield  {author} {\bibinfo {author} {\bibfnamefont {Y.}~\bibnamefont
  {Iwasaki}}\ and\ \bibinfo {author} {\bibfnamefont {T.}~\bibnamefont
  {Morinari}},\ }\href {\doibase 10.7566/JPSJ.87.033706} {\bibfield  {journal}
  {\bibinfo  {journal} {J. Phys. Soc. Japan}\ }\textbf {\bibinfo {volume}
  {87}},\ \bibinfo {pages} {8} (\bibinfo {year} {2018})}\BibitemShut {NoStop}%
\bibitem [{\citenamefont {Li}(2016)}]{Li2016c}%
  \BibitemOpen
  \bibfield  {author} {\bibinfo {author} {\bibfnamefont {H.-F.}\ \bibnamefont
  {Li}},\ }\href {\doibase 10.1038/npjcompumats.2016.32} {\bibfield  {journal}
  {\bibinfo  {journal} {NPJ Comput. Mater.}\ }\textbf {\bibinfo {volume} {2}},\
  \bibinfo {pages} {16032} (\bibinfo {year} {2016})}\BibitemShut {NoStop}%
\bibitem [{ffz()}]{ffz_mag}%
  \BibitemOpen
  \href@noop {} {}\bibinfo {note} {At $T=1.8$ K, the isothermal magnetization
  increases linearly and reaches 1.194 $\mu_B$ per Eu ion when the applied
  field $H_\perp=5$ T.}\BibitemShut {Stop}%
\bibitem [{\citenamefont {Nagamiya}\ \emph {et~al.}(1955)\citenamefont
  {Nagamiya}, \citenamefont {Yosida},\ and\ \citenamefont
  {Kubo}}]{Nagamiya1955}%
  \BibitemOpen
  \bibfield  {author} {\bibinfo {author} {\bibfnamefont {T.}~\bibnamefont
  {Nagamiya}}, \bibinfo {author} {\bibfnamefont {K.}~\bibnamefont {Yosida}}, \
  and\ \bibinfo {author} {\bibfnamefont {R.}~\bibnamefont {Kubo}},\ }\href
  {\doibase 10.1080/00018735500101154} {\bibfield  {journal} {\bibinfo
  {journal} {Adv. Phys.}\ }\textbf {\bibinfo {volume} {4}},\ \bibinfo {pages}
  {1} (\bibinfo {year} {1955})}\BibitemShut {NoStop}%
\bibitem [{\citenamefont {Rohrer}\ and\ \citenamefont
  {Thomas}(1969)}]{Rohrer1969}%
  \BibitemOpen
  \bibfield  {author} {\bibinfo {author} {\bibfnamefont {H.}~\bibnamefont
  {Rohrer}}\ and\ \bibinfo {author} {\bibfnamefont {H.}~\bibnamefont
  {Thomas}},\ }\href {\doibase 10.1063/1.1657515} {\bibfield  {journal}
  {\bibinfo  {journal} {J. Appl. Phys.}\ }\textbf {\bibinfo {volume} {40}},\
  \bibinfo {pages} {1025} (\bibinfo {year} {1969})}\BibitemShut {NoStop}%
\bibitem [{\citenamefont {Jin}\ \emph {et~al.}(2019)\citenamefont {Jin},
  \citenamefont {Meven}, \citenamefont {Sazonov}, \citenamefont {Demirdis},
  \citenamefont {Su}, \citenamefont {Xiao}, \citenamefont {Bukowski},
  \citenamefont {Nandi},\ and\ \citenamefont {Br{\"{u}}ckel}}]{Jin2019}%
  \BibitemOpen
  \bibfield  {author} {\bibinfo {author} {\bibfnamefont {W.~T.}\ \bibnamefont
  {Jin}}, \bibinfo {author} {\bibfnamefont {M.}~\bibnamefont {Meven}}, \bibinfo
  {author} {\bibfnamefont {A.~P.}\ \bibnamefont {Sazonov}}, \bibinfo {author}
  {\bibfnamefont {S.}~\bibnamefont {Demirdis}}, \bibinfo {author}
  {\bibfnamefont {Y.}~\bibnamefont {Su}}, \bibinfo {author} {\bibfnamefont
  {Y.}~\bibnamefont {Xiao}}, \bibinfo {author} {\bibfnamefont {Z.}~\bibnamefont
  {Bukowski}}, \bibinfo {author} {\bibfnamefont {S.}~\bibnamefont {Nandi}}, \
  and\ \bibinfo {author} {\bibfnamefont {T.}~\bibnamefont {Br{\"{u}}ckel}},\
  }\href {\doibase 10.1103/PhysRevB.99.140402} {\bibfield  {journal} {\bibinfo
  {journal} {Phys. Rev. B}\ }\textbf {\bibinfo {volume} {99}},\ \bibinfo
  {pages} {140402(R)} (\bibinfo {year} {2019})}\BibitemShut {NoStop}%
\bibitem [{\citenamefont {Shang}\ \emph {et~al.}(2013)\citenamefont {Shang},
  \citenamefont {Yang}, \citenamefont {Chen}, \citenamefont {Cornell},
  \citenamefont {Ronning}, \citenamefont {Zhang}, \citenamefont {Jiao},
  \citenamefont {Chen}, \citenamefont {Chen}, \citenamefont {Howard},
  \citenamefont {Dai}, \citenamefont {Thompson}, \citenamefont {Zakhidov},
  \citenamefont {Salamon},\ and\ \citenamefont {Yuan}}]{Shang2013}%
  \BibitemOpen
  \bibfield  {author} {\bibinfo {author} {\bibfnamefont {T.}~\bibnamefont
  {Shang}}, \bibinfo {author} {\bibfnamefont {L.}~\bibnamefont {Yang}},
  \bibinfo {author} {\bibfnamefont {Y.}~\bibnamefont {Chen}}, \bibinfo {author}
  {\bibfnamefont {N.}~\bibnamefont {Cornell}}, \bibinfo {author} {\bibfnamefont
  {F.}~\bibnamefont {Ronning}}, \bibinfo {author} {\bibfnamefont {J.~L.}\
  \bibnamefont {Zhang}}, \bibinfo {author} {\bibfnamefont {L.}~\bibnamefont
  {Jiao}}, \bibinfo {author} {\bibfnamefont {Y.~H.}\ \bibnamefont {Chen}},
  \bibinfo {author} {\bibfnamefont {J.}~\bibnamefont {Chen}}, \bibinfo {author}
  {\bibfnamefont {A.}~\bibnamefont {Howard}}, \bibinfo {author} {\bibfnamefont
  {J.}~\bibnamefont {Dai}}, \bibinfo {author} {\bibfnamefont {J.~D.}\
  \bibnamefont {Thompson}}, \bibinfo {author} {\bibfnamefont {A.}~\bibnamefont
  {Zakhidov}}, \bibinfo {author} {\bibfnamefont {M.~B.}\ \bibnamefont
  {Salamon}}, \ and\ \bibinfo {author} {\bibfnamefont {H.~Q.}\ \bibnamefont
  {Yuan}},\ }\href {\doibase 10.1103/PhysRevB.87.075148} {\bibfield  {journal}
  {\bibinfo  {journal} {Phys. Rev. B}\ }\textbf {\bibinfo {volume} {87}},\
  \bibinfo {pages} {075148} (\bibinfo {year} {2013})}\BibitemShut {NoStop}%
\bibitem [{\citenamefont {Masuda}\ \emph {et~al.}(2020)\citenamefont {Masuda},
  \citenamefont {Sakai}, \citenamefont {Takahashi}, \citenamefont {Yamasaki},
  \citenamefont {Nakao}, \citenamefont {Moyoshi}, \citenamefont {Nakao},
  \citenamefont {Murakami}, \citenamefont {Arima},\ and\ \citenamefont
  {Ishiwata}}]{Masuda2020}%
  \BibitemOpen
  \bibfield  {author} {\bibinfo {author} {\bibfnamefont {H.}~\bibnamefont
  {Masuda}}, \bibinfo {author} {\bibfnamefont {H.}~\bibnamefont {Sakai}},
  \bibinfo {author} {\bibfnamefont {H.}~\bibnamefont {Takahashi}}, \bibinfo
  {author} {\bibfnamefont {Y.}~\bibnamefont {Yamasaki}}, \bibinfo {author}
  {\bibfnamefont {A.}~\bibnamefont {Nakao}}, \bibinfo {author} {\bibfnamefont
  {T.}~\bibnamefont {Moyoshi}}, \bibinfo {author} {\bibfnamefont
  {H.}~\bibnamefont {Nakao}}, \bibinfo {author} {\bibfnamefont
  {Y.}~\bibnamefont {Murakami}}, \bibinfo {author} {\bibfnamefont
  {T.}~\bibnamefont {Arima}}, \ and\ \bibinfo {author} {\bibfnamefont
  {S.}~\bibnamefont {Ishiwata}},\ }\href {\doibase 10.1103/PhysRevB.101.174411}
  {\bibfield  {journal} {\bibinfo  {journal} {Phys. Rev. B}\ }\textbf {\bibinfo
  {volume} {101}},\ \bibinfo {pages} {174411} (\bibinfo {year}
  {2020})}\BibitemShut {NoStop}%
\bibitem [{\citenamefont {Qureshi}(2019)}]{Qureshi2019}%
  \BibitemOpen
  \bibfield  {author} {\bibinfo {author} {\bibfnamefont {N.}~\bibnamefont
  {Qureshi}},\ }\href {\doibase 10.1107/S1600576718016084} {\bibfield
  {journal} {\bibinfo  {journal} {J. Appl. Crystallogr.}\ }\textbf {\bibinfo
  {volume} {52}},\ \bibinfo {pages} {175} (\bibinfo {year} {2019})}\BibitemShut
  {NoStop}%
\end{thebibliography}%

\end{document}